\documentclass[11pt]{article}
\usepackage{latexsym,amsthm,amssymb,amsfonts,amsbsy,calrsfs,multibox,wasysym,stmaryrd}

\usepackage[dvips]{graphicx,epsfig}

\csname @addtoreset\endcsname{equation}{section}

\def\mso{\mathfrak{so}}
\def\miso{\mathfrak{iso}}

\def\msl{\mathfrak{sl}}

\def\msp{\mathfrak{sp}}

\def\mg{\mathfrak{g}}\def\ms{\mathfrak{s}}
\def\mm{\mathfrak{m}}
\def\mh{\mathfrak{h}}
\def\mD{\mathfrak{D}}
\def\mR{\mathfrak{R}}

\def\mC{\mathfrak{C}}

\def\mW{\mathfrak{W}}
\def\mI{\mathfrak{I}}

\def\mS{\mathfrak{S}}
\def\mT{\mathfrak{T}}

\def\mpe{\mathfrak{p}}
\def\mk{\mathfrak{k}}
\def\ml{\mathfrak{l}}

\def\Real{{\mathbb R}}
\def\Comp{{\mathbb C}}
\def\integ{{\mathbb Z}}
\def\1{1\hspace{-4pt}1}
\def\j1{\widetilde{1\hspace{-4pt}1}}
\def\bec{\begin{center}}
\def\ec{\end{center}}
\def\a{\alpha}

\def\b{\beta}  
\def\c{\gamma} 

\def\d{\delta} 
\def\D{\Delta}
\def\e{\epsilon}

\def\k{\kappa}
\def\l{\lambda}
\def\L{\Lambda}
\def\m{\mu}
\def\n{\nu}
\def\r{\rho}
\def\s{\sigma}
\def\S{\Sigma}

\def\th{\theta} 
\def\Th{\Theta}
\def\x{\xi}

\def\O{\Omega}
\def\o{\omega}

\def\nn{\nonumber}
\newcommand{\eq}[1]{(\ref{#1})}

\def\be{\begin{equation}}
\def\ee{\end{equation}}
\def\bea{\begin{eqnarray}}
\def\eea{\end{eqnarray}}
\def\ba{\begin{array}}
\def\ea{\end{array}}

\def\ft#1#2{{\textstyle{{\scriptstyle #1}
\over {\scriptstyle #2}}}}

\def\ket#1{|#1\rangle}

\usepackage{calc} 

\newsavebox{\carre}
\savebox{\carre}(3,3)[bl]{
\put(0,0){\line(0,1){3}}
\put(0,0){\line(1,0){3}}
\put(3,3){\line(0,-1){3}}
\put(3,3){\line(-1,0){3}}
}

\newsavebox{\carrebold}
\savebox{\carrebold}(3,3)[bl]{
\put(0,0){\line(0,1){3}}
\put(0,0){\line(1,0){3}}
\put(3,3){\line(0,-1){3}}
\put(3,3){\line(-1,0){3}}
\put(0.1,0.1){\line(0,1){2.8}}
\put(0.1,0.1){\line(1,0){2.8}}
\put(2.9,2.9){\line(0,-1){2.8}}
\put(2.9,2.9){\line(-1,0){2.8}}
\put(0.2,0.2){\line(0,1){2.6}}
\put(0.2,0.2){\line(1,0){2.6}}
\put(2.8,2.8){\line(0,-1){2.6}}
\put(2.8,2.8){\line(-1,0){2.6}
\put(-2.7,-2.7){$\blacksquare$}}
}

\newcounter{long}

\newcommand{\YoungU}[1]{
    \setcounter{long}{#1}
    \setcounter{long}{\value{long}*3}
    \begin{picture}(\value{long},3)
        \multiput(0,0)(3,0){#1}{\usebox{\carre}}
    \end{picture}
}

\newcommand{\YoungD}[2]{
    \setcounter{long}{#1}
    \setcounter{long}{\value{long}*3}
    \begin{picture}(\value{long},6)
        \multiput(0,0)(3,0){#1}{\usebox{\carre}}
        \multiput(0,-3)(3,0){#2}{\usebox{\carre}}
    \end{picture}
}

\newcommand{\YoungT}[3]{
    \setcounter{long}{#1}
    \setcounter{long}{\value{long}*3}
    \begin{picture}(\value{long},9)
        \multiput(0,0)(3,0){#1}{\usebox{\carre}}
        \multiput(0,-3)(3,0){#2}{\usebox{\carre}}
        \multiput(0,-6)(3,0){#3}{\usebox{\carre}}
    \end{picture}
 }

\newcommand{\YoungQ}[4]{
    \setcounter{long}{#1}
    \setcounter{long}{\value{long}*3}
    \begin{picture}(\value{long},12)
        \multiput(0,0)(3,0){#1}{\usebox{\carre}}
        \multiput(0,-3)(3,0){#2}{\usebox{\carre}}
        \multiput(0,-6)(3,0){#3}{\usebox{\carre}}
        \multiput(0,-9)(3,0){#4}{\usebox{\carre}}
    \end{picture}
 }

\newcommand{\YoungUbf}[1]{
    \setcounter{long}{#1}
    \setcounter{long}{\value{long}*3}
    \begin{picture}(\value{long},3)
        \multiput(0,0)(3,0){#1}{\usebox{\carrebold}}
    \end{picture}
}

\begin{document}

\begin{center}

\vspace*{30pt}

{\huge\sc Unfolding Mixed-Symmetry Fields in AdS\\and the BMV Conjecture:\\[15pt] I. General Formalism}

\vspace{40pt}

{\large Nicolas Boulanger\footnotemark
\footnotetext{Work supported by a ``Progetto Italia'' fellowship.
F.R.S.-FNRS associate researcher on leave from the
{\it{Service de M\'ecanique et Gravitation,
Universit\'e de Mons-Hainaut, Belgium.}}},
Carlo Iazeolla\footnotemark\footnotetext{Previous address:
{\it{Dipartimento di Fisica, Universit\`{a} di Roma ``Tor Vergata'' and INFN,
Sezione di Roma ``Tor Vergata'', Via
della Ricerca Scientifica 1, 00133 Roma, Italy}}} and Per Sundell}
\\[20pt]
{\it\small Scuola Normale Superiore\\
Piazza dei Cavalieri 7, 56126 Pisa, Italy} \vspace{15pt}


 {\sc\large Abstract}\end{center}

We present some generalities of unfolded on-shell dynamics that are useful in 
analysing the BMV conjecture for mixed-symmetry fields in constantly curved backgrounds. 
In particular we classify the Lorentz-covariant Harish-Chandra modules generated from 
primary Weyl tensors of arbitrary mass and shape, and in backgrounds with general values 
of the cosmological constant. 
We also discuss the unfolded notion of local degrees of freedom in theories with and without gravity and with and without massive deformation parameters, using the language of Weyl zero$\,$-form modules and their duals.

\vspace*{3cm}
\setcounter{page}{1}

\pagebreak
\tableofcontents

\newpage

\section{\sc \large Introduction}\label{sec:Into}

The theory of higher-spin gauge fields has witnessed two major
achievements with Vasiliev's formulation of fully nonlinear field
equations in four space-time dimensions~\cite{Vasiliev:1990en} and
more recently in $D$ space-time dimensions~\cite{Vasiliev:2003ev}.
For a review and further developments, see~\cite{Bekaert:2005vh,Sagnotti:2005ns}.
The equations are invariant under local non-abelian gauge
symmetries based on an infinite-dimensional, higher-spin Lie
algebra containing the anti-de Sitter algebra
$\mathfrak{so}(2,D-1)\,$ as its maximal finite-dimensional
subalgebra. The equations admit a simple exact solution in which
all fields vanish except a flat $\mso(2,D-1)\,$-valued connection.
The classical perturbative expansion around this solution yields
an infinite tower of totally symmetric massless spin-$s$ fields
with $s=0,1,2,3,\dots$. These carry a manifestly unitarizable
representation of the higher-spin algebra given by the tensor
product of two scalar singletons~\cite{Flato:1978qz},
as was initially checked in $D=4$ \cite{Konshtein:1988yg,Konstein:1989ij}
and later examined in the context of higher-spin
gauge theory in various dimensions
in~\cite{Sezgin:2001zs,Sezgin:2001ij,Vasiliev:2004cm,Engquist:2005yt}.

Vasiliev's formulation is manifestly diffeomorphism invariant
without any explicit reference to a metric --- although standard
minimal spin-2 couplings arise (albeit together with exotic
higher-derivative couplings) in the limit in which the
$\mso(2,D-1)\,$-valued part of the higher-spin connection one-form
is treated exactly while its remaining spin $s>2$ components
become weak fields together with all curvature zero$\,$-forms. Its
general covariance is instead incorporated into the principle of
unfolding~\cite{Vasiliev:1988xc,Vasiliev:1988sa,Vasiliev:1992gr}
whereby the concepts of space-time, dynamics and observables are
derived from infinite-dimensional free differential
algebras~\cite{Sullivan,D'Auria:1982nx,D'Auria:1982pm,vanNieuwenhuizen:1982zf}.
Roughly speaking, unfolded dynamics is an inclusion of local
degrees of freedom into topological field theories described
on-shell by flatness conditions on generalized curvatures, and
with the possibility of having infinitely many local zero$\,$-form
observables in the presence of a cosmological
constant~\cite{Sezgin:2005pv,Iazeolla:2007wt}.

Although a set of fully nonlinear equations of motion for
non-abelian totally symmetric gauge fields is now achieved, its
extension to non-abelian mixed-symmetry gauge fields is presently
unknown. Such fields must be considered in flat space-time as soon
as $D\geqslant 6$ and in constantly curved space-time as soon as
$D\geqslant 4$ (unitary massless mixed-symmetry two-row tensor
fields in $AdS_4$ decompose in the flat limit into topological
dittos plus one massless field in $\mathbb{R}^{1,3}\,$).

As far as free tensor gauge fields in flat space-time of dimension
$D\geqslant 4\,$ are concerned, a Lagrangian formulation was
proposed some time ago by Labastida \cite{Labastida:1987kw}. The
fact that the corresponding equations of motion indeed propagate
the proper massless degrees of freedom was understood later by analysing the corresponding generalized Bargmann-Wigner equations for Weyl tensors,
see~\cite{Bekaert:2006ix} for a review and references.

A frame-like equivalent of Labastida's formalism was given
recently by Skvortsov~\cite{Skvortsov:2008vs,Skvortsov:2008sh}.
His unfolded system consists of a set of $p\,$-forms with
$p\geqslant 0$ that are also traceless Lorentz tensors whose
symmetry type is determined by the Young diagram of the massless
field. The $p\,$-forms with fixed $p$ constitute
$\miso(1,D-1)$-modules that are finite-dimensional for $p>0$ and
infinite-dimensional for $p=0\,$. The system contains equations of motion in various form degrees: at degree zero one finds the generalized Bargmann-Wigner equations and in the highest form degree there is an equivalent equation of motion for a Labastida field that follows from a first-order action~\cite{Skvortsov:2008sh}. This action is the direct generalization to arbitrarily-shaped gauge fields of
Vasiliev's first-order action for Fronsdal fields in flat
space~\cite{Vasiliev:1980as}. 

We stress the fact that for the purpose of counting the local degrees of freedom of a gauge theory it is convenient to go from the on-shell gauge fields all the way down the Weyl tensors which in the free limit are made up on-shell entirely out of gauge invariant degrees of freedom. This approach is naturally incorporated into unfolded dynamics where potentials and curvatures are treated on a more equal footing than in the standard approach to field theory, though a completely democratic formulation off-shell leads to a rather radical deviation from the standard field theory.

In the case of anti-de Sitter space-time,
Metsaev~\cite{Metsaev:1995re,Metsaev:1997nj} has given the partially
gauge-fixed equations of motion for tensor fields $\varphi(\L;\Th)$ sitting in Lorentz irreps of arbitrary shapes $\Th$. A remarkable property that he found is that, due to the background curvature, residual gauge symmetries can only arise in one block of $\Th$ at a time, associated to different critical masses, unlike the case of flat spacetime where such residual symmetries arise simultaneously in all blocks in the limit of vanishing mass. He also found that all of these cases are nonunitary except if the gauge symmetry is symmetry is associated to the first block. The on-shell gauge fields
carry the lowest-weight $\mso(2,D-1)$-irreps $\mD(e_{_0};\Th)$, and in the unitary case $e_{_0}=s_{_1}+D-2-h_{_1}$ where $h_{_1}$ $(s_{_1})$ is the
height (width) of the first block of $\Th$.

Alkalaev, Shaynkman and Vasiliev (ASV)~\cite{Alkalaev:2003qv,Alkalaev:2005kw,Alkalaev:2006rw} have since
then proposed unfolded on-shell $\mso(2,D-1)$-modules for the unitary case consisting of a frame-like $h_1$-form $U^{\bf
h_1}(\L;\widehat\Th_{[h_1]})$ sitting in the tensorial
$\mso(2,D-1)$-irrep of shape $\widehat\Th_{[h_1]}\,$ obtained from
$\Th$ by deleting one column from its first block and then adding
one row of length $s_{_1}-1\,$.
ASV also anticipated the existence
of an infinite-dimensional Weyl zero$\,$-form module with
primary zero$\,$-form $C(\L;\overline\Th)$ sitting in the
Lorentz irrep of shape $\overline\Th$ obtained from $\Theta$
by extending the first row of its second block to the width
$s_{_1}$ of its first block, such that the Weyl zero$\,$-forms should be related
to each other by some differential equations giving Bianchi
identities for the expression of higher-spin curvatures in terms
of Weyl tensors and in such a way that a systematic analysis of these
relations would lead to the full unfolded formulation of
higher-spin dynamics for free mixed fields in $AdS_D\,$.

The main purpose of the present paper is to provide the basic algebraic setting for analysing the above proposal using unfolded free-field dynamics. In particular, in Section \ref{Sec:Twadj} we classify the Lorentz-covariant Harish-Chandra modules generated from primary Weyl tensors of arbitrary mass and shape, and in backgrounds with general values of the cosmological constant. In Section \ref{Sec:IntBI} we then discuss the unfolded integration of their Bianchi identities, leading to gauge potentials in various form degrees as well as St\"uckelberg fields.
A corresponding\footnote{At the free level, a given infinite-dimensional Weyl zero$\,$-form module can be integrated in many different ways. We shall work at the level of ``minimal'' unfolded systems whose variables are traceless Lorentz tensors and that do no take into account any Hodge-duality extensions, as discussed in Section \ref{Sec:Schemes}.} set of unfolded equations of motion are derived in a companion paper \cite{Boulanger:2008kw}, from now on referred to as Paper II, by radially reducing Skvortsov's equations using an explicit oscillator realization.

Although the ASV-system has been designed with the purpose of propagating the correct unitary degrees of freedom in $AdS_D\,$, its flat limit is nonetheless subtle in the $h_1$-form
sector~\cite{Alkalaev:2003qv,Alkalaev:2005kw,Alkalaev:2006rw}. On the other hand, on sheer group theoretic grounds, the
conjectured AdS Weyl zero$\,$-form module~\cite{Alkalaev:2003qv} has to decompose
in the flat limit into a direct sum of massless flat-space Weyl zero$\,$-form modules in
accordance with the conjecture of Brink, Metsaev and Vasiliev
(BMV)~\cite{Brink:2000ag}. Indeed, this follows manifestly from
the realization of the Weyl zero$\,$-form module to be given in
Paper II.

The BMV conjecture~\cite{Brink:2000ag} anticipates a
field-theoretic realization of the degrees of freedom in $\mD(e_{_0};\Th)$
in terms of an ``unbroken'' gauge field $\varphi(\L;\Th)$ plus a set of St\"uckelberg fields $\left\{\chi(\L;\Th')\right\}$. The latter break the gauge symmetries associated to the lower blocks of $\Th$ in such a way that the combined system has a smooth flat
limit in which the total number of local degrees of freedom
is conserved and given by the direct sum
$\varphi(\L\!\!=\!0;\Th)\oplus\bigoplus_{\Th'}\chi(\L\!\!=\!0;\Th')$
of irreducible gauge fields in flat spacetime. According to BMV, the set $\{\Th'\}$ should be given by the reduction of the
$\mso(D-1)$-tensor of shape $\Th$ under $\mso(D-2)$ subject to the
condition that there are no reductions made in the block to which the AdS gauge symmetry is associated.

The partially massive nature of mixed-symmetry gauge fields in
$AdS_D$~\cite{Metsaev:1995re,Metsaev:1997nj} and the dimensional
reduction leading to $\{\Th'\}$ suggest that the St\"uckelberg
fields can be incorporated explicitly via a suitable radial
reduction of an unbroken gauge field in $(D+1)$-dimensional flat
ambient space with signature $(2,D-1)\,$. The above procedure is
carried out using the unfolded language in Paper II with the aforementioned result.

The present paper is organized as follows: Section \ref{App:0} contains
some of our basic notation. Section \ref{Sec:FDA} provides some basic
notions of unfolded on-shell dynamics. Section \ref{Sec:line} presents their application to free fields in constantly curved space-times. Here we also spell out our strategy for
unfolding arbitrary tensor fields in $AdS_D$ using codimension one
foliations of Skvortsov's system that we shall then apply in Paper II to prove an unfolded on-shell version of the BMV conjecture.
In Section \ref{Sec:DOF} we present the treatment of local degrees of freedom in unfolded dynamics, that in particular is required in order to define the smoothness of the unfolded BMV limit. Finally come the conclusions in Section \ref{Sec:Conclusions}.

\section{\sc\large{Notation and Conventions}}\label{App:0}

The direct sum of two vector spaces is written as $\mathfrak
A\uplus \mathfrak B\,$. If $\ml$ is a Lie algebra (or more
generally an associative algebra) then the decomposition of an
$\ml$-module $\mR$ under a subalgebra $\mk\subseteq \ml$ is
denoted by $\mR|_{\mk}$. A module $\mathfrak R$ containing an
invariant subspace $\mathfrak I$, an ideal, is said to be either
(i) indecomposable if the complement of $\mI$ is not invariant in
which case one writes $\mathfrak R|_\ml=\mathfrak I\subsetplus
(\mathfrak R/\mathfrak I)\,$; or (ii) decomposable if both
$\mathfrak I$ and $\mathfrak R/\mathfrak I$ are invariant in which
case one writes $\mathfrak R|_\ml=\mathfrak I\oplus (\mathfrak
R/\mathfrak I)\,$.

Infinite-dimensional modules can be presented in many ways depending on how they
are sliced under various subalgebras.
If $\mk\subset \ml$ one refers to finite-dimensional $\mk$-irreps with non-degenerate bilinear forms as $\mk$-types, which we denote by
$\Th_\a$, $\Th_{\a_i}$ \emph{etc.} labeled by indices $\a$, $\a_i$ \emph{etc.}.
Correspondingly, if there exists a slicing $\mR|_\mk$ consisting of $\mk$-types
then we refer to such expansions as an $\mk$-typesetting of $\mR$. In particular,
we refer to finite-dimensional Lorentz-irreps as Lorentz types (that will be
tensorial in this paper). In unfolded dynamics one may view typesetting as local
coordinatizations of infinite-dimensional target spaces for unfolded sigma models.
We set aside issues of topology.

Young diagrams, or row/column-ordered shapes, with $m_i$ cells in the $i$th
row/column, $i=1,\dots,n$ are labeled by $(m_{_0},\dots,m_{n+1})$ and
$[m_{_0},\dots,m_{n+1}]$ where $m_i\geqslant m_{i+1}$ and $m_{_0}:=\infty$ and
$m_{n+1}:=0\,$. We let
$\mathbb P_{\Th}$ denote Young projections on shape $\Th\,$.
We also use the block-notation
\bea \left([s_{_1};h_{_1}],[s_{_2};h_{_2}],...,[s_{_B};h_{_B}]\right)\ :=
(\underbrace{m_1,\cdots,m_{h_1}}_{=s_{_1}},\underbrace{m_{h_1+1},
\dots,m_{h_1+h_2}}_{=s_{_2}}
\dots)\ ,\eea
for a shape with $B$ rectangular blocks of lengths $s_{_I}>s_{_{I+1}}$ and heights
$h_{_I}\geqslant 1$, $I=1,2,...,B\,$.
The space of shapes ${\cal S}$ forms a module,
the Schur module, for the universal Howe-dual algebra $\msl(\infty)\,$,
obtained as a formal limit of $\msl(\nu_\pm)$ acting in the spaces
${\cal S}^\pm_{\nu_\pm}$ of shapes with total height
$p_{_B}:=\sum_{I=1}^B h_{_I}\leqslant\nu_+$
($\msl(D)$-types in symmetric bases)
or widths $s_{_1}\leqslant \nu_-$ (($\msl(D)$-types in anti-symmetric bases).
Extension to traceless Lorentz tensors leads to Howe-dual algebras $\msp(2\nu_+)$ and
$\mso(\nu_-)\,$, with formal limits
$\msp(2\infty)$ and $\mso(2\infty)$, respectively.

The Schur module ${\cal S}$ can be treated \emph{explicitly} by using
``cell operators'' $\beta_{a,(i)}$ and $\bar\beta^{a,(i)}$ defined
(see Paper II) to act faithfully in ${\cal S}$ by removing or adding,
respectively, a cell containing the $\msl(D)$-index $a$ in the $i$th row.
Schematically,
\bea \bar\beta^{a,(i)}(m_1,\dots,m_i,\dots,m_n) & = &
(m_1,\dots,m_i+1,\dots,m_n) \ ,\nn\\[5pt]
\beta_{a,(i)}(m_1,\dots,m_i,\dots,m_n) & = &
(m_1,\dots,m_i-1,\dots,m_n) \ .\nn\eea
Similarly, $\beta_{a,[i]}$ and $\bar\beta^{a,[i]}\,$, respectively, remove and
add an $a$-labeled box in the $i$th column.

We let $\widehat\mg$ denote the real form of $\mso(D+1)$ with metric
$\eta_{AB}={\rm diag}(\s,\eta_{ab})$ where $\s=\pm 1$ and
$\eta_{ab}=(-1,\d_{rs})\,$, and with generators $\widehat M_{AB}$ obeying
the commutation rules
\bea [\widehat M_{AB},\widehat M_{CD}]&=&2i\,\eta_{C[B}\widehat
M_{A]D}-2i\,\eta_{D[B}\widehat M_{A]C}\ .\eea
We let $\mm:=\mso(1,D-1)$ and $\ms:=\mso(D-1)$ denote the ``canonical''
Lorentz and spin subalgebras, respectively, with generators $M_{ab}$ and
$M_{rs}\,$. We let $\mg_\l:=\mm\subsetplus \mpe$ where $\mpe$ is spanned by the
transvections\footnote{We are here abusing a standard terminology
used in the context of symplectic algebras, the only point being to make clear
the distinction between the cases where the generators
$\{P_a\}$ are commuting or not.} obeying
\bea [P_a,P_b]\ =\
i\l^2M_{ab}\ ,\quad [M_{ab},P_c]\ =\ 2i\eta_{c[b}P_{a]}\ .\eea
If $\l^2=0$ then $\mg_\l\cong \miso(1,D-1)$ and if
$\l^2\neq 0$ then $\mg_\l\cong \widehat\mg\,$ with
$\sigma=-\lambda^2/|\lambda^2|\,$,
the isometry algebras of $AdS_D$ ($\s=-1$) and $dS_D$ ($\s=1$) with
radius $L_{\rm AdS}:=L$ and $L_{\rm dS}:=-i\,L\,$, respectively,
where $L:=\lambda^{-1}$ is assumed to be real for $AdS_D$ and
purely imaginary for $dS_D\,$. The $\mg_\l$-valued connection
$\Omega$ and curvature $\cal R\,$ are defined as follows
\bea \O& :=& e+\o  :=  -i(e^a\,P_a+\ft12\, \o^{ab}\,M_{ab})\
,\label{canonicalconnection}\\
{\cal R} & := & d\O+\O^2 \ = \ -i\left[T^a P_a +\ft12
(R^{ab}+\l^2e^a e^b)\,M_{ab}\right]\ ,\\[5pt]
T^a & := & de^a+ \o^a{}_b~e^b \ , \quad R^{ab} \ := \
d\o^{ab}+\o^a{}_c~\o^c{}_b\ ,\eea
and are associated with a cosmological constant $\L=-\frac{(D-1)(D-2)}{2}\,\l^2\,$.
The Lie derivative along a vector field $\xi\,$ is
${\cal L}_{\xi} := d~i_\xi + i_\xi~d$ and we use conventions where the exterior
total derivative $d$ and the inner derivative $i_\xi$ act from the left.
If the frame field $e^a$ is invertible we define the inverse frame
field $\th^a$ by $i_{\th^a} e^b=\eta^{ab}\,$.

We use weak equalities $\approx$ to denote equations that hold on the constraints
surface.
In the maximally symmetric backgrounds ${\cal R}\approx 0\,$
the connection $\Omega$ can be frozen to a fixed background value,
breaking the diffeomorphisms down to isometries
$\d_{\e(\x)}$ with Killing parameters $\e(\x)=i_\xi(e+\o)$ obeying
$\d_{\e(\x)}(e+\o)\approx {\cal L}_\x(e+\o)=0$
(one has ${\cal L}_\x e^a=\d_{\e(\x)}e^a+i_\x T^a$ where
$\d_{\e(\x)}e^a=\nabla \e^a-\e^{ab}e_b$ with $\e^a=i_\x e^a\,$,
$\e^{ab}=i_\x \o^{ab}\,$ and $\nabla:=d- \frac{i}{2}\,\omega^{ab}M_{ab}\,$).

We use
$\mD^{\pm}(\pm e_{_0};\Th_{_0})$ to denote lowest-weight ($+$) and
highest-weight ($-$) modules of $\mg_\l$ that are sliced under its maximal compact
subalgebra $\mh\cong\mso(2)\oplus\mso(D-1)$ into $\mh$-types
$\ket{e;\th}^{\pm}\,$. In compact basis, the $\mso(2,D-1)$ algebra reads
\bea M_{0r}&=& \ft12(L^+_r+L^-_r)\ ,\quad P_r\ =\ \ft{i\l}2(L^+_r-L^-_r)\ ,
\quad E\ =\ \l^{-1} P_0\ ,\label{mp}\\[5pt]
[L^-_r,L^+_s]&=& 2iM_{rs} +2\d_{rs} E\ ,\quad [E,L^\pm_r]\ =\ \pm
L^\pm_r\ , \quad [M_{rs},L^\pm_t]\ =\ 2i\delta_{t[s}L^\pm_{r]}\
.\label{algd-1}\eea
By their definition, the modules $\mD^{\pm}(\pm e_{_0};\Th_{_0})$ are the irreps
obtained by factoring out all proper ideals in the generalized Verma module
generated from a unique lowest-energy ($+)$ or highest-energy ($-$) state
$\ket{\pm e_{_0};\Th_{_0}}^{\pm}$ with $E$-eigenvalue $\pm e_{_0}\,$.
We let $\mD( e_{_0};\Th_{_0}):= \mD^+( e_{_0};\Th_{_0})$ and
$\ket{e;\th}:=\ket{e;\th}^+$. The generalized Verma module is irreducible for
generic values of $e_{_0}$, \emph{i.e.} singular vectors arise only for certain
critical values related to $\Th_{_0}$.

In unfolded field theory the mass-square $M^2$ of an unfolded Lorentz tensor field
$\phi(\Th)$ (dynamical field, Weyl tensor, ...) carrying an $\mg_\l$-irrep
($\L\neq 0$) with representation $\rho$, is the eigenvalue of
\bea -\rho(P^a P_a)\ \equiv \ \l^2
\rho(\frac12M_{AB}M^{AB}-\frac12M_{ab}M^{ab})\ .\eea
In the case of $\L<0$ one sometimes deals with harmonic expansions involving
 lowest-weight spaces where
\bea C_{_2}[\mg_\l|\mD(e_{_0};\Th_{_0})]&=&e_{_0}[e_{_0}-2(\e_{_0}+1)]+
C_{_2}[\ms|\Th_{_0}]\
,\quad \ms\ :=\ \mso(D-1)\ ,\quad \e_{_0}~:=~\frac12 (D-3)\ \quad\eea
leading to the mass formula
\bea L^2 M{}^2&=& e_{_0}[e_{_0}-2(1+\e_{_0})]+C_{_2}[\ms|\Th_{_0}]-C_{_2}[\mm|\Th]\ .
\label{masses}\eea

We let ${\cal T}^\pm_{(i)}(\Th^\pm)$ denote $\miso(1,D-1)$-irreps with (a) largest
and smallest $\mm$-types $\Th^+$ and $\Th^-$, respectively; and (b) translations
represented by $\rho^+_{(i)}(P_a)=\beta_{a,(i)}$ and
$\rho^-_{(i)}(P_a)=\bar\c^{a,(i)}\,$ (the trace-corrected cell creation operator) for
fixed $i\geqslant 1\,$.
As a special case ${\cal T}^-_{(1)}(\Th^-)\cong {\cal T}^\ast(\L\!\!=\!0;\overline M{}^2\!\!=\!0;\Th^-)\,$,
the dual of the twisted-adjoint representation containing a strictly massless
primary Weyl tensor\footnote{In a similar context, see also the very
recent work~\cite{Alkalaev:2008gi}.}.
We also let ${\cal T}^\pm_{(0)}(\Th):=\Th$, the irrep consisting of a single 
$\mm$-type $\Th$ annihilated by $P_a\,$.

The translations are nilpotent in ${\cal T}^\pm_{(i)}(\Th^\pm)$ for $i\geqslant 2$
and in ${\cal T}^+_{(1)}(\Th^+)$. Factoring out ideals yields ``cut''
finite-dimensional modules ${\cal T}^\pm_{(i),N}(\Th^\pm)$ of ``depth''
$N\geqslant 0$ such that $\left(\rho^{\pm}_{(i),N}(P_a)\right)^{n}\equiv\!\!\!\!\!/\ 0$
iff $n\leqslant N\,$. For $i\geqslant 2$ the duals $\left({\cal T}^\pm_{(i)}(\Th^\pm)\right)^\ast\cong{\cal T}^\mp_{(i),N}(\Th^{\prime \mp})$
for some $N$ and $\Th^{\prime \mp}$ determined from the shape of $\Th^\pm\,$.
In particular,
$({\cal T}^\pm_{(i)}(\Th^\pm))^\ast\cong {\cal T}^\mp_{(i)}(\Th^\mp)$
iff the $i$th row does not form a block of its own in $\Th^+$ nor $\Th^-\,$.

The $\miso(1,D-1)$-irreps ${\cal T}^\pm_{(i)}(\Th^\pm)$ with $i\geqslant 2$
and ${\cal T}^+_{(1)}(\Th^+)$ are contractions of
$\mso(2,D-1)$-types as follows: the $\mso(2,D-1)$-type
$\widehat\Th$ with its canonical representation $\widehat M_{AB}$ is
isomorphic to twisted representations $\widehat\Th^\pm_{(i),\k;\l}$
with canonical $\rho^\pm_{(i),\k;\l}(M_{ab}):=\widehat M_{ab}$ and non-canonical
$\rho^+_{(i),\k;\l}(P_a):=\l\,\widehat\x^B \widehat M_{Ba}+\k\,\beta_{a,(i)}$ and
$\rho^-_{(i),\k;\l}(P_a):=\l\,\widehat\x^B \widehat M_{Ba}+\k\,\bar\gamma_{a,(i)}$
where $\widehat\x^2=-1$ (these are representations for $[P_a,P_b]=i
\l^2 M_{ab}$ for all values of $\k$, $\l$ and $i$). The limit
$\l\rightarrow 0$ at fixed $\k$ yields a reducible $\miso(1,D-1)$
representation that decomposes into
${\cal T}^{\pm}_{(i)}$-plets if $\k\neq 0$ and ${\cal
T}^{+}_{(0)}$-plets if $\k=0\,$.

\section{\sc \large Generalities of Unfolded On-shell Dynamics}
\label{Sec:FDA}

\subsection{\sc Preamble: Free differential algebras and unfolded dynamics}

The notion of unfolded dynamics was introduced by 
Vasiliev \cite{Vasiliev:1988xc,Vasiliev:1988sa,Vasiliev:1992gr} 
who realized that the full dynamics of general gauge theories can be cast into a free 
differential algebra for locally defined variables of form degree $p\geqslant0$, 
including infinite towers of ``twisted-adjoint'' zero$\,$-forms.

The topological usage of free differential algebras dates back to the works of Cartan 
and de Rham, and also of Chevalley and Eilenberg \cite{ChevalEilen} who studied 
equivariant cohomologies on manifolds carrying actions of Lie groups, leading to the 
notion of Chevalley--Eilenberg cocycles that will be important in what follows. The usage 
of cocycles to probe more general topological spaces was then developed by Sullivan 
\cite{Sullivan}, leading to the notion of the Sullivan map $X\mapsto S(X)$ sending a 
topological manifold $X$ to a free differential algebra $S(X)$ in strictly positive form 
degree whose cocycles capture various topological features of $X$. Moreover, as a lemma, 
Sullivan characterized a free differential algebra in strictly positive degree (and with 
each degree being finite-dimensional) as being the semi-direct sum of a ``minimal'' 
algebra with nonlinear cocycles, and a ``contractible'' ideal with linear cocycles.

Sullivan's ideas were then transformed to fit the context of field theories where 
(i) locally defined variables carry local degrees of freedom; (ii) the local translations 
can be softly broken leading to the identification of the vielbein as the soldering 
one-form; and (iii) observables are given by integrals of globally defined, conserved, 
composite variables. An initial step was taken by d'Auria, Fre and van Nieuwenhuizen 
\cite{vanNieuwenhuizen:1982zf,D'Auria:1982nx,D'Auria:1982my,D'Auria:1982pm} 
who adapted free differential algebras to supergravities albeit within a hybrid context, 
aimed at the standard Lagrangian formulation of lower-derivative theories, without 
infinite towers of zero$\,$-forms, and where not all on-shell properties of the dynamics 
are encoded into the algebra.

Later Vasiliev realized that by abandoning the standard Lagrangian formulation and 
introducing infinite towers of zero$\,$-forms all the on-shell properties of general 
(possibly higher-derivative) field theories (with local degrees of freedom) can be 
systematically encoded into first-order equations of motion on universal base manifolds 
(containing standard spacetime as a submanifold). The resulting unfolded dynamics thus 
consists of two ``layers'' of free differential algebras: fundamental algebras consisting 
of locally defined $p$-form variables with $p\geqslant 0$ (including infinite towers of 
zero$\,$-forms) constituting the independent fields; and observable algebras consisting 
of globally defined, composite $p$-forms with $p\geqslant 0$ (possibly also including 
infinite towers of zero$\,$-forms) whose integrals over cycles constitute conserves 
charges that can be used as observables in (noncompact) gauge theory.

\subsection{\sc Overview}\label{Sec:Preamble}

Free differential algebras are sets $\{X^\a\}$ of \emph{a priori} independent
variables that are differential forms obeying first-order equations
of motion whereby $dX^\a$ are equated on-shell to algebraic functions, say
$-Q^\a(X)$, of all the variables expressed entirely using the exterior
algebra, \emph{viz.} $R^\a:=dX^\a+Q^\a(X)\approx 0\,$ with
$Q^\a(X):=\sum_{n}Q^\a_{\b_1\dots \b_n}X^{\b_1}\wedge \cdots
\wedge X^{\b_n}$ (the symbol $\wedge$ will be omitted henceforth
and we use weak equalities for equations that hold on-shell). The
nilpotency of $d$ requires $Q:=Q^\a\partial_\a$, an odd vector
field of degree one on the space of differential forms, to be
nilpotent, that is $Q^2\equiv 0$ or $Q^\b\partial_\b Q^\a\equiv
0\,$. As a result, the constraint surface $\{R^\a\approx 0\}$ is
left invariant under generalized gauge transformations $\d_\e
X^\a=d\e^\a-\e^\b\partial_\b Q^\a$.

The $Q\,$-$\,$cohomology~\cite{Vasiliev:2005zu} is related to a special class of
gauge-invariant charges, namely integrals of algebraic functions ${\cal C}[X]$ that
are on-shell closed, that is $d{\cal C}\approx 0$, \emph{and} globally
defined on the base manifold. Exact zero$\,$-form charges have been
given~\cite{Sezgin:2005pv,Iazeolla:2007wt} for higher-spin gauge
theories. These charges are non-local on-shell, \emph{i.e.} functions on the
infinite jet space of the physical on-shell fields. Their
existence depends crucially on the presence of a massive deformation parameter (the
cosmological constant $\L$ in the case of higher-spin gauge theory). Zero-form
charges are, roughly speaking, unfolded analogs of topological vertex operators
(with vanishing conformal weights) in two-dimensional topological theories.

Invariants ${\cal C}$ of form-degree $\geqslant 1$ require a split of the variables
$X^\a$ with form-degrees $p_\a\geqslant 1$ into a generalized vielbein one-form and
a fiber connection. The latter ``gauges'' a subalgebra of the free-differential
gauge algebra that leaves ${\cal C}$ identically invariant and whose parameters can therefore
be taken to be locally defined. The remaining local translations, which do not
leave ${\cal C}$ invariant, are instead ``softly broken'' and converted by the vielbein
into infinitesimal diffeomorphisms along the base manifold, which leave the charges
invariant. This facilitates the geometric realization of non-compact gauge
symmetries in a suitable ``spacetime'' with local properties following algebraic
properties of the free differential algebra.

Another consequence that we shall exploit here is that the \emph{local}
degrees of freedom of a \emph{classical} free differential algebra
are given by the on-shell values of its zero$\,$-form charges. These are gauge-
invariant integration constants
that parameterize the space of field configurations that cannot be
gauged away locally.

Perturbative expansions around classical solutions yield
linearized $Q\,$-structures $\d Q=(\s_{_0})^\a_\b
\d X^{\b}\partial_\a$ where $\d X^{\a}$ are linearized
fields and the background-dependent matrix $(\s_{_0})^\a_\b$ obeys a
non-abelian ``flatness'' condition (see Eq.\eq{BGfe}). The zero$\,$-form charges
for the \emph{free} theory is coordinatized by the integration constants for all
$\d X^{\a}$ of form degree $p_\a=0$ that cannot be gauged away by
means of St\"uckelberg shift symmetries. These zero$\,$-forms constitute a
representation of the free-differential gauge algebra referred to as
the Weyl zero$\,$-form module.

In expansions around maximally symmetric backgrounds with isometry algebras
$\mg\,$,
the Weyl zero$\,$-form module is built from $\mg$-irreps that are
infinite-dimensional for generic masses (including critically massless cases in
backgrounds
with non-vanishing $\L$) in which case we refer to them as twisted-adjoint
$\mg$-modules.
The twisted-adjoint zero$\,$-forms consist of primary Weyl tensors -- such
as scalar fields $\phi$, Faraday tensors $F_{ab}$ and spin-2 Weyl tensors
$C_{ab,cd}$ -- and secondary, or descendant, Weyl tensors given on-shell by
derivatives of the primary Weyl tensors.

In the case of non-vanishing $\L$ there exist special ``subcritical'' masses for
which there arise finite-dimensional (non-unitarizable) Weyl zero$\,$-form modules.
These are topological sectors with finite sets of integration constants. The basic
example is the scalar field $\phi$ obeying $(\nabla^2-M^2)\phi\approx 0$ on the
$D$-sphere minus a point (or some points) where singularities are tolerated. The
harmonic expansion of $\phi$ yields infinite-dimensional $\mso(D+1)$ modules except
for $M^2=-\ell(\ell+D-3)$ ($\ell=0,1,\dots$) where $\phi$ contains the totally
symmetric rank-$\ell$ tensor.

The Weyl zero$\,$-forms obey various Bianchi identities: the primary Weyl tensor
may obey independent primary identities, which requires vanishing mass in flat
space and critical mass if $\L\neq 0$, while the descendant Weyl tensors always
obey secondary identities that follow from either primary identities or the fact that $d^2\equiv 0$. The local integration of Bianchi identities introduces new modules
in form-degrees
$p\geqslant 0$ consisting of (i) dynamical fields in
various ``dual pictures''; (ii) auxiliary fields; (iii) contractible
St\"uckelberg pairs; and (iv) finite-dimensional topological Weyl zero$\,$-forms.

We stress that, besides the zero-modes in the zero$\,$-forms in (iv), the integration does not introduce
any new local on-shell degrees of freedom. Put differently, the dynamical
fields, although being algebraically independent variables on-shell, do actually ``propagate'' local degrees
of freedom only if the unfolded system contains a corresponding twisted-adjoint
infinite-dimensional Weyl zero$\,$-form module.
In particular, a dynamical field is unitarizable only if there
exists an equivariant map from its associated (manifestly Lorentz-covariant)
twisted-adjoint module to a unitary compactly sliced representation of $\mg$.

A dynamical field that is not sourced by a Weyl tensor may be referred to as
``frozen''. Such fields may acquire finite ``expectation values'' that break the
unfolded gauge symmetries, including diffeomorphisms. The basic example is a
background vielbein $e^a$ and Lorentz connection $\o^{ab}$, obeying the manifestly
diffeomorphism invariant ``topological'' spin-2 field equations $T^a\approx 0$ and $R^{ab}+\l^2 e^a e^b\approx 0\,$, that on the one hand imply Einstein's equation for $g_{\m\n}:=e_\m{}^a e_\n{}^b\eta_{ab}$, and on the other hand imply that $C_{ab,cd}\approx 0$. The diffeomorphisms are thus broken spontaneously by finite solutions for $e_\m{}^a$, which is a remark that of course applies equally well to the case where $C_{ab,cd}$ is no longer constrained on-shell. Indeed, although at the expense of introducing a heavier formalism which lies beyond the scope of this paper, it is possible to treat unfolded dynamics perturbatively in a Hamiltonian system with an expansion around the manifestly diffeomorphism invariant ``empty-space'' vacuum $e^a=0$.

The aforementioned linearized $Q\,$-structure $\d Q=\s_{_0} \d X$ extends to a
``triangular'' gauge/Bianchi module consisting of gauge
parameters, fields, curvatures and Bianchi identities, organized
into modules of the gauge algebra and the (nilpotent) algebra
of massive shift symmetries. In maximally symmetric backgrounds
the linearized field content can be assigned an additional
perturbatively defined $\mathbb N$-valued quantum number referred to as the grade,
that essentially counts the number derivatives used to express the auxiliary fields
in terms of the dynamical fields. Correspondingly, the component of $\s_{_0}$ of
lowest grade, namely $(\s_{_0})^-$ of grade $-1$, extends to a nilpotent matrix
$\s^-$ acting on the triangular module, now a bi-graded complex under the action of
$\s^-$ and the Lorentz-covariant exterior derivative $\nabla\,$,
with (grade, degree) given by $(-1,+1)$ and $(0,+1)$, respectively. Remarkably, the $\s^-$-cohomology fetches dynamical fields, equations of motion, corresponding differential gauge parameters and Bianchi/Noether
identities~\cite{Skvortsov:2008vs,Vasiliev:2005zu,Shaynkman:2004vu}.

In what follows we shall exhibit in more detail some of the topics discussed above,
starting with more general background independent features in the present Section
\ref{Sec:FDA}, and pointing to key differences in the behavior of massless fields
in flat versus constantly curved spacetimes in Section \ref{Sec:line} after which we spell out the BMV conjecture in Subsection \ref{Sec:BMV}. We shall then
digress in more detail into the notion of local degrees of freedom in unfolded
dynamics in Section \ref{Sec:DOF}.
%
\subsection{\sc{On-shell sigma-models and $Q\,$-structure}}\label{Sec:Q}
%
The on-shell formulation of unfolded dynamics is in terms of a sigma-model with
worldvolume ${\cal M}$ covered by coordinate charts $U$ and target space $\mR$ coordinatized by $\{X^\a\}_{\alpha\in{\cal S}}$ where
${\cal S}$ is an indexation set.
The sigma-model map $f:{\cal M}\rightarrow \mR$
induces the pull-back
\bea
f^\ast X&:=&\sum_{p\leqslant {\rm dim}\,{\cal M}} X^{\bf p}\ ,\qquad
X^{\bf p}\ :=\
\sum_{\tiny{\ba{c}\\[-12pt]\alpha\in{\cal S}\\[-3pt]p_\a =\, p \ea}}
f^\ast X^\a\Th_\alpha \ \in \ \O^{\bf p}(U)\otimes  {\cal T}(p)\ ,\eea
where $\O^{\bf p}(U)$ is the space of $p\,$-forms on $U$, $\Th_\a$ are types, finite-dimensional vector spaces, and
\bea {\cal T}(p)&:=&\sum_{\tiny{\ba{c}\\[-15pt]\alpha\in{\cal S} \\[-3pt] p_\a =\, p\ea }} \Th_\alpha\ .\label{sumoftypes}\eea
We suppress $f^\ast$ when confusion cannot arise and use the notation
$X^\alpha \equiv X^{\alpha^{{\bf p}_\alpha}}\equiv  X^{{\bf p}_\alpha}(\Th^{\ast\alpha})$.
We use $\approx$ to denote equations that hold on-shell.
The field equations then read
\bea R^\alpha&:=& dX^\alpha+Q^\alpha(X^\beta)\ \approx\ 0\ ,\eea
where $R^\a$ are referred to as the generalized curvatures, and
$Q:=Q^\alpha \frac{\partial}{\partial X^{\alpha}}=Q^\alpha \partial_{\alpha}\,$ is
an identically nilpotent vector field of degree~$1$,
\bea
Q^2& =& \frac12\{Q,Q\}\ =\ \frac12 {\cal L}_Q Q\ \equiv\ 0\ \Leftrightarrow\
Q^\a\partial_\a Q^\b\ \equiv 0\ ,
\eea
referred to as the $Q\,$-structure. In our conventions the exterior derivative $d$ and the vector fields $\partial_\alpha={\partial\over\partial X^\alpha}$
act from the left.

The $Q$-structure in principle contains all the local information about the classical equations of motion in the ``duality picture'' defined by the coordinates $X$. More generally, additional contractible and dual sectors can be added, as we shall discuss below.

The generalized curvatures have the following two key properties:

(i) The generalized Bianchi identities are
\bea Z^\a&:=& dR^\a-R^\b\partial_\b Q^\a\ \equiv\ 0\ .\label{BI1}
\eea
Note that the extended system consisting of the variables
$\{W^I\}:=\{ X^{\alpha} \,, R^{\alpha}\}$ with structure functions
$\{ Q^I \}=\{ Q^{\alpha}(X) \,, Q^{\alpha}_1(X,R)\}\,$,
$Q^{\alpha}_1(X,R):=-R^{\beta}\partial_{\beta}Q^{\alpha}\,$, is consistent.
Put in equation,
\begin{eqnarray}
Q^J\partial_J Q^I\equiv 0\,.
\label{littlelemma}
\end{eqnarray}
In order to verify this identity,
we split the left-hand side into two groups, the first group reproducing
$Q^{\beta} \partial_{\beta} Q^{\alpha}$ which is identically zero by
assumption, the other yielding
$[Q^{\beta} \frac{\partial}{\partial X^{\beta}}
+Q_1^{\beta} \frac{\partial}{\partial R^{\beta}}] \,Q_1^{\alpha}(X,R)\,$.
The latter expression is identically zero by using the
definition of $Q^{\alpha}_1(X,R)$ and by making use of the identity
\begin{eqnarray}
\partial_{\alpha}Q^{\c}\partial_{\c}Q^{\beta}
+(-1)^{\alpha}Q^{\c}\partial_{\c}\partial_{\alpha}Q^{\beta}\equiv 0
\end{eqnarray}
which is obtained upon differentiating
$Q^{\beta} \partial_{\beta} Q^{\alpha}\equiv 0\,$ and where we use the notation $(-1)^\a:=(-1)^{p_\a}$.

(ii) The constraint surface $\Sigma\equiv R^\a\approx 0 $ is invariant under Cartan
gauge transformations
\bea \d_\e X^\a&:=& G^\a\ := \;d\e^\a-\e^\b \partial_\b Q^\a\ ,
\qquad \d_\e R^\a\ =\ (-1)^\b\e^\b R^\c \partial_\c\partial_\b Q^\a\ ,
\label{gauge1}\eea
where the parameters
$\e^\a\in\O^{{\bf p}_\a-\bf 1}(U)\otimes \Th^\a$ ($:= 0$ if ${p}_\a=0\,$).
The closure relation reads
\bea [\d_{\e_1},\d_{\e_2}]X^\a&=& \delta_{\e_{12}} X^\a+(-1)^\c \e_1^\b \e_2^\c
R^\d \partial_\d \partial_\c\partial_\b Q^\a\ ,\qquad \e^{\a}_{12}\ :=\
(-1)^{\b+1}\e_1^\b \e_2^\c \partial_\c \partial_\b Q^\a\ .\eea

The symmetry $G^\alpha(\e,X)=Z^\a(R,X)|_{R^\a\rightarrow\e^\a}$
actually extends to the full ``tower'' of higher Bianchi identities and the ``basement'' of deeper gauge symmetries, both of which are related to one and the same ``triangular'' extension of $(\mR,Q)$ to be described in more detail below.


\subsection{\sc Contractible and dual cycles}\label{Sec:cycles}


Since there is a gauge parameter for each $p\,$-form with $p>0\,$,
all local degrees of freedom of the system are actually contained in
the space of zero$\,$-forms.
A consequence of this basic lemma is that if $(\mR,Q)$ admits a consistent
truncation to $(\mR',Q')$, then the systems are equivalent locally in $U\subset{\cal M}$ provided the complement $\mR\setminus \mR'$ does not contain ``too many'' zero$\,$-forms.

We refer to $\mS=\mR\setminus\mR'$ as a \emph{contractible cycle} if
$\mS$ contains finitely many zero$\,$-forms and if for $Z\in \mS$ and $X'\in\mR'$
it is the case that
\bea
\mbox{contractible cycle}&:& R^{X'}\ =\ dX'+ Q^{X'}(X',Z)\ ,\quad R^{Z}\ = \
dZ+Q^Z(X',Z)\ ,\\[5pt] &&
Q^Z(X',0)\ =\ 0\ .\qquad \eea
Then there exists a consistent truncation of $\mR$ to $\mR'$ in which the elements in $\mS$ are trivialized (\emph{i.e.} setting $Z=0$ is one valid solution to the flatness conditions).
More generally, we refer to $\mS$ as a \emph{perturbatively contractible cycle}
in the background $\stackrel{(0)}{X}{}'$ if
\bea R^{X'}(\stackrel{(0)}{X}{}'+\d X',\d Z)&=& d\d X'
+\d X'\cdot \partial_{X'} Q^{X'}
(\stackrel{(0)}{X}{}',0)+\d Z\cdot \partial_Z Q^{X'}(\stackrel{(0)}{X}{}',0)
+{\cal O}(\d X^2)\,,\quad\label{contractible}\\
[5pt]R^{Z}(\stackrel{(0)}{X}{}'+\d X',\d Z)
&=&d\d Z+ \d Z\cdot \partial_Z Q^Z(\stackrel{(0)}{X}{}',0)+{\cal O}(\d X^2)\ ,
\qquad\eea
which we denote by
\bea \mR&=&\left\{\ba{ll}\mR'\supsetplus \mS&\mbox{if $\partial_Z Q^{X'}(\stackrel{(0)}{X}{}',0)\neq 0$}\ ,\\[5pt]\mR'\oplus \mS&\mbox{if $\partial_Z Q^{X'}(\stackrel{(0)}{X}{}',0)= 0$\ ,}\ea\right. \eea
referred to as indecomposable and reducible contractible cycles, respectively. 
A perturbatively contractible cycle decomposes into
\bea &\mbox{contractible pairs $(\chi,z)$}~:\  d\chi+z~\approx~0\ ,\quad dz~\approx~0\ ;\qquad \mbox{closed forms $y$}~:\  dy\ \approx\ 0\ .&\eea
Contractible pairs and closed forms with degree $p(y)>0$ carry no local degrees of freedom\footnote{Contractible pairs may become non-trivial at the quantum level due to ghost zero-modes.}, while a closed form with $p(y)=0$ carries one local degree of freedom (a constant valued in the type which contains $y$). 

Instead, if
\begin{itemize}
\item[(i)]
$\mR$ contains a subset $\mR'$ that forms a free differential algebra
of its own;
\item[(ii)]
the complement $\widetilde\mR=\mR\setminus\mR'$ contains finitely many
zero$\,$-forms; and
\item[(iii)]
there does \emph{not} exist a consistent truncation from $\mR$ to
$\mR'\,$,
\end{itemize}
then we shall refer to $\widetilde\mR$ as a \emph{dual cycle}.
Thus, if $X'\in \mR'$ and $\widetilde X\in\widetilde\mR$ then
\bea \mbox{dual cycle}&:& R^{X'}\ =\ dX'+ Q^{X'}(X')\ ,\quad
R^{\widetilde X}\ =\ d \widetilde X + Q^{\widetilde X}(X',\widetilde X)\ ,
\label{dualcycle}\\[5pt]
&&Q^{\widetilde X}(X',0)\ \neq\ 0\ ,\eea
which we write as
\bea \mR&=& \mR'\subsetplus \widetilde\mR\ .\eea
In general, a given submodule $\mR'$ can be ``glued'' to several dual cycles 
(see Fig. \ref{Fig:Dualcocycle}). 

If the free differential algebra $(\mR,Q)$ is a nonlinear deformation of a Lie algebra 
$\mg$ and a set of its representations, then its linearization around an $\mg$-invariant 
``vacuum'' equips $\mR$ with the structure of a $\mg$-module, that is, if 
$X=\stackrel{(0)}{X}+\d X$ where $\stackrel{(0)}{X}$ contains a $\mg$-valued 
Maurer--Cartan form in degree $1$, then the fluctuation fields $\d X$ span a $\mg$-module 
isomorphic as a vector space to $\mR$. This $\mg$-module is indecomposable 
(as a Lie-algebra module) if the full $Q$-structure contains dual or indecomposable 
contractible cycles. Indecomposable $\mg$-modules are characterized by ``gluings'' of 
$\mg$-submodules via Chevalley--Eilenberg cocycles. Their existence is determined by the 
properties of $\mg$ and of the $\mg$-submodules. 

In particular, if $\mg$ is semi-simple, as in the case of non-vanishing cosmological 
constant, then any finite-dimensional $\mg$-module is fully reducible, as follows from a 
well-known theorem due to Weyl, or equivalently, from its dual free differential algebra 
version due to Chevalley and Eilenberg \cite{ChevalEilen}. 
This means that two $\mg$-submodules can be glued only if one of them is 
infinite-dimensional.

On the other hand, if $\mg$ is reductive then there are no such restrictions anymore in 
the case of finite-dimensional modules. Indeed, such $\miso(D-1,1)$-cocycles arise in the 
unfolding of mixed-symmetry massless fields in flat spacetime \cite{Skvortsov:2008vs}, as 
we shall discuss in Sections \ref{Sec:444} and \ref{Sec:445}, and implement explicitly 
using oscillators and cell operators in Paper II.

\begin{figure}[!h]
\begin{picture}(120,100)(-11,0)
\put(0,100){\vector(1,0){140}~~~{degree}}\put(0,100){\vector(0,-1){95}}
\put(-10,0){grade}
\put(0,100){\multiput(40,-40)(10,10){4}{\put(0,0){$\bullet$}\put(1,1){\line(0,-1){9.5}}}
\multiput(90,-30)(10,10){3}{\put(0,0){$\bullet$}\put(1,1){\line(0,-1){9.5}}}
\multiput(40,-40)(10,10){3}{\multiput(1,1)(1,0){10}{\line(1,0){.3}}}
\multiput(80,-40)(10,10){3}{\multiput(1,1)(1,0){10}{\line(1,0){.3}}}
\multiput(51,-39)(1,0){30}{\line(1,0){.3}}
\put(41,-39){\line(0,-1){40}}
\put(40,-37){$C$}\put(70,-7){$\varphi_{_1}$}\put(110,-7){$\varphi_{_2}$}
\put(10,-60){$\mC^{\bf 0}~\left\{\ba{l}{}\\[80pt]{}\ea\right.$}
\put(10,-20){$\widetilde\mR_1~\left\{\ba{l}{}\\[40pt]{}\ea\right.$}
\put(120,-20){$\left.\ba{l}{}\\[40pt]{}\ea\right\}~\widetilde\mR_2$}}
\end{picture}
\caption{\small An unfolded module of the form $\mR=\mR'\subsetplus \widetilde\mR_2$ where (i) $\mR'=\mC^{\bf 0}\subsetplus \widetilde\mR_1$ is a submodule consisting of a Weyl zero$\,$-form module $\mC^{\bf 0}$ with primary Weyl tensor $C$ and dual subcycle $\widetilde \mR_1$ (``potential module'') with dynamical field $\varphi_{_1}$; and (ii) $\widetilde\mR_2$ is a dual cycle (``dual potential module'') with dynamical field $\varphi_{_2}$ (``dual potential'). The dashed lines indicate ``gluings'' by non-trivial generators in $\s^-_{_0}$ (see Section \ref{Sec:sminus}) whose existence conditions depend on the nature of the underlying symmetry Lie algebra $\mg$ (see Section \ref{Sec:cycles}).}\label{Fig:Dualcocycle}
\end{figure}


\subsection{\sc Couplings and homotopy Lie algebras}\label{Sec:Linfty}


Expanding the structure functions $Q^\a$ in $X^\b$ yields graded $n$-ary
products, \emph{viz.}\footnote{More formally, the $n$-ary products
$Q_{(n)}\in \mR\otimes\mR^\ast\wedge\cdots\wedge \mR^\ast\; : \;
\mR\wedge\cdots\wedge\mR\rightarrow \mR$.}
\bea Q^\a(X)&=&\sum_{n\geqslant 0} Q^\a_{(n)}(X)\ ,\quad
Q^\a_{(n)}(X)\ =\ \sum_{{p}_{\b_1}+\cdots+{p}_{\b_n}
={p}_\a} Q^\a_{\b_1\dots \b_n} X^{\b_1}\cdots X^{\b_n}\ ,\eea
whose structure coefficients obey generalized graded Jacobi identities
\bea \sum_{n_1+n_2=n} (n_{_2}+1)Q^\b_{[\c_1\dots \c_{n_1}|}
Q^\a_{\b|\c_{n_1+1}\dots \c_n]}&\equiv &0\ ,\eea
defining a homotopy Lie algebra (see e.g. \cite{Lada:1992wc} and references therein
and also \cite{Sullivan} for the classification and topological usage of finite-dimensional free differential algebras containing no zero$\,$-forms).

Splitting $\{X^{\a}\}_{\a\in\cal{S}(\mathfrak{R})}$ into 0-forms $\{\Phi^{\a^{\bf 0}}\}$,
1-forms $\{A^{\a^{\bf 1}}\}$ and higher-degree forms $\{B^{\a^{\bf p}}\}$ with
$p\geqslant 2\,$, assuming that $Q^\a_{(0)}=0$ (\emph{i.e.} that there are
no field-independent $(p_\a+1)$-forms), and expanding the generalized curvatures
to first order in $B^{\a^{\bf p}}$ yields
\bea R^{\a^{\bf 0}}&=&d\Phi^{\a^{\bf 0}}+T^{\a^{\bf 0}}_{\b^{\bf 1}}(\Phi)A^{\b^{\bf 1}}\ ,\\[5pt]
R^{\a^{\bf 1}}&=&dA^{\a^{\bf 1}}+f^{\a^{\bf 1}}_{\b_1^{\bf 1}\b_2^{\bf 1}}(\Phi)A^{\b_1^{\bf 1}}
A^{\b_2^{\bf 1}}+N^{\a^{\bf 1}}_{\b^{\bf 2}}(\Phi)B^{\b^{\bf 2}}\ ,\\[5pt]
R^{\a^{\bf p}}&=&dB^{\a^{\bf p}}+T^{\a^{\bf p}}_{\b^{\bf 1},\c^{\bf p}}(\Phi)A^{\b^{\bf 1}}B^{\c^{\bf p}}+
N^{\a^{\bf p}}_{\b^{\bf p+1}}(\Phi)B^{\b^{\bf p+1}}+\S^{\a^{\bf p}}_{\b_1^{\bf 1}\dots\b^{\bf 1}_{p+1}}(\Phi)
A^{\b_1^{\bf 1}}\cdots
A^{\b_{p+1}^{\bf 1}}\\[5pt]&&+\sum_{q=2}^{p-1}
\S^{\a^{\bf p}}_{\b_1^{\bf 1}\dots\b^{\bf 1}_{p+1-q},\c^{\bf q}}(\Phi)
A^{\b_1^{\bf 1}}\cdots A^{\b_{p+1-q}^{\bf 1}}B^{\a^{\bf q}}+{\cal O}(B^2)\ .\eea
Expanding further in zero$\,$-forms yields
\bea R^{\a^{\bf 0}}&=&D\Phi^{\a^{\bf 0}}+N^{\a^{\bf 0}}_{\b^{\bf 1}} A^{\b^{\bf 1}}
+{\cal O}(A\Phi^2)\label{Ralfa0}\ ,
\\[5pt]
R^{\a^{\bf 1}}&=& F^{\a^{\bf 1}}+N^{\a^{\bf 1}}_{\b^{\bf 2}} B^{\b^{\bf 2}}
+\S^{\a^{\bf 1}}_{\b^{\bf 1}_1 \b^{\bf 1}_2;\c^{\bf 0}} A^{\b^{\bf 1}_1} A^{\b^{\bf 1}_2} \Phi^{\c^{\bf 0}}
+{\cal O}(A^2\Phi^2)+{\cal O}(B\Phi)\ ,\label{Ralfa1}
\\[5pt]
R^{\a^{\bf p}}&=& DB^{\a^{\bf p}}+N^{\a^{\bf p}}_{\b^{\bf p+1}}
B^{\b^{\bf p+1}}+
\S^{\a^{\bf p}}_{\b^{\bf 1}_1\dots \b^{\bf 1}_{p+1}} A^{\b^{\bf 1}_1}\cdots
A^{\b^{\bf 1}_{p+1}}
+\S^{\a_{p}}_{\b^{\bf 1}_1\dots \b^{\bf 1}_{p+1};\c^{\bf 0}}
A^{\b^{\bf 1}_1}\cdots A^{\b^{\bf 1}_{p+1}}\Phi^{\c^{\bf 0}}
\nn\\[5pt]
&&
+\sum_{q=2}^{p-1}
\S^{\a^{\bf p}}_{\b^{\bf 1}_1\dots \b^{\bf 1}_{p+1-q};\c^{\bf q} }
A^{\b^{\bf 1}_1}\cdots A^{\b^{\bf 1}_{p+1-q}} B^{\c^{\bf q}}
+{\cal O}(A^{p+1}\Phi^2,\Phi B,B^2) \ ,\label{Ralfap}\eea
where the $f$ and $\S$-couplings are generalized (integrated) Chevalley--Eilenberg
cocycles, the $N$-couplings represent massive deformations,
and where the Yang-Mills-like constructs
\bea
F^{\a^{\bf 1}}&:=&dA^{\a^{\bf 1}} +
f^{\a^{\bf 1}}_{\b_1^{\bf 1}\b_2^{\bf 1}} A^{\b_1^{\bf 1}} A^{\c_2^{\bf 1}}\ ,
\\[5pt]
D\Phi^{\a^{\bf 0}} &:=& d\Phi^{\a^{\bf 0}} +
(T_{\b^{\bf 1}})^{\a^{\bf 0}}_{\c^{\bf 0}} A^{\b^{\bf 1}} \Phi^{\c^{\bf 0}}\ ,
\\[5pt]
DB^{\a^{\bf p}} &:=& dB^{\a^{\bf p}}+(T_{\b^{\bf 1}})^{\a^{\bf p}}_{\c^{\bf p}}
A^{\b^{\bf 1}} B^{\c^{\bf p}}\ .
\eea
The higher-order couplings in \eq{Ralfa0}-\eq{Ralfap} contain zero$\,$-form
deformations, including deformations by the physical scalar fields, which we shall refer to as generalized curvature couplings. The generalized Jacobi identities imply that
\bea && 2\,f^{\delta^{\bf 1}}_{[\a^{\bf 1} \b^{\bf 1}|}\,
f^{\kappa^{\bf 1}}_{\delta^{\bf 1} |\gamma^{\bf 1}]}
+N^{\kappa^{\bf 1}}_{\a^{\bf 2}} \,f^{\a^{\bf 2}}_{\a^{\bf 1} \b^{\bf 1} \gamma^{\bf 1}}\ =\ 0\ ,\\[5pt]
&&2\,(T_{[\a^{\bf 1}|})^{\a^{\bf 0}}_{\b^{\bf 0}}\, (T_{|\b^{\bf 1}]})^{\b^{\bf 0}}_{\gamma^{\bf 0}}
+f^{\gamma^{\bf 1}}_{\a^{\bf 1} \b^{\bf 1}} (T_{\gamma^{\bf 1}})^{\a^{\bf 0}}_{\gamma^{\bf 0}}
+ N^{\b^{\bf 0}}_{[\a^{\bf 1}}(P_{\b^{\bf 1}]})^{\a^{\bf 0}}_{{\b^{\bf 0}} {\gamma^{\bf 0}}}\ =\ 0\ ,
\\[5pt]
&&2\,(T_{[\a^{\bf 1}|})^{\a^{\bf p}}_{\gamma^{\bf p}} (T_{|\b^{\bf 1}]})^{\gamma^{\bf p}}_{\b^{\bf p}} +
f^{\gamma^{\bf 1}}_{\a^{\bf 1} \b^{\bf 1}} (T_{\gamma^{\bf 1}})^{\a^{\bf p}}_{\beta^{\bf p}}
+ \mbox{$N$-terms}\ =\ 0\ .\eea
If $N^{\a^{\bf 1}}_{\a^{\bf 2}}=0$ and $N^{\a^{\bf 0}}_{\a^{\bf 1}}=0$,
or more generally, if there exists a projector
${\mathbb P}^{\a^{\bf 1}}_{\b^{\bf 1}}$ such that
${\mathbb P}^{\a^{\bf 1}}_{\b^{\bf 1}}\,N^{\b^{\bf 1}}_{\a^{\bf 2}}=0\,$,
${\mathbb P}^{\b^{\bf 1}}_{\a^{\bf 1}}\,N^{\a^{\bf 0}}_{\b^{\bf 1}}=0\,$,
$(\1-{\mathbb P})^{\a^{\bf 1}}_{\b^{\bf 1}} f^{\b^{\bf 1}}_{\gamma^{\bf 1}\delta^{\bf 1}}
{\mathbb P}^{\gamma^{\bf 1}}_{\varepsilon^{\bf 1}}=0\,$ and
${\mathbb P}^{\a^{\bf 1}}_{\b^{\bf 1}} f^{\b^{\bf 1}}_{\gamma^{\bf 1}\delta^{\bf 1}}
(\1-{\mathbb P})^{\gamma^{\bf 1}}_{\varepsilon^{\bf 1}}=0\,$,
then the 1-form
$\widetilde{A}^{\a^{\bf 1}}:={\mathbb P}^{\a^{\bf 1}}_{\b^{\bf 1}} A^{\b^{\bf 1}}\,$,
which we shall refer to as the connection, takes values in a
Lie algebra $\widetilde \mg$ which we shall refer to as the
\emph{gauge Lie algebra}.

We refer to the free differential algebra as $D$-dimensionally
\emph{Riemannian} if: i) $\widetilde \mg\supset \mg=\mm\subsetplus \mpe$
where $\mm\cong \mso(D;\Comp)$ and $\mpe$ are $D$-dimensional transvections; ii) $\mR|_{\mm}$ consists of $\mm$-tensors; and iii) the
$\mm$-valued connection $\o$ occurs in $R^\o$ only via the Riemann tensor
$R:=d\o+\o^2$ and in the remaining generalized curvatures only via the
covariant derivative $\nabla:=d+\omega\,$.
The types $\Th_{\a}\,$ can then be taken to be irreducible Lorentz tensors
which we label by Young diagrams, sometimes referred to as shapes, and we
shall say the $X^\a$ sits in the $\mm$-type $\Th_\a\,$.

%
\subsection{\sc{Triangular gauge-Bianchi module}}\label{Sec:Triang}
%
Repeated exterior differentiations of the Bianchi identity \eq{BI1} yield
an infinite tower of Bianchi identities
\bea Z^\a_{q+1}&:=& dZ^\a_{q}+Q^\a_{q}(\{Z^\b_{q'}\}_{q'=0}^q)\ \equiv\  0\ ,
\qquad q=1,2,\dots\ ,\eea
which we refer to as higher if $q>2$ and where $Z^\a_2:=Z^\a\,$,
$Z^\a_1:=R^\a\,$ and $Z^\a_0:=X^\a\,$.
The structure functions are given by $Q^\a_0:=Q^\a$ together with
the recursion relation
\bea Q^\a_q&=&-\sum_{q'=0}^{q-1} Z^\b_{q'+1}\partial^{q'}_\b Q^\a_{q-1}\ ,
\qquad q=1,2,\dots\ ,
\label{recu1}
\eea
where $\partial^{q'}_\a:=\partial/\partial Z^\a_{q'}$ ($q'\geqslant 0\,$).
By virtue of the reasoning used in order to obtain \eq{littlelemma},
the structure functions obey the integrability conditions
\bea \sum_{q'=0}^{q}Q^\b_{q'}\partial^{q'}_\b Q^\a_{q}&\equiv &0\ .\eea
Indeed,
with $Q^\a(X^\b)=:Q^\a_0(X^\b)$ obeying $Q^\b\partial_\b Q^\a\equiv 0\,$,
the Bianchi identity $dR^\a\equiv R^\b\partial_\b Q^\a\,$ can be
rewritten as $Z^\a_2\equiv 0$ provided $Q^\a_1= -Z^\b_1\partial_\b Q^\a_0\,$.
This function, as we have shown with \eq{littlelemma},
obeys the integrability condition
$(Q^\b_0\partial^0_\b+Q^\b_1\partial^1_\b)Q^\a_1= 0\,$.
Induction implies that
\bea dZ^\a_{q+1}&\equiv& \sum_{q'=0}^q(Z^\b_{q'+1}-Q^\b_{q'})\partial^{q'}_\b
Q^\a_q\ =\ \sum_{q'=0}^q Z^\b_{q'+1}\partial^{q'}_\b Q^\a_q\ ,\eea
amounting to $Z^{\alpha}_{q+2}\equiv 0$ provided that
 $Q^\a_{q+1}= -\sum_{q'=0}^{q} Z^\b_{q'+1} \partial^{q'}_\b Q^\a_q\,$,
which is the recursion formula \eq{recu1}.

\vspace*{.3cm}

\noindent Thus, the tower of Bianchi identities is related to the triangular
$Q\,$-structure $(\mT^+,Q^+)$ coordinatized by variables $W^\a_q$ ($q\geqslant 0$)
in
\bea \mT^+&:=& \bigoplus_{q\in\mathbb{N}} \mR_q\ ,\quad \mR_q\ :=\
\O^{{\bf p}_\a +{\bf q}}(U)\otimes {\cal T}(p)\ ,\eea
where ${\cal T}(p)$ is defined in \eq{sumoftypes}. The odd integrable vector field
\bea Q^+&:=& \sum_{q\in\mathbb{N}}  Q^\a_q\partial^{q}_\a\ ,
\qquad (Q^+)^2\ \equiv\ 0\ ,
\eea
where the structure functions
$Q^\a_q=Q^\a_q(\{W^\b_{q'\leqslant q}\})$ ($q\geqslant 0$)
are given by
\bea Q^\a_q&=& (-1)^q \prod_{q'=1}^q \sum_{q''=1}^{q'} L^{(-1)}_{q''} Q^\a\
=\ (-1)^q P_q(\{\ell_{q'}\}_{q'=1}^q)Q^\a\ ,\eea
where $P_q$ are polynomials in $\ell_q:=L^{(-q)}_q\,$, for
$L^{(n)}_{q}:=W^\a_{q}\partial^{q+n}_\a$ which have Grassmann parity $(-1)^n$
and obey
\bea&& L^{(m)}_q L^{(n)}_{q'}-(-1)^{mn}L^{(n)}_{q'}L^{(m)}_q \ =\
\d_{q',q+m}L^{(m+n)}_{q}-(-1)^{mn}\d_{q,q'+n}L^{(m+n)}_{q'}\ ,\\[5pt] &&L^{(m)}_q
Q^\a\ =\ \d_{q,-m}\ell_{q}Q^\a\ .\eea
In particular, $L^{(-1)}_q \ell_{q-1}=(-1)^{q-1}\ell_{q-1}L^{(-1)}_q+\ell_q\;$,
$\quad\ell_q\ell_{q'}=(-1)^{qq'}\ell_{q'}\ell_q\,$ and one finds
\bea &&P_1=\ell_1\ ,\quad P_2=\ell_2\ ,\quad P_3=\ell_1\ell_2+\ell_3\ ,\quad
P_4=(\ell_2)^2+\ell_4\ ,\label{P1}\\[5pt]
&&P_5=\ell_1((\ell_2)^2+\ell_4)+2\ell_3\ell_2+\ell_5\ ,\quad
P_6=(\ell_2)^3+3\ell_2\ell_4+\ell_6\ ,\\[5pt]
&&P_7=\ell_1((\ell_2)^3+\ell_6)+3\ell_3(\ell_2)^2+3\ell_3\ell_4+\ell_7\ ,\\[5pt]
&&P_8=(\ell_2)^4+6(\ell_2)^2\ell_4+4\ell_2\ell_6+3(\ell_4)^2+\ell_8\ .
\label{P8}\eea
The tower of Bianchi identities arises upon imposing the constraints
\bea W^\a_q\ =\ (d+Q^+)W^\a_{q-1}\ \quad {\rm for}\quad \ q=1,2,\dots\quad
\Rightarrow\quad W^\a_q\ \equiv\ 0\ \mbox{for}\ q=2,3,\dots\ ,\eea
and identifying $W^\a_q=Z^\a_q$.

\vspace*{.3cm}

If $p_\a\geqslant 2$ the Cartan gauge symmetry \eq{gauge1} is
accompanied by reducibility transformations
\begin{eqnarray}
\delta \epsilon^{\alpha}_{q+1} &=&
d \epsilon_{q} + (-1)^{q}\,\epsilon^{\beta}_{q}\,\partial_{\beta}Q^{\alpha}\;,
\quad q=-2\,,-3\,, \ldots, -p_{\alpha}
\label{lingauge4gauge}
\end{eqnarray}
such that
\begin{eqnarray}
\delta_{\e_q} (\delta\epsilon^{\alpha}_{q+1})
\;=\; d\,(\delta {\e_q^{\alpha}})+(-1)^{q}\,
\delta\e_q^{\beta}\;\partial_{\beta}Q^{\alpha}
&=&(-1)^{\beta+1}\e^{\beta}_{q-1}R^{\delta}
\partial_{\delta}\partial_{\beta}Q^{\alpha}
\;\approx \;0\;.
\end{eqnarray}
Note that, in general,
one can write the transformations that leave invariant the constraint
surface $\Sigma$ as well as more shallow gauge orbits, \emph{viz.}
\begin{eqnarray}
\d\e^\a_{q}&=&G^\a_q:= d\e_{q-1}^{\alpha} +
Q^{\alpha}_{q-1}(\e_{q-1},\e_{q},\e_{q+1},\ldots,\e_{_{-1}},\e_{_0})
\,,
\label{gaugeforgauge}
\\
\d_{\e_q} G^\a_{q+1}\ &=&\
 \sum_{q'=q}^0 G^\b_{q'+1}\;T^{q',\a}_{\;q\,,\b}\ ,\qquad q=0,-1,\dots,-p_\a\ ,
 \label{triangcond}
\end{eqnarray}
with parameters
$\e^\a_{q}\in\O^{{\bf p}_\a+{\bf q}}(U)\otimes \Th^{*\alpha}\,$ and where we temporarily
tolerate terms nonlinear in the $\e_{q}$'s.
We use the notation
$\e^\a_0:=X^\a$, $\e^\a_{-1}:=\e^\a$, $G^\a_1:= R^\a$, $G^\a_0:= G^\alpha\,$,
$G^\a_{-p_{\alpha}}\equiv 0\,$.
%
One can show that the structure functions $Q^\a_{q}\,$ with $q<0\,$
are related to those in the Bianchi identities by
\bea
q<0&:& Q^\a_{q}\ =\ Q^\a_{-q}|_{Z^\a_{q'}\rightarrow \e^\a_{-q'}}
\label{shift}
\eea
and that the rotation matrices are explicitly given by
\bea T^{q',\a}_{\;q\,,\b}&=&\partial^{q'}_\b
Q^\a_{q-1}(\{\e_{\tilde{q}}\}_{\tilde{q}=q-1}^0)\ ,\qquad
\partial^{q'}_\a:=\partial/\partial \e^\a_{q'}\;,\quad
q \leqslant q'\leqslant 0\;.
\label{repmatr}
\eea

\vspace*{.3cm}

\noindent In order to demonstrate the above assertions,
let us first summarize the notation ($q\leqslant 0$):
\bea \{ \e^{\a}_q\}_{-q=0}^{\infty} &=&
\{ \e_0^{\a}\,,\e_{-1}^\a\,,\ldots\} = \{ X^\a,\e^\a,\dots \} \ ,
\qquad \d \e^\a_q\ =\ G^\a_{q}\ ,\\[5pt]
\{ G^\a_{q+1}\}_{-q=0}^{\infty} &=&
\{ G^\a_{1}:=R^{\a},G^\a_{0},G^\a_{-1},\ldots \}\,, \quad
G^{\a}_{q+1}:=\ d\e^\a_q+Q^\a_q(\{\e^\b_{q'}\}_{q'=q}^0)\ .\eea
Then, for any given $q\leqslant 0\,$, the $(1-q)$th level of gauge transformations
$\d \e^\a_{q}=G^{\a}_q\,$ must by definition transform
$\{G^\a_{q'}\}_{q'=q+1}^1$ into themselves. This is trivial for
$q'\geqslant (q+2)\,$ since the corresponding $G^\a_{q'}$'s are independent of
$\e^\a_q\,$, while
\bea \d G^\a_{q+1}&=& d(\d \e^\a_q)
+\d\e^\b_{q}\;\partial^{q}_\b Q^\a_q\ =\
dG^\a_q + G^\b_{q}\;\partial^{q}_\b Q^\a_q\ =\
\sum_{q'=q-1}^0(G^\b_{q'+1}-Q^\b_{q'})\partial^{q'}_\b Q^\a_{q-1}
+ G^\b_{q}\partial_\b^{q} Q^\a_q\nn\\[5pt]&=&
G^\b_q(\partial^{q-1}_\b Q^\a_{q-1}+\partial_\b^q Q^\a_q)
-\sum_{q'=q-1}^0 Q^\b_{q'}\;\partial^{q'}_\b Q^\a_{q-1}
+ \sum_{q'=q}^0 G^{\beta}_{q'+1}\,\partial_{\beta}^{q'}Q_{q-1}^{\alpha} \,,
\eea
where the last term can be written
$\sum_{q'=q}^0 G^{\beta}_{q'+1}\,\partial_{\beta}^{q'}Q_{q-1}^{\alpha} $
$ = \sum_{q'=q}^0 G^{\beta}_{q'+1}\, T^{q',\a}_{\;q\,,\b}\,$, with the
definition of the matrix $T$ given in \eq{repmatr}.
Canceling the first two terms requires $Q^\a_{q-1}$ to obey $\partial^{q-1}_\b
Q^\a_{q-1}+\partial_\b^q Q^\a_q=0$ and the integrability condition
$\sum_{q'=q-1}^0 Q^\b_{q'}\partial^{q'}_\b Q^\a_{q-1}=0\,$.
For example, for $q=0$ one has
$\d R^\a= G^\b(\partial^{-1}_\b Q^\a_{-1}+\partial_\b Q^\a)-\sum_{q'=-1}^0
Q^\b_{q'}\partial^{q'}_\b Q^\a_{-1}+R^\b\partial_\b Q^\a_{-1}$
which is admissible iff $Q^\a_{-1}=-\e^\b\partial_\b Q^\a\,$.
For $q\leqslant 0$ this solution generalizes to
\bea Q^\a_{q-1}=-\sum_{q'=q}^{0}\e^\b_{q'-1}\partial^{q'}_\b Q^\a_q\ ,\eea
which we identify as the transformation \eq{shift} of \eq{recu1}.

\noindent Linearizing the expression \eq{gaugeforgauge} in the parameters
$\e^{\alpha}_q$ with $q<0\,$, one recovers \eq{lingauge4gauge}.
%
\subsection{\sc{Foliations}}\label{Sec:Folia}
%
In this subsection, by the symbol
$L$ we mean either $L_{\rm AdS}$ or $L_{\rm dS}\,$.
We consider a Riemannian free differential algebra $\widehat\mR$
with generalized curvatures
\bea \widehat T^{\widehat\a} \ := \ d\widehat
W^{\widehat\a}+\widehat Q^{\widehat\a}(\widehat W)
\eea
over a base
manifold $\widehat{\cal M}$ with a smooth foliation
$i:\widehat{\cal M}\times \Real \rightarrow \widehat{\cal
M}_i\subseteq\widehat{\cal M}$ where $\widehat{\cal M}_i$ is a
region of $\widehat{\cal M}$ foliated with leaves
${\cal M}_L:=i_L(\widehat{{\cal M}}):=i(\widehat{{\cal M}},L)$ of codimension $1$ and
a non-vanishing normal 1-form $N=d\phi\,$, where
$\phi:\widehat{\cal M}_i\rightarrow \Real$ is defined by
$\phi({\cal M}_L)=L\,$. The normal vector field $\x$ is parallel
to $N$ and normalized such that $i_\xi N=1\,$.

Defining ($n\geqslant 0$)
\bea ({\cal L}_\x)^n\widehat W^{\widehat\a}&:=& \widehat U_n^{\widehat\a}
+N\widehat V_n^{\widehat\a}\ ,\qquad
i_\x \widehat U^{\widehat\a}_n \; := \ 0\ =: \;i_\x \widehat V^{\widehat\a}_n\ ,
\\[5pt]
\widehat X^{{\widehat\a}}&:=& \widehat U_{_0}^{\widehat\a}\ ,\quad \widehat
Y^{\widehat\a}\ :=\ \widehat V^{{\widehat\a}}_{_0}\ ,\qquad \widehat
U^{\widehat\a}\ :=\ \widehat U^{\widehat\a}_{_1}\ ,\quad \widehat V^{\widehat\a}\
:=\ \widehat V^{\widehat\a}_{_1}\ , \eea
where $\widehat V^{\widehat\a}_n\equiv0$ if $p_{\widehat\a}=0\,$, it follows that
\bea \widehat U^{\widehat\a}_n&=&({\cal L}_\x)^n\widehat X^{\widehat\a}\ ,\qquad
\widehat V^{\widehat\a}_n\ =\ ({\cal L}_\x)^n\widehat Y^{\widehat\a}\ .\eea
Defining
$\widehat R^{\widehat\a}_n:=(1-Ni_\x)({\cal L}_\x)^n\widehat T^{\widehat\a}$ and
$\widehat S^{\widehat\a}_n:=-i_\x({\cal L}_\x)^n\widehat T^{\widehat\a}\,$,
the constraints take the form
\bea \widehat R^{\widehat\a}_n&=&(d-N{\cal L}_\x)\widehat U^{\widehat\a}_n
+\widehat f^{\widehat\a}_n(\{\widehat U_m\}_{m=0}^n)\ \approx\ 0\ ,
\\[5pt]
\widehat S^{\widehat\a}_n&=&(d-N{\cal L}_\x)\widehat V^{\widehat\a}_n
+\widehat g^{\widehat\a}_n(\{\widehat U_m,\widehat V_m\}_{m=0}^n)\ - \widehat U^{\widehat\a}_{n+1}\approx\ 0
\qquad \mbox{for $\;\;p_{\widehat\a}\geqslant1$}\ ,\eea
where the structure functions are given by
\bea \widehat f^{\widehat\a}_n&:=&
(1-N i_\x)({\cal L}_\x)^n \widehat Q^{\widehat\a}
(\widehat X+N\widehat Y)\ =\ ({\cal L}_\x)^n \widehat Q^{\widehat\a}(\widehat X)\ ,
\\[5pt]
\widehat g^{\widehat\a}_n&:=&
- i_\x({\cal L}_\x)^n \widehat Q^{\widehat\a}
(\widehat X+N\widehat Y)\ =\ -({\cal L}_\x)^n
\left(\widehat Y^{\widehat\b}\partial_{\widehat\b}
\widehat Q^{\widehat\a}(\widehat X)\right)\qquad
\mbox{for $\;\;p_{\widehat\a}\geqslant1$}\ .\eea
Defining $(U^{\widehat\a}_n,V^{\widehat \a}_n;
R^{\widehat\a}_n,S^{\widehat \a}_n):=i_{L}^\ast(\widehat U^{\widehat\a}_n,
\widehat V^{\widehat \a}_n;\widehat R^{\widehat \a}_n,
\widehat S^{\widehat \a}_n)\,$, the reduced constraints read
\bea R^{\widehat\a}_n &=&  dU^{\widehat\a}_n +
\widehat f^{\widehat\a}_n(\{U_m\}_{m=0}^n)\ \approx \ 0\ ,
\\[5pt]
S^{\widehat\a}_n      &=&  dV^{\widehat\a}_n-U^{\widehat\a}_{n+1}+
\widehat g^{\widehat\a}_n(\{U_m,V_m\}_{m=0}^n)\ \approx\ 0
\qquad \mbox{for $\;\;p_{\widehat\a}\geqslant1$}\ .\eea
Define $f^{\widehat\a}(X):=\widehat Q^{\widehat\a}(X)$ and
$g^{\widehat\a}(X,Y):=-Y^{\widehat\b}\partial_{\widehat\b} f^{\widehat\a}(X)\,
$.
The closed subsystem
\bea R^{\widehat\a} &:=&dX^{\widehat\a} + f^{\widehat\a}(X)\ \approx\ 0\ ,
\\[5pt]
S^{\widehat\a}&:=& d Y^{\widehat\a}+g^{\widehat\a}(X,Y)- U^{\widehat\a}\
\approx\ 0\qquad \mbox{for $\;\;p_{\widehat\a}\geqslant1$}\ ,
\\[5pt]
P^{\widehat\a} &:=& d U^{\widehat\a}-g^{\widehat\a}(X, U)\ \approx\ 0\ ,
\eea
contains three sets of zero$\,$-forms, namely $\{\Phi^{\widehat\a^{\bf 0}}\}\,$,
$\{U^{\widehat\a^{\bf 0}}\}=\{i^\ast_{L}{\cal L}_\xi\widehat\Phi^{\widehat\a^{\bf 0}}\}\,$
and $\{Y^{\widehat \a^{\bf 0}}\}=\{i^\ast_L i_\xi \widehat A^{\widehat\a^{\bf 1}}\}\,$.

\noindent An irreducible model may arise from subsidiary constraints on:
\begin{itemize}
\item[i)]
the normal Lie derivatives
\bea U^{\widehat\a}&\approx & -\D^{\widehat\a}(X,Y)\ ,\label{Qconstraints}\eea
where the functions $\D^{\widehat\a}$ thus assign scaling weights to the fields
under rescalings in $L\,$; and
\item[ii)] zero$\,$-forms
\bea \Xi^{R^{\bf 0}}(X^{\widehat{\a}^{\bf 0}},Y^{\widehat{\a}^{\bf 0}})&\approx &0\ ,
\label{Pconstraints}\eea
where $\Xi^{R^{\bf 0}}$ denotes a set of functions.
\end{itemize}
Cartan integrability requires that
\bea d\D^{{\widehat\a}}-g^{\widehat\a}(X,\D)&\equiv&
(R^{\widehat\b}\partial^{(X)}_{\widehat\b}+S^{\widehat\b}
\partial^{(Y)}_{\widehat\b})\D^{\widehat\a}\ ,
\label{DeltaCI}
\\[5pt]
d\Xi^{R^{\bf 0}}&\equiv&(R^{\widehat{\a}^{\bf
0}}\partial_{\widehat{\a}^{\bf 0}}^{(X)} +S^{\widehat{\a}^{\bf
0}}\partial^{(Y)}_{\widehat{\a}^{\bf 0}})\Xi^{R^{\bf 0}}\ .
\label{XiCI}\eea
where the exterior derivatives on the left-hand sides are expanded using the chain
rule.
The former condition ensures the integrability of the constrained curvature
constraint
\bea S^{\widehat\a}|_{U=-\D}&=&dY^{\widehat\a}+\D^{\widehat\a}(X,Y)
+g^{\widehat\a}(X,Y)\ \approx\ 0\ ,\eea
since the $U$-dependent terms in $dS^{\widehat\a}$ cancel separately prior to
imposing (\ref{Qconstraints}). The subsidiary constraints can equivalently be
imposed directly on $\widehat{\cal M}$ as
\bea \left(\widehat U^{\widehat\a},\widehat V^{\widehat\a}\right)&\approx &
\left(\D^{\widehat\a}(\widehat X,\widehat Y),\Upsilon^{\widehat\a}(\widehat X,
\widehat Y)\right)\ ,\quad \Xi^{R^{\bf 0}}(\widehat X^{\widehat{\a}^{\bf 0}},
\widehat Y^{\widehat{\a}_1})\
\approx \ 0\ ,\eea
where the functions $\Upsilon^{\widehat\a}$ can be determined from
$\D^{\widehat\a}$ using Cartan integrability. This is the approach we shall use
below.

There may exist many consistent sets of subsidiary constraints (it is, for example,
always consistent to set the normal Lie derivatives equal to zero). In the case of
free mixed-symmetry fields,
as we shall examine in Paper II,
unitarity ultimately selects non-trivial scaling
weights
$\D^\a(X,Y)\equiv\D_{[p_\a]} X^\a+\mbox{h.o.t.}$ and a subsidiary constraint
$(\widehat\x)^{\widehat\a^{\bf 0}}_{\widehat \b^{\bf 0}}\Phi^{\widehat\b^{\bf 0}}\approx 0$, where $\widehat\x$ is a differential operator in the fiber whose image is an ideal, such that the complement $\Phi^{\widehat{\a}^{\bf 0}}$, in a non-trivial coset,
belongs to the unitarizable partially-massive Weyl zero$\,$-form module.

\section{\sc \large Unfolded free fields in constantly curved spacetimes}\label{Sec:line}

\subsection{\sc Linearization and $\s$-map}

The expansion of the generalized curvatures $R^\a:=dX^\a+Q^\a(X)$
around a consistent background, \emph{viz.}
\bea &&X^\a\ :=\ \stackrel{(0)}{X}{}^\a+\d X^{\a}\ ,\qquad
d\stackrel{(0)}{X}{}^\a+Q^\a(\stackrel{(0)}{X}{}^\b)\ = 0\ ,\eea
yields a linear map $\s_{_0}:\mR_{_0}\rightarrow \mR_{_1}$ with
matrix elements $(\s_{_0} \d X)^\a:=(\sigma_{_0})^\a_\b \d X^{\b}$
defined by
\bea Q^\a(X^\b)\ =\ Q^\a(\stackrel{(0)}{X})+(\s_{_0})^\a_\b \d
X^{\b}+ {\cal O}((\d X)^2)\ , \qquad (\s_{_0})^\a_\b\ :=\
(-1)^{\a\b}\partial_\b Q^\a|_{\stackrel{(0)}{X}}\ .\eea
This map has the expansion
\bea \s_{_0} &=&\sum_{p'\leqslant p+1}(\s_{_0})^{{\bf p}+{\bf
1}}_{{\bf p}'}\ ,\qquad
 (\s_{_0})^{\bf p+\bf 1}_{\bf p'}:\mR^{\bf p '}_{_0}\rightarrow \mR^{\bf p +\bf 1}_{_1}\ ,
\eea
where $(\s_{_0})^{\bf p +\bf 1}_{\bf p+\bf 1}$ are ``massive''
constants, $(\s_{_0})^{\bf p+\bf 1}_{\bf p}$ are related to
representation matrices of the gauge Lie algebra $\widetilde\mg$,
and $(\s_{_0})^{\bf p+\bf 1}_{\bf p'}$ with $p'\leqslant p-1$ are
integrated cocycles of $\widetilde\mg\,$.
%
%
%

The linearized Bianchi identities, constraints and gauge symmetries
can now be written as
\bea q\geqslant 2&:& \d Z_{q}\ :=\ (d+\s_{q-1})\d Z_{q-1}\ \equiv\ 0\ ,\\[5pt]
q=1&:& \d R\ :=\ (d+\s_{_0})\d X\ \approx\ 0\ ,\\[5pt]
q\leqslant 0&:& \d G_q\ :=\ (d+\s_{q-1})\e_{q-1}\ ,\eea
where the maps $\s_q:\mR_q\rightarrow\mR_{q+1}$ ($q\in\integ$) have
the expansions
\bea \s_q&=&\sum_{p'\leqslant p+1}(\s_q)^{\bf p+\bf q+\bf 1}_{\bf
p'+\bf q}\ ,\qquad
 (\s_q)^{\bf p+\bf q+\bf 1}_{\bf p'+\bf q}:\mR^{\bf p'+\bf q}_{q}\rightarrow \mR^{\bf p+\bf q+\bf 1}_{q+1}\ .\eea
%
%
%
The resulting triangular module $\mT$ and extended map
$\s:\mT\rightarrow \mT$ are defined by
\bea \mT&:=&\bigoplus_{q\in\integ} \mR_q\ ,\qquad \s\ :=\
\sum_{q\in\integ}\s_q\ ,\eea
where $\mR_q:=\bigoplus_{p\in\mathbb{N}}\O^{\bf p+\bf q}(U)\otimes {\cal
T}(p)$. The consistency of the linearization procedure implies that
$(d+\s)(d+\s)\equiv0$, that is ($q\in\integ$)
\bea &(d+\s_{q+1})(d+\s_q)\ \equiv\ d\s_q+
\s_{q+1}\s_q+(\s_{q+1}+(-1)^{\s_q}\s_q)d\ \equiv\
0&\\&\Leftrightarrow&\\&\s_q\ \equiv \ (-1)^{q(1+\s_\circ)}\s_{_0}\
,\qquad (\s_q)^{\bf p+\bf q+\bf 1}_{\bf p'+\bf q}\ \equiv\
(-1)^{q(p+p')}(\s_{_0})^{\bf p+\bf 1}_{\bf p'}&,\eea
since the lower identity implies $d\s_q+ \s_{q+1}\s_q\equiv 0$ by
virtue of the background field equations, which can be written as
\bea d\s_{_0}+\left[(-1)^{1+\s_\circ}\s_{_0}\right]\s_{_0}&=&0\
.\label{BGfe}\eea
%
%
%
The maps $(\s_q)^{\bf p+\bf q+\bf 1}_{\bf p+\bf q}=(\s_{_0})^{\bf
p+\bf 1}_{\bf p}$ are actually representations of $\widetilde\mg$,
and the maps $\s_q:\mR_q\rightarrow\mR_{q+1}$ are given in matrix
notation by $(\s_q \d W_q)^\a=(\s_q)_\b^\a \d
W^{\b}_q=(-1)^{q(\a+\b)}(\s_{_0})_\b^\a \d W^{\b}$.


\subsection{\sc Grading and $\s^-$-$\,$cohomology}\label{Sec:sminus}


The perturbative scheme may admit an ordering of the types,
\emph{i.e.} a surjective
$\mathbb N$-grading~\cite{Shaynkman:2004vu,Vasiliev:2005zu,Skvortsov:2008vs}
\bea g&:&\mT\rightarrow \mathbb{N}\ ,\qquad g\left(\mR^{\bf p_\a+\bf
q}_q(\Th_\a)\right)\ =\ g(\a)\ ,\eea
such that $\s_q$ has a grading bounded from below by $-1$, that is
\bea \s_q\ =\ \sum_{k\geqslant -1}\s^{(k)}_q\ ,\qquad g\circ \s^{(k)}_q\
=\ \s^{(k)}_q\circ(g+k)\ , \eea
and consequently $g\circ\s^{(k)}_{q+1}\circ\s^{(k')}_q=
\s^{(k)}_{q+1}\circ\s^{(k')}_q\circ(g+k+k')$. The extended
triangular module can then be arranged into a bi-graded complex
\bea \mT&=&\bigoplus_{{\ba{c}{}\\[-30pt]{\scriptstyle k\in\mathbb N}\\
[-10pt]{\scriptstyle q\in\integ}\ea}}T_{k,q}\ ,\qquad T_{k,q}\ :=\
g^{-1}(k)\cap\mR_q\ =\ \bigoplus_{\a \; | \; g(\a)=k} \mR^{\bf
p_\a+\bf q}_q(\Th_\a)\ ,\eea
in which $\s^\pm:=\sum_{q\in\integ}\s^{(\pm1)}_q$ and
$\widetilde\nabla:=d+\sum_{q\in\integ}\s^{(0)}_q$ act as follows:
\bea \widetilde\nabla&:&T_{k,q}\rightarrow T_{k,q+1}\ ,\qquad \widetilde\nabla^2 + \{\s^+,\s^-\}\ =\ 0\ ,\\[5pt]
\s^\pm&:&T_{k,q}\rightarrow T_{k\pm 1,q+1}\ ,\qquad (\s^-)^2\ =\ 0\
.\eea
Each entry $T_{k,q}$ is a direct sum over contributions from
different degrees, which we write as
\bea T_{k,q}&=&\bigoplus_{p\in\mathbb N}T^{\bf p}_{k,q}\ ,\qquad
T^{\bf p}_{k,q}\
:=\ \bigoplus_{\ba{c}{}\\[-30pt]{\scriptstyle\a}
\\[-10pt]
{\scriptstyle g(\a)=k}\\[-10pt]{\scriptstyle p_\a=p}\ea}
\mR^{\bf p+\bf q}_q(\Th_\a)\ .\eea
The complex $\mT$ decomposes under the action of $\s^-$ into finite chains. The
resulting $\s^-$-cohomology is governed by simple ``counting'' of Lorentz irreps
provided that candidate $\s^-$-trivial pairs are actually connected by nonzero
matrix elements. This holds for massless theories in
flat spacetime while it does not hold in general for critically massless theories
in constantly curved backgrounds (the examples of the unitary massless
$(2,1)$ and $(3,1)$ fields in $AdS_D$ will be presented in Paper II).

The resulting cohomological groups $H^q(\s^-|\mT)$ have the
following meanings~\cite{Shaynkman:2004vu,Vasiliev:2005zu,Skvortsov:2008vs}:
\bea H^{q<0}(\s^-|\mT)&:&\mbox{differential gauge parameters}\ ,\\
[5pt]H^{q=0}(\s^-|\mT)&:&\mbox{dynamical fields}\ ,\\[5pt]
H^{q=1}(\s^-|\mT)&:&\mbox{dynamical field equations}\ ,\\[5pt]
H^{q=2}(\s^-|\mT)&:&\mbox{Noether, or Bianchi, identities}\ ,\eea
where the dynamical fields are thus all the variables in $\mR_{_0}$ that cannot be gauged away by any of the shift symmetries in $({\rm Im} ~\s^-_{-1})\cap \mR_{_0}$ nor eliminated algebraically by solving any of the curvature constraints in $({\rm Im} ~\s^-_{_0})\cap \mR_{_1}$. If the dynamical fields sit in $T_{k,0}$ and their
equations of motion in $T_{k',1}$ then the latter contains up to
$k'-k+1$ derivatives.

We stress that unfolded dynamics distinguishes between the notion of dynamical fields as defined above, and that of local degrees of freedom which we shall outline in Section \ref{Sec:DOF}. Thus, a dynamical field may be ``frozen'', half-flat, and in general share Weyl tensor with dual dynamical fields.

In the application to constantly curved backgrounds the massive St\"uckelberg shift-symmetry generators can be assigned grade $-1\,$, thereby extending the range of
the $g$-grading.
Gauging away the St\"uckelberg fields from $\mR^{\bf 0}_{_0}$ leaves
the Weyl zero$\,$-form $\widetilde\mg$-module
\bea \mC^{\bf 0}_{_0}&:=& \frac{\mR^{\bf 0}_{_0}}{(\s_{_{-1}})^{\bf
0}_{\bf 0}~\mR^{\bf 0}_{_{-1}}}\ .\label{Sberg}\eea
We refer to its elements as the Weyl zero$\,$-forms and
denote them by $X^{\bf 0}\,$. Their constraint $(d+(\s_{_0})^{\bf 1}_{\bf 0})X^{\bf 0}\approx 0$ constitutes a
free differential subalgebra of $\mR$ with associated
triangular module
\bea \mT_{\rm Weyl}&:=& \bigoplus_{q\in\mathbb N} \mC^{\bf q}_q\
.\eea
We refer to the elements of its $\s^-$-$\,$cohomology at degree
$q=0$ as the primary Weyl tensors. In the following,
we shall write $\mC^{\bf 0}$ for $\mC^{\bf 0}_{_0}$.

%

\subsection{\sc Weyl zero$\,$-forms}\label{Sec:Twadj}


\subsubsection{\sc  Twisted-adjoint module and its dual}

In a Riemannian unfolded system (see Section \ref{Sec:Linfty}) the Weyl zero$\,$-form module $\mC^{\bf 0}$ decomposes under $\mg_\l$ into a ``spectrum'' of manifestly $\mm$-covariant $\mg_\l$-modules $\left\{{\cal T}_\ell\right\}$, \emph{viz.}
\bea \left.\mC^{\bf 0}\right|_{\mg_\l}&:=&\bigoplus_{\ell} \mC^{\bf 0}_\ell,\qquad \mC^{\bf 0}_\ell\ :=\ \Omega^{\bf 0}(U)
\otimes {\cal T}_\ell(0)\ ,\label{Weyl0f0rm}\eea
where each ${\cal T}_\ell:= {\cal T}_\ell(0)$ decomposes further under $\mm$ into a basis $\left\{\Th_{\a_r}\right\}_{\a_r\in{\cal S}_\ell}$ consisting of $\mm$-types, that is
\bea \left.{\cal T}_\ell\right|_{\mm}\ :=\ \bigoplus_{\a_r\in{\cal S}_\ell}\Th_{\a_r}\ ,\qquad \rho_\ell(Q)\Th_{{\a_r}}\ :=\ (\rho_\ell(Q))_{\a_r}{}^{\b_s}\Th_{\b_s}\ ,\eea
where $\rho_\ell(Q)$ denotes the representation of $Q\in\mg_\l$ in ${\cal T}_\ell$ and $(\rho_\ell(Q))_{\a_i}{}^{\b_j}$ the representation matrix with respect to the chosen basis. The dual representation
\bea \left.{\cal T}^\ast_\ell\right|_\mm&=&\bigoplus_{\a_r\in{\cal S}_\ell} \Th^{\ast{\a_r}}\ ,\quad \rho^\ast_\ell(Q)\Th^{\ast{\a_r}}\ =\ (\rho^\ast_\ell(Q))^{\a_r}{}_{\b_s} \Th^{\ast\b_s}\ .\eea
is defined by ($S^\ast\in{\cal T}^\ast$, $S\in{\cal T}$)
\bea (\rho^\ast_\ell(Q)S^\ast)S+S^\ast(\rho_\ell(Q)S)\ :=\ 0\quad\Rightarrow\quad (\rho^\ast_\ell(Q))^{\a_r}{}_{\b_s} \ =\ -(\rho_\ell(Q))_{\b_s}{}^{\a_r}\ .\eea
We use the indexation
\bea |\Th^{\ast\a_r}|\ =\ |\overline\Th^\ast|+\a\ ,
\quad \a\in\mathbb N\ ,\ r=1,\dots, n_\a\ ,\quad n_{_0}\ =\ 1\ ,
\quad\mbox{\emph{idem} $\Th_{\a_r}$}\ ,\eea
where $\overline\Th^\ast$ is the type of the primary Weyl tensor corresponding
to $\ell\,$, $|\Th|$ denotes the rank of an $\mm$-type, the subindex $r$ takes into
account degeneracies at fixed rank, and $n_\a\geqslant 0$ for $\a\geqslant 1\,$.
Since $(\rho_\ell(P_a))_{\b_s}{}^{\a_r}$ vanishes if $\a\neq \b\pm 1$ it follows that if $n_\a=0$ for some $\a\geqslant 1$ then $n_{\a'}=0$ for all
$\a'\geqslant \a$ and the module has finite dimension.
We refer to the remaining infinite-dimensional cases as twisted-adjoint modules.

The ${\cal T}_{\ell}$-valued Weyl zero$\,$-form $\mathbf{X}_{\ell}^{\bf 0}:=
\sum_{{\a_r}}{X}_{\ell}^{\bf 0} (\Th^{\ast\a_r}) \Th_{\a_r}$
(from now on we drop the index $\ell$) has vanishing $\mg_\l$-covariant derivative
\bea \mathbf{R}^{\bf 1}&:=& {\cal D} \mathbf{X}^{\bf 0}\ :=\
\left[\nabla+(\s_{_0})^{\bf 1}_{\bf 0}\right]
\mathbf{X}^{\bf 0}\ \approx\ 0\ ,\qquad (\s_{_0})^{\bf 1}_{\bf 0}\ :=\ -ie^a
\r(P_a)\ .\label{unf} \eea
In components ${\cal D} \mathbf{X}^{\bf 0} :=\sum_{\a_r} ({\cal D}
{X}^{\bf 0})(\Th^{\ast\a_r})\Th_{\a_r}$, so that
\bea ({\cal D}X^{\bf 0})( \Th^{\ast{\a_r}})&=& \nabla X^{\bf 0}( \Th^{\ast{\a_r}})+i e^a (\rho^\ast(P_a))^{\a_r}{}_{\b_s} X^{\bf 0}( \Th^{\ast{\b_s}})\ \approx 0\ .\label{0formcomp}\eea
Using Howe-dual notation (see Section \ref{App:0} and Paper II)
the above matrix representation of the transvections on column vectors can be mapped to %
\bea {\cal T}^\ast_D&:=&{\cal T}^\ast\otimes {\cal S}_D\ ,\eea
\emph{i.e.} column vectors with components in ${\cal S}_D$, the Schur module consisting of  $\mm$-types, and decomposed as (suppressing type-indices)
\bea \mbox{symmetric basis}&:& \bar P^{\ast(i)}\ :=\
\bar\c^{(i)}_a\rho^\ast(P^a)\ ,\quad P^\ast_{(i)}\ :=\
\beta^a_{(i)}\rho^\ast(P_a)\ ,\\[5pt]
\mbox{anti-symmetric basis}&:& \bar P^{\ast[i]}\ :=\
\bar\c^{[i]}_a \rho^\ast(P^a)\ ,\quad P^\ast_{[i]}\ :=\
\beta^a_{[i]}\rho^\ast(P_a)\ ,\eea
where $\bar\c^{(i)}_a$ and $\bar\c^{[i]}_a$, respectively, are cell operators adding one cell with $\mm$-index $a$ in the $i$th row and column of a Schur state (and subtracting traces), and $\beta_{(i)}^a$ and $\beta_{[i]}^a$ are dittos removing one cell (which automatically preserves tracelessness).
Assuming the vielbein to be invertible the zero$\,$-form constraint thus splits into the Howe-dual components
\bea \overline\nabla^{(i)} \mathbf{X}^{\bf 0} +i\bar P^{\ast(i)}
\mathbf{X}^{\bf 0} &\approx&0\ ,\quad
\nabla_{(i)} \mathbf{X}^{\bf 0}+i P^\ast_{(i)}\mathbf{X}^{\bf 0} \approx0\ ,\label{0formdual}\eea
where $\overline\nabla^{(i)}=\bar\c^{(i)}_a\nabla^a$ and
$\nabla_{(i)}=\c_{(i)}^a\nabla_a$, and it is understood that now
$\mathbf{X}^{\bf 0}\in \Omega^{\mathbf{0}}(U)\otimes{\cal T}^\ast_D\,$, a column vector with components in
${\cal S}_D\,$.

\subsubsection{\sc Bargmann-Wigner equations}

In what follows we consider $\mg_\l$-modules
${\cal T}(\L;\overline M{}^2;\overline \Th)$ --- referred to as smallest
$\mm$-type $\mg_\l$-modules or simply smallest-type spaces when there
is no ambiguity --- whose duals
\bea {\cal T}^\ast&:= & \frac{{\cal V}^\ast}{{\cal B}^\ast}\ ,\eea
where ${\cal V}^\ast$ and ${\cal B}^\ast$ are the generalized Harish-Chandra
modules defined by
\bea {\cal V}^\ast &= &\left\{~(\sum\prod \bar P^\ast)~
\overline\Th^\ast~\right\}\ \supset\ {\cal B}^\ast \ =\
\left\{~(\sum\prod \bar P^\ast)~ \mathbb B^\ast~\overline\Th^\ast\right\}\ ,
\label{calVast}\eea
generated from the primary $\mm$-type $\overline \Th^\ast$ obeying
\bea P^\ast_{(i)}\overline\Th^\ast&\approx &0\ ,\quad \left(\rho^\ast(P^a P_a)+\overline M{}^2\right)\overline \Th^\ast~\approx~0\ ,\label{primconstr}
\eea
and a set of \emph{primary-Bianchi} singular vectors
\bea \mbox{symmetric basis}&:& \mathbb B^{\ast+} ~\overline\Th^\ast\ :=\
\mathbb B^+(\bar P^{\ast(j)}) ~\overline\Th^\ast\ ,\\[5pt]
\mbox{anti-symmetric basis}&:&  \mathbb B^{\ast-} ~\overline\Th^\ast\ :=\
\mathbb B^-(\bar P^{\ast[j]}) ~\overline\Th^\ast\ ,\eea
where $\left\{\mathbb B^\pm\right\}$ are monomials obeying the consistency
conditions
\bea P^\ast_{(i)}~\mathbb B^{\ast+}~\overline\Th^\ast
  \ \approx\ & 0 & \approx \
P^\ast_{[i]}~
\mathbb B^{\ast-}~\overline\Th^\ast .\label{primBIcons}\eea
In \eq{calVast} the transvections $\rho^\ast(P^a)$ by definition  act freely on
$\overline\Th^\ast$ subject only to the commutation rules and the primary
constraints \eq{primconstr}. The resulting bases elements are then embedded into
the $\mm$-invariant subspace
\bea {\cal V}^{\ast}_{\rm diag}&:=& \left\{ ~V^\ast~\in~{\cal V}^\ast_D~:\
(\rho^\ast(M_{ab})+\widehat M_{ab})V^\ast\ =\ 0~\right\}\ ,
\quad {\cal V}^\ast_D \ :=\ {\cal V}^\ast\otimes {\cal S}_D\ ,\eea
where $\bar P^{\ast(i)}=\bar\c^{(i)}_a\rho^\ast(P^a)\,$ act faithfully.

The resulting dual indecomposable structures read
\bea {\cal V}^\ast & \cong & {\cal T}^\ast\supsetplus{\cal B}^\ast\quad\Rightarrow
\quad {\cal V}\ \cong\ {\cal T}\subsetplus{\cal B}\ .\eea
In $\Omega^{\mathbf{0}}(U)\otimes{\cal T}^\ast$ hold the generalized \emph{Bargmann-Wigner equations} for the
\emph{primary Weyl tensor}:
\bea \nabla_{(i)} C&\approx &0\ ,\quad (\nabla^2-\overline M{}^2)C\ \approx\ 0\ ,
\quad \mathbb B^\pm (\overline\nabla)C\ \approx\ 0\ ,\quad C\ :=\ X^{\bf 0}
(\overline \Th^\ast)\ .\label{EoMWeyl}\eea
We note that for generic masses there are no primary Bianchi
identities. Such identities arise only for critical masses, in which
case their combination with $\nabla_{(i)} C\approx 0$ implies the
mass-shell condition\footnote{As we shall see, the Bianchi
identities may involve more than one derivative of the primary Weyl
tensor. In such cases, their combination with $\nabla_{(i)} C\approx
0$ implies the mass-shell condition of descendants of the primary
Weyl tensor. Nevertheless, since the space of Weyl 0-form fills a
$\mg_\l$-module, the mass-shell condition for the lowest-type $C$ is
implied by that of any of its descendants, as they all share the
same value of the quadratic Casimir operator
$C_2[\mg_\l]=C_2[\mm]-L^2 P^a P_a\,$. In flat space, on the other
hand, the situation is more subtle, due to the completely
indecomposable structure of the twisted-adjoint module (see
\eq{complind}).}.

\subsubsection{\sc Canonical bilinear form: self-duality versus strict
masslessness}

More explicitly, using Howe-dual notation the canonical basis for
${\cal V}^\ast_{\rm diag}$ reads
\bea \mbox{symmetric basis}&:&\Th^{\ast\{n_J\}} \ :=\ \prod_{J=1}^{\overline{B}+1} (\bar P^{\ast(\overline{p}_{J-1}+1)})^{n_J} \overline \Th^\ast\ ,\label{symmbasis}\\[5pt]
\mbox{anti-symmetric basis}&:& \Th^{\ast\{n_J\}} \ :=\
\prod_{i=1}^\infty (\bar
P^{\ast[i]})^{\d_i(\{n_J\})}\overline\Th^\ast\
,\label{antisymmbasis}\eea
where (i) $J=0,1,\dots,\overline{B}+1$ labels the blocks of the
shape $\overline\Th$ parametrized as\footnote{In the following, we
shall frequently suppress the labels $\overline{s}_{_0}$,
$\overline{s}_{_{\overline{B}+1}}$, $\overline{h}_{_0}$ and
$\overline{h}_{_{\overline{B}+1}}$ in the presentation of the
zero-form types.}
\bea
\overline\Th&=&\Big([\overline{s}_{_0};\overline{h}_{_0}],[\overline{s}_{_1};\overline{h}_{_1}],[\overline{s}_{_2};\overline{h}_{_2}],\dots,[\overline{s}_{_{\overline
B}};\overline{h}_{_{\overline
B}}],[\overline{s}_{_{\overline{B}+1}};
\overline{h}_{_{\overline{B}+1}}]\Big)\ ,\\[5pt]
\overline{s}_{_0}&:=&\infty\ >\ \overline{s}_{_1}\ >\ \cdots\ >\ \overline{s}_{_{\overline{B}}}\ >\ \overline{s}_{_{\overline{B}+1}}\ :=0\ ,\\[5pt] \overline{h}_{0}&:=0&\ ,\quad \overline{h}_{_1}\ \geqslant\ 1\ ,\quad
\overline{h}_2\ \geqslant\ 1\ ,\ \dots\ ,\
\overline{h}_{_{\overline{B}+1}}\ :=\ \infty\ ;\eea
(ii) $n_{_J}\in\{0,\dots,\overline{s}_{J-1,J}\}$
($J=1,\dots,\overline{B}+1$) are the number of cells added to the
first row of the $J$th block, that is, to the
$(\overline{p}_{J-1}+1)$st row of $\overline\Th$, where
\bea \overline{s}_{_{J,K}}&:=&\overline{s}_{_J}-\overline{s}_{_K}\
,\qquad \overline{p}_{_J}\ :=\ \sum_{K=0}^J \overline{h}_{_K}\ ;\eea
(iii) $\d_i(\{n_{_J}\})\in\{0,1\}$ are dual parameters for the anti-symmetric basis.
The $n$th level of the module consists of the states ${\cal
V}^{\ast\{n\}}:=\bigoplus_{\sum_I n_I=n}\Th^{\ast\{n_I\}}$. Notice
that, in particular, $\Th^{\ast\{0\}}:=\overline \Th ^\ast$. The
canonical $\mg_\l$-invariant bilinear form $(\cdot,\cdot)_{{\cal
V}^\ast}$, which is equivalent to a canonical ditto on ${\cal
V}^\ast_{\rm diag}$, is defined by ($Q\in \mg_\l$)
\bea (\overline\Th^\ast,\overline \Th^\ast)_{{\cal V}^\ast_{\rm diag}}&:=&1\ ,\qquad (\r^\ast(Q)S^\ast,S^{\ast\prime})_{{\cal V}^\ast}+(S^\ast,\r^\ast(Q)S^{\ast\prime})_{{\cal V}^\ast}\ :=\ 0\ .\eea
Given $S^\ast=\sum\prod \rho^\ast(M_{AB})
\overline\Th^\ast=:\rho^\ast (Q(M_{AB}))\overline\Th^\ast$ now with
$Q\in{\cal U}[\mg_\l]$, \emph{idem} $S^{\ast\prime}\,$, the inner
product $(S^*,S^{*'})_{{\cal V}^\ast}=(\overline
\Th^\ast,\tau(Q)Q'\overline \Th^\ast)_{{\cal
V}^\ast}=\left.\left[\tau(Q)Q'\overline\Th^\ast\right]\right|_{\overline\Th^\ast}$,
the coefficient of $\overline\Th^\ast$ in the expansion of
$\tau(Q)Q'\overline\Th^\ast$ in the canonical basis, and where
$\tau(Q):=Q(-M_{AB})$ (the enveloping-algebra counterpart of matrix transposition)
is the canonical
anti-automorphism  of ${\cal U}[\mg_\l]$. The matrix elements
$\left.\left[\tau(Q)Q'\overline\Th^\ast\right]\right|_{\overline\Th^\ast}$
are diagonal in the canonical basis. Writing
$\Th^{\ast\{n_I\}}=Q^{\{n_I\}}\overline \Th^\ast$ one has
(symmetric basis, $n:=\sum_I n_{_I}$)
\bea
\left.\left[\tau(Q^{\{n_I\}})Q^{\{n'_I\}} \overline\Th^\ast \right]
\right|_{\overline\Th^\ast}&=&
\left.(-1)^n \left[ \prod_{J=\overline{B}+1}^1
(P^{\ast}_{(\overline{p}_{J-1}+1)})^{n_J}\prod_{K=1}^{\overline{B}+1}
(\bar P^{\ast(\overline{p}_{K-1}+1)})^{n'_K}\;\overline\Th^\ast
\right]
\right|_{\overline
\Th^\ast}\ .\eea
Obviously, in order for the above quantity to be non-vanishing,
one should have $n_{_J}=n'_{_J}$ for all $J\,$.
A general ``divergence''
\bea P^\ast_{(i)}\Th^{\ast\{n_I\}}&=&\sum_{\{n'_J\}|\sum_J n'_J=n-1}\left(A^{\{n_I\}}_{(i),\{n'_J\}}\overline M{}^2+\l^2 B^{\{n_I\}}_{(i),\{n'_J\}}\right)\Th^{\ast\{n'_J\}}\ ,\label{divergence}\eea
and there exists at least one
$i\in\{1+\overline{p}_{J-1}\}_{J=1}^{\overline{B}+1}$ such that one
of the matrix elements $A^{\{n_I\}}_{(i),\{n'_J\}}$ is
non-vanishing. It follows that there are two very distinct classes
of lowest-type spaces:
\begin{itemize}
\item[$\bullet$] the \emph{self-dual (or massively deformed) spaces}
\bea \mbox{$|\L|+|\overline M{}^2|> 0$}&:& {\cal T}\ \cong {\cal T}^\ast\ ,\label{selfdual}\eea
for which (i) the canonical inner product is non-degenerate on ${\cal T}^\ast$ and (ii) the primary Bianchi identities are completely fixed by $\overline\Th$ and $\overline M{}^2$;
\item[$\bullet$] the \emph{completely indecomposable (or strictly massless) spaces}\footnote{If $\overline\Th=[\overline{s}_{_1};\overline{h}_{_1}]$ with $\overline{h}_{_1}=\ft D2$ then ${\cal T}(\L\!\!=\!0;\overline M{}^2\!\!=\!0;[\overline{s}_{_1};\ft D2])$ is an $\mso(2,D)$-module where $\rho(K_a)\overline \Th=0$ and $(\rho(D)-\D(\overline{s}_{_1}))\overline \Th=0$ with $\D(\overline{s}_{_1})=\overline{s}_{_1}+D-2$. This module is self-dual with respect to an $\mso(2,D)$-invariant bilinear that is inequivalent to the $\miso(1,D-1)$-invariant bilinear form for general strictly massless mixed-symmetry fields.}
\bea \mbox{$\L=\overline M{}^2=0$}&:& \left.{\cal
T}\right|_{\mg_{_0}}\ =\ \overline \Th\subsetplus
\Th_{1_r}\subsetplus\cdots\ ,\quad  \left.{\cal
T}^\ast\right|_{\mg_{_0}}\ =\ \overline \Th^\ast\supsetplus \Th^{\ast
1_r}\supsetplus\cdots\ , \label{complind}\eea
for which (i) the canonical inner product is completely degenerate and (ii) the primary Bianchi identities can be chosen arbitrarily;
\end{itemize}
Thus, if $\L=0$ then massive lowest-type spaces must have trivial primary Bianchi identities, \emph{viz.}
\bea \mbox{$\L=0$ and $\overline M{}^2\neq 0$}&:& {\cal B}^\ast\ =\ \emptyset\ ,\quad {\cal T}^\ast\ =\ {\cal V}^\ast\ ,\eea
and hence these spaces are necessarily twisted-adjoint (infinite-dimensional), while the strictly massless lowest-type spaces are completely degenerate in the sense that
\bea \mbox{$\L=\overline M=0$}&:&{\cal V}^\ast \ =\ \overline \Th^\ast\supsetplus \Th^{\ast\{1\}}\supsetplus
\Th^{\ast\{2\}}\supsetplus\cdots\ ,\eea
so that any set of excited states can be taken to generate ${\cal B}^\ast$ (in the absence of any extended symmetry principle).

\subsubsection{\sc Critical masses for $\L\neq 0$}

If $\L\neq 0$ then ${\cal B}^\ast$ is generated by the singular
vectors $\mathbb B^\ast_N\overline \Th^\ast$ (which can always be
taken to have fixed rank) obeying
\bea N&:=&|\mathbb B^\ast_N\overline \Th^\ast|-|\overline \Th^\ast|\
>\ 0\ ,\qquad P^\ast_{(i)} \mathbb B^\ast_N\overline \Th^\ast\ =\
0\ ,\quad
i\in\left\{1+\overline{p}_{J-1}|J=1,\dots,\overline{B}+1\right\}\
.\eea
{}From \eq{divergence} it follows that demanding a fixed
$\Th^{\ast\{n_I\}}$ to become singular in general overdetermines
$\overline M{}^2$. We focus on the special
\bea \mbox{critical masses $\L\neq 0$}&:& \overline M{}^2\ =\
\overline M{}^2_{I,N}\ ,\qquad I=0,\dots,B\ ,\quad
N\in\{1,\dots,\overline{s}_{I,I+1}\}\ ,\eea
for which ${\cal B}^\ast$ contains the singular vector\footnote{In general ${\cal B}^\ast$ may contain more than one singular vector. It is known that such ``multiple critical phenomena'' do not occur in what we refer to as the massless cases below \cite{Brink:2000ag}.}
\bea \overline\mathbb B^\ast_{I,N}\overline \Th^\ast&=&(\bar
P^{(1+\overline{p}_I)})^N\overline \Th^\ast\ =\ \bar
P^{\ast[1+\overline{s}_{I+1}]}\cdots \bar
P^{\ast[N+\overline{s}_{I+1}]} \overline \Th^\ast\ .\eea
This state has only one non-trivial divergence for general
$\overline M{}^2$ (in the $(1+\overline{p}_{_I})$th row,
\emph{i.e.}, the first row of the $(1+I)$th block) that hence
vanishes iff $\overline M{}^2$ assumes a critical value. Factoring
out ${\cal B}^\ast$ corresponds to imposing the primary Bianchi
identities
\bea (\overline\nabla^{(1+\overline{p}_I)})^N C\ =\
\overline\nabla^{[1+\overline{s}_{I+1}]}\cdots \overline
\nabla^{[N+\overline{s}_{I+1}]} C\ \approx\ 0\ .\eea
Summarizing the results of the analysis carried on in the present
paper and in Paper II \cite{Boulanger:2008kw}, the above critical
cases consist of
\begin{itemize}\label{nomenclature}
\item[(i)] \emph{tensorial modules} for $I=0$ and $N\geqslant 1$;
\item[(ii)] \emph{cut twisted-adjoint modules} for $I=1$ and
$N\geqslant 1$ if $\overline{h}_{_1}=1$; \item[(iii)] \emph{massless
twisted-adjoint modules} for:
   \begin{itemize}
   \item[(a)] $I=1,\dots,\overline{B}-1$, $1\leqslant N\leqslant \overline{s}_{_{I,I+1}}$ and $\overline{h}_{_I} > 1$
   \item[(b)]  $I=\overline{B}$, $N=1$ and $\overline{h}_{_{\overline B}}> 1$;
   \end{itemize}
   these cases are of special interest to us and we denote the corresponding critical masses by
    \bea\mbox{``massless'' critical masses}&:& \overline M{}^2_I:= \overline M{}^2_{I,N}\ ;\label{masslessmasses}\eea
\item[(iv)] \emph{partially massless twisted-adjoint modules} for:
\begin{itemize}
   \item[(a)] $I=2,\dots,\overline{B}-1$, $1\leqslant N\leqslant
   \overline{s}_{_{I,I+1}}$ and $\overline{h}_{_I}=1$
   \item[(b)] $I=\overline{B}$, $N=1$ and $\overline{h}_{_{\overline{B}}}=1$.
   \end{itemize}
\end{itemize}
%

The tensorial ${\cal T}^\ast$-modules consist of tensorial
harmonics in $AdS_D$ or $dS_D$ obtained from tensorial harmonics
on $S^D$ by Wick rotation in the $(D+1)$-dimensional ambient
space. Writing $\overline\Th=(\overline{s}_{_1},\Xi)$, one
has~\footnote{Upon harmonic expansion the Young tableaux
re-surface in the compact weight spaces. If $\l^2>0$ then
$\big(\overline{s}_{_1}+N-1,\Xi\big)|_{\mg_\l}\cong
\mD^+(1-\overline{s}_{_1}-N;\overline\Th)$ which is the shadow of
a massive unitary module. If $\l^2<0$ then
$\big(\overline{s}_{_1}+N-1,\Xi\big)|_{\mso(1,D)}$ is an ideal
subspace of the compact $\mso(D)'$-slicing of ${\cal
V}(\L;\overline M{}^{2}_{0,N};\overline \Th)$ whose complement is
a unitary partially massless representation.}
\bea& L^2\overline M^2_{0,N}\ =\ (N+\overline{s}_{_1}-1)(N+\overline{s}_{_1}+D-2)+C_{_2}[\ms|\overline{s}_{_1}+N-1,\Xi]-C_{_2}[\mm|\overline\Th]\ ,\\[5pt]& {\cal T}^\ast(\L;\overline M{}^2_{0,N};\overline\Th)\ \cong\ \big(\overline{s}_{_1}+N-1,\overline\Th\big)|_{\mg_\l}\ ,&\\[5pt] &{\cal B}^\ast(\L;\overline M{}^2_{0,N};\overline\Th)\ \cong\ {\cal T}^\ast(\L;\overline M^{2}_{2,1};(\overline{s}_{_1}+N,\Xi))\ ,&\eea
where the ideal is a cut twisted-adjoint module since
$(\overline{s}_{_1}+N,\Xi)$ has first block of height $1$ and
width $\overline{s}_{_1}+N$ and singular vector given by the first
excitation of the first row of the second block.

The rationale behind the statements on the
gauge-field cases $(\mbox{iii})$ and $(\mbox{iv})$ of the above
classification becomes clear upon integration of the Bianchi
identities, as explained in Section \ref{Sec:integration}.

\subsubsection{\sc Strictly massless case}

Returning to $\L=\overline M{}^2=0$ we note that the translations $\rho^\ast(P_a)$ acting in the strictly massless smallest-type modules have by definition Howe-dual projections of only type $\bar P^{\ast(i)}$ (that is $P^\ast_{(i)}\equiv 0$). Their action on $\overline\Th^\ast$ generates ${\cal V}^\ast(\overline\Th):={\cal V}^\ast(\L\!\!=\!0;\overline M{}^2\!\!=\!0;\overline\Th)$. Factoring out ${\cal B}^\ast$ yields the module ${\cal T}^\ast$. Its dual ${\cal T}$ has translations $\rho(P_a)$ (``dual derivatives'') with Howe-dual projections only of type $P_{(i)}$. In the strictly massless case the submodule ${\cal B}^\ast$ can be chosen arbitrarily. Note the recent work \cite{Alkalaev:2008gi} in the
same context. We are interested in
\begin{itemize}
\item[(i)] \emph{finite-dimensional $\mg_{_0}$-modules};
\item[(ii)] \emph{strictly massless twisted-adjoint $\mg_{_0}$-modules} in which the only translation is $P_{(1)}$;
\end{itemize}
If a strictly massless smallest-type space ${\cal T}^\ast(\overline\Th)$ is a proper submodule of a larger ditto ${\cal T}^\ast(\overline\Th')$ (with shape $\overline\Th'\subset \overline\Th$) then one refers to ${\cal T}(\overline\Th)$ as being cut. Else one refers to ${\cal T}(\overline\Th)$ as being maximal, in which case the primary Bianchi identities (generating ${\cal B}^\ast$) read
\bea &\overline\nabla^{[i]}C(\overline{\Theta})\ \approx\ 0\
,\quad i=1,\dots,s_{_1}\ .& \eea
The cut twisted-adjoint modules arise as strictly massless limits of St\"uckelberg sectors of massive twisted-adjoint modules in flat spacetime as well as critical dittos in constantly curved spacetime.

\subsubsection{\sc Primary and secondary Bianchi identities}

Returning to the zero$\,$-form constraints \eq{0formdual} one has
\bea &P^\ast_{(i)}\mathbf{X}^{\bf 0}\ \in\ {\rm Im}~\s^+\cap {\cal T}^{\ast}_D\ ,\quad \bar P^{\ast(i)}\mathbf{X}^{\bf 0}\ \in\ {\rm Im}~\s^-\cap {\cal T}^{\ast}_D\ ,\eea
where $P^\ast_{(i)}\mathbf{X}^{\bf 0}$ contains separate massive contributions from $\overline M{}^2$ and $\L$ as given in \eq{divergence}, while on the other hand
\bea \nabla_{(i)} \mathbf{X}^{\bf 0}&=& \nabla_{(i)} C+\left.\nabla_{(i)}
\mathbf{X}^{\bf 0}\right|_{{\cal T}^{\ast}_D}\ ,\\[5pt]
\overline\nabla^{(i)}\mathbf{X}^{\bf 0}&=& \underbrace{\left.\overline\nabla^{(i)}
\mathbf{X}^{\bf 0}\right|_{{\cal T}^{\ast}_D}+\left.\overline\nabla^{(i)}
\mathbf{X}^{\bf 0}\right|_{{\cal B}^{\ast}_D}}_{\in {\cal V}^{\ast}_D}+\left.\overline\nabla^{(i)} \mathbf{X}^{\bf 0}\right|_{{\cal N}^{\ast}_D}
\label{Weyldecomp}\eea
where, more precisely, here $C=C(\overline{\Th}^*)\ket{\overline{\Th}}$
$\in$  ${\cal{T}}^*\otimes {\cal S}_D$ and
${{\cal N}^{\ast}_D}$
is the content of $\overline\nabla^{(i)} \mathbf{X}^{\bf 0}$
outside ${\cal V}^{\ast}_D\,$. The zero$\,$-form constraint thus amounts to
\begin{itemize}
\item[(i)] $\left.\overline\nabla^{(i)} \mathbf{X}^{\bf 0}\right|_{{\cal T}^{\ast}_D}+i\bar P^{\ast(i)}\mathbf{X}^{\bf 0}\approx0$
which are algebraic equations for auxiliary fields;
\item[(ii)] $\left.\overline\nabla^{(i)} \mathbf{X}^{\bf 0}\right|_{{\cal B}^{\ast}_D}\approx 0$ which comprise the primary Bianchi identities (that are the components lying in $H^1(\s^-)$) and some (but not all) of their descendants which are Bianchi identities for auxiliary fields that hold by virtue of the primary Bianchi identities;
\item[(iii)] $\nabla_{(i)}C\approx 0$ (which lie in $H^1(\s^-)$) which are the primary divergence conditions on $C$;
\item[(iv)] $\left.\nabla_{(i)} \mathbf{X}^{\bf 0}\right|_{{\cal T}^{\ast}_D}+i\, P^{\ast}_{(i)}\mathbf{X}^{\bf 0}\approx 0$ which are (all) the descendants of the primary divergence conditions, containing mass-shell conditions for $C$ as well as auxiliary fields;
\item[(v)] $\left.\overline\nabla^{(i)} \mathbf{X}^{\bf 0 }\right|_{{\cal N}^{\ast}_D}\approx 0$ which are secondary Bianchi identities.
\end{itemize}
The primary Bianchi identities, divergence conditions and corresponding mass-shell condition on $C$ are the Bargmann-Wigner equations.
Roughly speaking, the integration of primary and secondary Bianchi identities, respectively, yield dynamical gauge fields and St\"uckelberg fields.

\subsection{\sc{Unfolded integration of Weyl zero$\,$-form}}
\label{Sec:IntBI}

\subsubsection{\sc Integration schemes and dimensional reduction}\label{Sec:Schemes}

The Weyl zero$\,$-form module $\mC^{\bf 0}$ described by \eq{0formcomp} or
equivalently \eq{0formdual}, can be glued to $\mg_\l$-modules $\mR^{\bf p}$ in
various form-degrees to form chains, or branches, where each link, or
subbranch, is a separately contractible cycle (see Eq. \eq{contractible} and
Fig. \ref{Fig:Dualcocycle}). The systematic integration yields a tree with
trunk given by a common Weyl zero$\,$-form $\mC^{\bf 0}$ connected via branches
and subbranches to ``leaves'' given by a spectrum of dynamical fields $\{\varphi\}$
in various duality pictures.

The basic mechanism for growing a branch is to integrate a Bianchi identity in
\eq{0formdual}. In the strictly massless cases the primary Bianchi identities and
their Hodge duals initiate primary chains that are non-contractible. In the
massless self-dual cases the issue of contractibility is more subtle
(see Paper II).
In the genuinely massive cases, where there are no primary branches, the lowest
secondary branch contains the gauge potential and all the St\"uckelberg fields,
forming a massively contractible cycle, since the primary Weyl tensor and the
dynamical gauge field share the same Lorentz type (see the example of massive
spin-1 below).
More generally, extended secondary integration schemes induce infinite towers of
dual dynamical potentials~\cite{CPN2}.

The ``thickness'' of a given branch can be varied by replacing the $\mm$-types in
the finite-dimensional
$\mg_\l$-irreps $\mR^{\bf p}$ by $\msl(D)$-types, leading to $\mg_\l$-reducible
subbranches and trace-unconstrained metric-like dynamical fields
$\varphi|_{\msl(D)}$~\cite{Bekaert:2003az,Bekaert:2006ix}
which carry the local degrees of freedom coming
from $\mC^{\bf 0}\,$ (see also \cite{Engquist:2007yk}).
Such trace-unconstrained formulations thus activate extended
patterns of shift symmetries whose gauge fixing lead back to the trace-constrained,
or ``minimal'', formulations, \emph{i.e.} one has the following commuting diagram:
\bea \ba{ccc}\ba{c}\mbox{trace-unconstrained}\\[-4pt]\mbox{frame-like
formulation}\ea&\longrightarrow & \ba{c}\mbox{trace-
constrained}\\[-4pt]\mbox{frame-like formulation}\ea\\
\downarrow&&\downarrow\\
\ba{c}\mbox{trace-unconstrained}\\[-4pt]\mbox{metric-like
formulation}\ea&\longrightarrow & \ba{c}\mbox{trace-
constrained}\\[-4pt]\mbox{metric-like formulation}\ea\ea\eea

The ``double-dimensional'' reduction of strictly massless systems with fiber
algebra $\widehat\mg_{_0}=\miso(1,D)$ yields systems with fiber algebra $\mg_\l$ and
mass parameters $\overline M{}$. The parameters $\l$ and $\overline M$,
respectively, originate from the reductions of the fiber and the base-manifold (see
scheme below). Our working hypothesis is that starting from Skvortsov's minimal
frame-like formulation of free gauge fields $\widehat\varphi$ in $D+1$ dimensions
with fiber algebras $\widehat \mg_{_0}$ of various tangent-space signatures it is
possible to reach the minimal dittos in $D$ dimensions with \emph{all} possible
values for $\l$ and $\overline M{}$ (or $M$, the critical gauge-field mass
which is determined by the critical mass given in \eq{masslessmasses} for its
primary Weyl tensor).
Schematically,
\begin{picture}(200,60)(10,-8)
\put(50,40)
{$$\mbox{minimal frame-like scheme for $\widehat\mR(\widehat\varphi)|_{\widehat\mg_{_0}}$}$$}
\put(70,35){\vector(0,-1){10}}\put(74,30)
{${\cal L}_{\widehat\x_\l}\widehat\varphi~\stackrel{!}{=}~\D(\l,M) \widehat\varphi$}
\put(50,17){$$\mbox{minimal frame-like schemes for $\uplus_I ~\mR(\varphi_{_I}(\L;
M{}^2_I))|_{\mg_\l}$}$$}
\put(70,12){\vector(0,-1){10}}\put(74,7){$\mbox{projection to irreducible submodules}$}
\put(50,-5){$$\mbox{minimal frame-like scheme for $\mR(\varphi(\L; M{}^2))|_{\mg_\l}$}$$}
\end{picture}
where ${\cal L}_{\widehat\x_\l}$ are Lie derivatives along vectors fields $\widehat\x_\l$, bringing in the parameter $\l$, and $\D(\l, M)$ are scaling dimensions.
The relation $\D\leftrightarrow C_{_2}[\mg_\l]$ is actually
$2\leftrightarrow 1$ that for $\L\neq 0$ implies two roots $\D_\pm$ with dual indecomposable structures, say
\bea \D_+&:& \supsetplus_I \mR(\varphi_{_I}(\L; M{}^2_I))|_{\mg_\l}\ ,\qquad
\D_-\ :\quad \subsetplus_I \mR(\varphi_{_I}(\L; M{}^2_I))|_{\mg_\l}\ .\eea

\subsubsection{\sc Remarks on metric-like
integration}\label{Sec:integration}

In the case of $\L=0=\overline M{}^2$ it was shown in
\cite{Bekaert:2002dt} that the primary Bianchi identity for a generalized
Riemann tensor sitting in an $\msl(D)$-type $\overline\Th$ and
obeying
\bea \overline\nabla^{[1]}{\cal K}(\overline \Th)&\equiv&0\
,\label{BIK}\eea
has the general solution
\bea {\cal K}(\overline\Th)&\equiv&\overline\nabla^{[s_{_1}]}\cdots
\overline\nabla^{[1]}\varphi(\Th)\ ,\label{Knablaphi}\eea
where $\varphi(\Th)$ is a metric-like tensor gauge field sitting in
the $\msl(D)$-type of shape $\Th\,$. It was then shown
\cite{Bekaert:2003az,Bekaert:2003zq} that if $s_{_1}\geqslant 2$
then the on-shell constraint
\bea T_{_{[12]}}{\cal K}(\overline \Th)&\approx& 0\
,\label{T12K0}\eea
\emph{i.e.} ${\cal K}\approx C\,$, together with the usual boundary
conditions (no runaway solutions, the fields and all their
derivatives vanish at infinity),
induce carriage of $\mD(M=0;\Th)\,$. Moreover, it was shown in
\cite{Bekaert:2003az,Bekaert:2006ix} that integration of
\eq{BIK}--\eq{T12K0} yields a compensator version of the Labastida
equation. This equation reduces to the Labastida equation upon
fixing shift symmetries --- for totally symmetric $\Th$ the compensator
form of the Fronsdal equation had previously been given in
\cite{Francia:2002aa}\footnote{Totally symmetric $\msl(D)$-tensor
gauge fields were first considered in \cite{Francia:2002aa} though
the dynamical field equation was not of the form (\ref{T12K0}). The
field equation (\ref{T12K0}) for arbitrary $\msl(D)$-tensor gauge
fields was first proposed in \cite{Bekaert:2002dt} and then shown in
\cite{Bekaert:2003az} to be equivalent to that of
\cite{Francia:2002aa} upon restricting to the totally symmetric
cases. Finally, in the general bosonic case, the equation
(\ref{T12K0}) was proven to propagate the correct massless physical
degrees of freedom in \cite{Bekaert:2003zq}. For a review and other
results on those issues, see \cite{Bekaert:2006ix}. See the very recent
work \cite{Campoleoni:2008jq} for related results.}. Thus
the Bargmann-Wigner equations \eq{EoMWeyl} are equivalent
modulo boundary conditions to the Labastida equation once all
intermediate shift symmetries are fixed.

The above on-shell integration generalizes to critically massless
Weyl tensors when $\L\neq0\,$, since antisymmetric combinations of
$\mg_{\l}$-covariant derivatives only introduce pure trace terms
that are removed by the overall traceless projection the Weyl
zero-forms are subject to. We can now give the rationale behind
the classification of Section \ref{nomenclature}. We parametrize
the dynamical fields as\footnote{As for the zero-form types, in
the following, we shall frequently suppress the labels $s_{_0}$,
$s_{_{B+1}}$, $h_{_0}$ and $h_{_{B+1}}$, in the presentation of
Young diagrams associated to dynamical fields.}
\bea
\Th&=&\Big([s_{_0};h_{_0}],[s_{_1};h_{_1}],\dots,[s_{_B};h_{_B}],[s_{_{B+1}};
h_{_{B+1}}]\Big)\ ,\\[5pt]
s_{_0}&:=&\infty\ >\ s_{_1}\ >\ \cdots\ >\ s_{_B}\ >\ s_{_{B+1}}\ :=0\ ,\\[5pt] h_{0}&:=0&\ ,\quad h_{_1}\ \geqslant\ 1\ ,\quad
h_2\ \geqslant\ 1\ ,\ \dots\ ,\ h_{_{B+1}}\ :=\ \infty\eea
and define the quantities
\bea s_{_{J,K}} \ = \ s_{_J}-s_{_K} \ , \qquad p_{_J}\ :=\
\sum_{K=0}^J h_{_K} \ , \qquad J=0,\dots,B+1 \ .\eea
Then, the integration of the Bianchi identities in the cases (iii)
and (iv) listed above proceeds as follows. We further distinguish
the subcases $N=1$ and $N>1$.
\begin{itemize}
\item[(iii)] \emph{massless case}\,:
\begin{itemize}
\item[$\bullet$] \quad $N=1$, \,\, $1\leqslant I \leqslant \overline{B}$,
\,\, $\overline{h}_{_I}>1\,$:
\bea
\overline{\nabla}^{(\overline{p}_{_I}+1)}\,C(\overline{\Th}_{_I})
\ = \ 0 \qquad \Rightarrow \qquad C(\overline{\Th}_{_I}) \ = \
(\overline{\nabla}^{(\overline{p}_{_I})})^{\overline{s}_{I,I+1}}\,\varphi_{_I}(\Th)
\ ,\eea
leading to a metric-like massless dynamical field
$\varphi_{_I}(\Th)\equiv\varphi(\L;M{}^2_I;\Th)$ with shape $\Th$
characterized by $s_{_J}=\overline{s}_{_J}$ for all $J=1,\dots,B$,
$B=\overline{B}$, $h_{_J}=\overline{h}_{_J}$ for $J\neq I, I+1\,$,
and $h_{_I}=\overline{h}_{_I}-1\,$,
$h_{_{I+1}}=\overline{h}_{_{I+1}}+1$, \emph{i.e.} obtained from
$\overline{\Th}$ by subtracting one row to its $I$th block and
adding one to the $(I+1)$st block:
\bea
\Th&=&\Big([\overline{s}_{_1};\overline{h}_{_1}],\dots,[\overline{s}_{_{I-1}};\overline{h}_{_{I-1}}],[\overline{s}_{_I};\overline{h}_{_I}-1],
[\overline{s}_{_{I+1}};\overline{h}_{_{I+1}}+1],[\overline{s}_{_{I+2}};\overline{h}_{_{I+2}}],\dots,[\overline{s}_{_{\overline
B}};\overline{h}_{_{\overline B}}]\Big)\ ,\hspace{1cm}\eea
with gauge symmetry
\bea \d\varphi_{_I}(\Th) \ = \
\overline{\nabla}^{(\overline{p}_I-1)}\,\e_{_I}(\Th') \ ,\eea
where $\Th'$ is obtained from $\Th$ by deleting one box in the
$(\overline{p}_I-1)$st row. Equivalently, the Weyl tensor type can
be parametrized with the dynamical field labels as
\bea
\overline{\Th}_I&=&\Big([s_{_1};h_{_1}],\dots,[s_{_{I-1}};h_{_{I-1}}],[s_{_I};h_{_I}+1],
[s_{_{I+1}};h_{_{I+1}}-1],[s_{_{I+2}};h_{_{I+2}}],\dots,[s_{_B};h_{_B}]\Big)\
,\hspace{1cm}\eea
and expressed as $C(\overline{\Th}_{_I}) \ = \
(\overline{\nabla}^{(p_{_I}+1)})^{s_{I,I+1}}\,\varphi_I(\Th)$,
with $\d\varphi_{_I}(\Th) \ = \
\overline{\nabla}^{(p_I)}\,\e_{_I}(\Th')$. The primary divergence
condition implies that $\varphi(\L;M{}^2_I;\Th)$ obeys
Lorentz-like first-order divergence conditions in the blocks
$J\neq I$.

\item[$\bullet$]   \qquad $1< N\leqslant\overline{s}_{_{I,I+1}}$, \,\,
$1\leqslant I \leqslant \overline{B}-1$, \,\,
$\overline{h}_{_I}>1\,$:
\bea
(\overline{\nabla}^{(\overline{p}_{_I}+1)})^N\,C(\overline{\Th}_{_I})
\ = \ 0 \qquad \Rightarrow \qquad C(\overline{\Th}_{_I}) \ = \
(\overline{\nabla}^{(\overline{p}_{_I})})^{\overline{s}_{I,I+1}-N+1}\,\varphi_{_I}(\Th)
\ ,\eea
corresponding to the primary Weyl tensor of a metric-like massless
dynamical field of shape $\Th$ obtained by cutting off one row
from the $I$th block of $\overline{\Th}_{_I}$ and inserting one
extra block of height one and length $\overline{s}_{_{I+1}}+N-1$
inserted between the $I$th and the $(I+1)$st blocks,
\bea
\Th&=&\Big([\overline{s}_{_1};\overline{h}_{_1}],[\overline{s}_{_2};\overline{h}_{_2}],\dots,[\overline{s}_{_I};\overline{h}_{_I}-1],[\overline{s}_{_{I+1}}+N-1;1],
[\overline{s}_{_{I+1}};\overline{h}_{_{I+1}}],\dots,[\overline{s}_{_{\overline
B}};\overline{h}_{_{\overline B}}]\Big)\ ,\hspace{1cm}\eea
\emph{i.e.}, $B=\overline{B}+1$, $s_{_J}=\overline{s}_{_J}$ for
$J=1,\dots,I$, $s_{_{I+1}}=\overline{s}_{_{I+1}}+N-1$ and $s_{_J}=
\overline{s}_{_{J-1}}$ for $J=I+2,\dots,B\,$, while
$h_{_J}=\overline{h}_{_J}$ for $J=1,\dots,I-1$,
$h_{_I}=\overline{h}_{_I}-1$ $h_{_{I+1}}=1$ and $h_{_J}=
\overline{h}_{_{J-1}}$ for $J=I+2,\dots,B\,$. The gauge symmetry
still involves only one derivative,
\bea \d\varphi_{_I}(\Th) \ = \
\overline{\nabla}^{(\overline{p}_I-1)}\,\e_{_I}(\Th') \ .\eea
\end{itemize}
Notice that for the gauge symmetry to exist and to be the standard
one associated to massless fields it is crucial that
$\overline{h}_{_I}>1$.
\end{itemize}

In the massless case with $N=1\,$, see Fig. \ref{Table:FigCar}
for a pictorial representation of the integration precedure.
\begin{figure}
$\overline{\Th}_2 \ = \ $ \ \begin{picture}(30,30)(0,0)
\multiframe(0,1.5)(3.5,0){1}(30,12){}
\multiframe(0,-7.7)(3.5,0){1}(24,9){}
\multiframe(0,-13.9)(3.5,0){1}(9,6){}
\multiframe(0,-17.1)(3.5,0){1}(6,3){}
\end{picture} \hspace{1cm}\,, \hspace*{2cm}
$\mathbb{B}_{_{2,1}}(\overline{\Th}_{_2}) \ = \ $ \
\begin{picture}(30,30)(0,0) \multiframe(0,1.5)(3.5,0){1}(30,12){}
\multiframe(0,-7.7)(3.5,0){1}(24,9){}
\multiframe(0,-13.9)(3.5,0){1}(9,6){}\multiframe(9.2,-10.9)(3.5,0){1}(3,3){\tiny
$\nabla$} \multiframe(0,-17.1)(3.5,0){1}(6,3){}
\end{picture} \ $ \ = \ 0 \hspace{1cm}\Rightarrow $ \\[20pt]
\hspace*{4cm} $C(\overline{\Th}_{_2}) \ = \
(\overline{\nabla}^{(p_2+1)})^{\overline{s}_{23}} \varphi_{_2}(\Th)\
= \ $ \
\begin{picture}(30,30)(0,0)
\multiframe(0,1.5)(3.5,0){1}(30,12){}
\multiframe(0,-7.7)(3.5,0){1}(24,9){}\multiframe(9.1,-7.7)(3.5,0){1}(3,3){\tiny
$\nabla$}
\multiframe(12.3,-7.7)(3.5,0){1}(8.5,3){$\ldots$}\multiframe(21.0,-7.7)(3.5,0){1}(3,3){\tiny
$\nabla$} \multiframe(0,-13.9)(3.5,0){1}(9,6){}
\multiframe(0,-17.1)(3.5,0){1}(6,3){}
\end{picture} \ ,\hspace{0.5cm}
\\[20pt]
\hspace*{4cm} $\d \varphi_{_2}(\Th) \ = \
\overline{\nabla}^{(p_2)}\epsilon_{_2}(\Th') \ = \ $ \
\begin{picture}(30,30)(0,0)
\multiframe(0,1.5)(3.5,0){1}(30,12){}
\multiframe(0,-4.7)(3.5,0){1}(24,6){}\multiframe(21,-4.6)(3.5,0){1}(3,3){\tiny
$\nabla$}
\multiframe(0,-13.9)(3.5,0){1}(9,9){}\multiframe(0,-17.1)(3.5,0){1}(6,3){}
\end{picture}\vspace{2cm}
\caption{\small By means of the integration lemma, the primary Weyl tensor
$C(\overline{\Th}_{_2})$ with Bianchi identity
$\mathbb{B}_{_{2,1}}(\overline{\Th}_{_2})$ is shown to correspond
to a massless gauge field $\varphi_{_2}(\Th)$ whose shape is
obtained from $\overline{\Th}_{_2}$ by cutting off one row from
its second block and by adding one to its third block. It
possesses a one-derivative gauge symmetry with parameter
$\epsilon_{_2}(\Th')$, obtained from $\Th$ by deleting one cell in
the second block.}
\label{Table:FigCar}\end{figure}

On the other hand, partially massless dynamical fields arise for:
\begin{itemize}
\item[(iv)] \emph{partially massless fields}\,:
\begin{itemize}
\item[$\bullet$]  \qquad $N=1$, \,\, $2\leqslant I \leqslant
\overline{B}$,\,\, $\overline{h}_{_I}=1\,$ \,\,($2\leqslant
k=\overline{s}_{_{I-1,I}}+1\leqslant \overline{s}_{_{I-1,I+1}}$):
\bea
\overline{\nabla}^{(\overline{p}_{_I}+1)}\,C(\overline{\Th}_{_{I,k}})
\ = \ 0 \qquad \Rightarrow \qquad C(\overline{\Th}_{_{I,k}}) \ = \
(\overline{\nabla}^{(\overline{p}_{_I})})^{\overline{s}_{I,I+1}}\,\varphi_{_{I,k}}(\Th)
\ ,\eea
leading to a metric-like partially massless dynamical field with
shape $\Th$ which can be obtained from $\overline{\Th}$ by cutting
off the $I$th block and by adding one row to the $I+1$st block,
\bea
\Th&=&\Big([\overline{s}_{_1};\overline{h}_{_1}],\dots,[\overline{s}_{_{I-1}};\overline{h}_{_{I-1}}],[\overline{s}_{_{I+1}};\overline{h}_{_{I+1}}+1],\dots,[\overline{s}_{_{\overline
B}};\overline{h}_{_{\overline B}}]\Big)\ ,\eea
\emph{i.e.}, characterized by $B=\overline{B}-1$,
$s_{_J}=\overline{s}_{_J}$ and $h_{_J}=\overline{h}_{_J}$ for all
$J=1,\dots,I-1$, $s_{_J}=\overline{s}_{_{J+1}}$,
$h_{_I}=\overline{h}_{_{I+1}}+1$ and
$h_{_J}=\overline{h}_{_{J+1}}$ for $J=I+1,\dots, B\,$. Due to the
fact that $\overline{h}_{_I}=1$, the gauge symmetry this time is a
higher-derivative one,
\bea \d\varphi_{_{I,k}}(\Th) \ = \
(\overline{\nabla}^{(\overline{p}_I-1)})^k\,\varphi_{_{I,k}}(\Th)
\ = \
(\overline{\nabla}^{(\overline{p}_{I-1})})^k\,\e_{_{I,k}}(\Th') \
,\label{Hdergauge}\eea
where now $\Th'$ is obtained from $\Th$ by deleting $k$ boxes from
the $(\overline{p}_{I-1})$th row.

\item[$\bullet$] \qquad $1< N\leqslant\overline{s}_{_{I,I+1}}$,
\,\,$2\leqslant I \leqslant \overline{B}-1$,
\,\,$\overline{h}_{_I}$,\,\, ($2\leqslant
k=\overline{s}_{_{I-1,I}}+1\leqslant
\overline{s}_{_{I-1,I+1}}-N+1$):
\bea
(\overline{\nabla}^{(\overline{p}_{_I}+1)})^N\,C(\overline{\Th}_{_{I,k}})
\ = \ 0 \qquad \Rightarrow \qquad C(\overline{\Th}_{_{I,k}}) \ = \
(\overline{\nabla}^{(\overline{p}_{_I})})^{\overline{s}_{I,I+1}-N+1}\,\varphi_{_{I,k}}(\Th)
\ ,\eea
corresponding to the primary Weyl tensor of a metric-like
partially massless dynamical field of shape $\Th$ that can be
obtained from $\overline{\Th}$ by shortening the $I$th block (of
height one) from length $\overline{s}_{_I}$ to
$\overline{s}_{_{I+1}}+N-1$, while all other lengths and heights
remain untouched,
\bea
\Th&=&\Big([\overline{s}_{_1};\overline{h}_{_1}],\dots,[\overline{s}_{_{I-1}};\overline{h}_{_{I-1}}],[\overline{s}_{_{I+1}}+N-1;1],[\overline{s}_{_{I+1}};\overline{h}_{_{I+1}}],\dots,[\overline{s}_{_{\overline
B}};\overline{h}_{_{\overline B}}]\Big)\ .\eea
The higher-derivative gauge symmetry is as in \eq{Hdergauge}.
\end{itemize}
\end{itemize}

Our classification and definition of partially massless fields
generalizes to the mixed-symmetry cases the results of \cite{Skvortsov:2006at}
for totally symmetric fields in the framework on unfolding.
Totally symmetric partially massless fields were first discussed
in \cite{Deser:1983mm} and further studied in \cite{Deser:2001us,Zinoviev:2001dt}
(see also \cite{Deser:2006zx} and references therein).

In the partially massless case, a pictorial representation of the
integration procedure is given in Fig.~\ref{Table:PMCa}.

\begin{figure}
$\overline{\Th}_2 \ = \ $ \ \begin{picture}(33,30)(0,0)
\multiframe(0,1.5)(3.5,0){1}(33,12){}
\multiframe(0,-1.7)(3.5,0){1}(24,3){}
\multiframe(0,-7.9)(3.5,0){1}(9,6){}
\multiframe(0,-11.1)(3.5,0){1}(6,3){}
\end{picture}\hspace*{1cm}\,, \hspace{2cm}
$\mathbb{B}_{_{2,1}}(\overline{\Th}_{_2}) \ = \ $ \
\begin{picture}(33,30)(0,0)
\multiframe(0,1.5)(3.5,0){1}(33,12){}
\multiframe(0,-1.7)(3.5,0){1}(24,3){}\multiframe(9.2,-4.9)(3.5,0){1}(3,3){\tiny
$\nabla$} \multiframe(0,-7.9)(3.5,0){1}(9,6){}
\multiframe(0,-11.1)(3.5,0){1}(6,3){}
\end{picture} \ $ \ = \ 0 \hspace{1cm}\Rightarrow $ \\[20pt]
\hspace*{2cm} $C(\overline{\Th}_{_2}) \ = \
(\overline{\nabla}^{(p_2+1)})^{\overline{s}_{23}} \varphi_{_2}(\Th)\
= \ $ \
\begin{picture}(33,30)(0,0)
\multiframe(0,1.5)(3.5,0){1}(33,12){}
\multiframe(0,-1.7)(3.5,0){1}(24,3){}\multiframe(9.1,-1.7)(3.5,0){1}(3,3){\tiny
$\nabla$}
\multiframe(12.3,-1.7)(3.5,0){1}(8.5,3){$\ldots$}\multiframe(21.0,-1.7)(3.5,0){1}(3,3){\tiny
$\nabla$}  \multiframe(0,-7.9)(3.5,0){1}(9,6){}
\multiframe(0,-11.1)(3.5,0){1}(6,3){}
\end{picture} \ ,\\[20pt]
\hspace*{2cm} $\d \varphi_{_2}(\Th) \ = \
(\overline{\nabla}^{(p_2)})^{\overline{s}_{12}+1}\epsilon_{_2}(\Th')
\ = \ $ \
\begin{picture}(33,30)(0,0)
\multiframe(0,1.5)(3.5,0){1}(33,12){}
\multiframe(21.0,1.5)(3.5,0){1}(3,3){\tiny $\nabla$}
\multiframe(24.2,1.5)(3.5,0){1}(5.6,3){\tiny
$\ldots$}\multiframe(29.9,1.5)(3.5,0){1}(3,3){\tiny $\nabla$}
\multiframe(0,-7.8)(3.5,0){1}(9,9){}
\multiframe(0,-11.1)(3.5,0){1}(6,3){}
\end{picture}\vspace{2cm}\caption{\small Through the integration lemma explained
above, the primary Weyl tensor
$C(\overline{\Th}_{_2})$ with second block of height one and
Bianchi identity $\mathbb{B}_{_{2,1}}(\overline{\Th}_{_2})$ is
shown to correspond to a partially massless gauge field
$\varphi_{_2}(\Th)$ whose shape is obtained from
$\overline{\Th}_{_2}$ by cutting off its second block
and by adding one row to its third block. It possesses a
higher-derivative gauge symmetry with parameter
$\epsilon_{_2}(\Th')$, obtained from $\Th$ by deleting
$\overline{s}_{_{12}}+1$ cells in the second block.}
\label{Table:PMCa}
\end{figure}

The cases (i) and (ii) of the classification given in Section
\ref{nomenclature} do not involve gauge symmetries.

%
%

In what follows we leave the details of the above metric-like integration scheme
aside, and instead focus on minimal frame-like integration schemes.

\begin{figure}
\begin{tabular}{|l|l|l|l|l|l|l|}
  \hline
   {\rm grade}& $\mR_{_{-1}}$ & $\mR_{_0}$ & $\mR_{_1}$ & $\mR_{_2}$ & $\mR_{_3}$ & $\mR_{_4}$ \\\hline \hline&&&&&&\\[-39pt]
  $g=0$ & --- & $\cdot_\chi$ & $~\YoungU{1}$ & $~\YoungD{1}{1}$ & $~\YoungT{1}{1}{1}$ & $~\YoungQ{1}{1}{1}{1}$ \\[25pt]\hline&&&&&&\\[-39pt]
  $g=1$ & $\cdot_\e$ & $\ba{l}\YoungUbf{1}~_A\\ \YoungU{1}~_C\ea$ & $\ba{l}\YoungD{1}{1}\\ \YoungU{2}~\YoungD{1}{1}~\bullet\ea$ & $\ba{l}\YoungT{1}{1}{1}\\\YoungD{2}{1}~\YoungT{1}{1}{1}~\YoungU{1}\ea$ & $\ba{l}\YoungQ1111\\{} \YoungT211~\YoungQ1111~\YoungD11\ea$ & \\[55pt]\hline&&&&&&\\[-39pt]
  $g=2$ & ---& $\ba{l}\YoungD11\\\YoungU2\ea$ & $\ba{l}\YoungD21~\YoungT111~\YoungUbf1\\[5pt]\YoungU3~\YoungD21~\YoungU1\ea$ & $\ba{l}\YoungD22~\YoungT211~\YoungQ1111~\YoungU2~\YoungD11~\bullet \\[5pt]\YoungD31~\YoungT211~\YoungU2~\YoungD11\ea$ & & \\[45pt]\hline&&&&&&\\[-30pt]
  $g=3$ & --- & $\ba{l}\YoungD21\\\YoungU3\ea$ & $\ba{l} \YoungD31~\YoungD22~\YoungT211~\YoungU{2}~\YoungD{1}{1} \\[5pt]\YoungU4~\YoungD31~\YoungU2\ea$ & & & \\[28pt]\hline
  $g=4$ & --- & $\ba{l}\YoungD31\\\YoungU4\ea$& & & & \\\hline
\end{tabular}
\caption{\small Some entries of the bi-graded triangular module for the
massive spin-1 field in flat spacetime. The $\s^-$-cohomology
contains the massive gauge field $\blacksquare$ at $g=1$, the
massive gauge condition $\bullet$ at $g=1$, the Proca equation
$\blacksquare$ at $g=2$ and the Noether identity $\bullet$ at
$g=2$.}\label{Table:Mass1}
\end{figure}

\subsubsection{\sc Example of massive spin 1 in flat spacetime}

The ``minimal'' unfolded $\miso(1,D-1)$-module of a massive
spin-$1$ field in $\Real^{1,D-1}$ can be obtained by dimensional
reduction of a strictly massless spin-$1$ in one higher
dimension. Equivalently, it may be obtained in a more pedestrian
way by integration of the Weyl zero$\,$-form module. The latter
is spanned by
\bea  \overline{\Th}&=&(1)\ ;\qquad {\Th}_{\a_r}\
=\ (\a+2-r,r-1)\ ,\quad \a\geqslant 1\ ,\quad r=1,2\ . \eea
The
first two levels of the Weyl zero$\,$-form constraint read
\bea
\nabla C_a+e^b\Phi_{ab}+\frac {\overline{M}}2\, e^b\Phi_{a,b}
&\approx & 0 \quad \quad (\a=0)\quad\; , \\
\nabla\Phi_{ab}+e^c\Phi_{abc}+\frac{\overline{M}}{4}e^c
\Phi_{ab,c} -\frac{\overline{M}^2}{(D-1)}\,e_{(a}C_{b)} &\approx
& 0 \quad \quad (\a=1_1)\quad , \\
\nabla\Phi_{a,b}+e^c\Phi_{c[a,b]}+\frac{2\overline{M}}{D-1}e_{[a}C_{b]}
& \approx & 0 \quad \quad (\a=1_2) \quad . \eea
There are no
primary Bianchi identities, while there is a secondary one at the
first level, \emph{viz.} $\nabla_{[a}\Phi_{b,c]}\approx0\,$. Its
integration yields $dA+\frac12 e^a e^b\Phi_{a,b} \approx 0\,$.
Revisiting the zeroth level, its totally anti-symmetric part
reads $\nabla_{[a}C_{b]}+{\overline{M}}\,\nabla_{[a}A_{b]}
\approx 0\,$, which can be integrated using a 0$\,$-form $\chi$,
obtaining
\bea dA+\frac12 e^a e^b\Phi_{a,b} \approx 0
 \ &,&\qquad d\chi+\overline{M}A+e^a C_a\ \approx\ 0 \ .
\label{Massdefo} \eea
The $\s^-$-cohomology is given in Fig. \ref{Table:Mass1}.
For $\overline{M}>0$ the $A$ and $\chi$ fields form the contractible cycle
\bea d\chi+Z&\approx&0\ ,\qquad dZ\ \approx\ 0\ ,\qquad Z\ :=\
\overline{M}A+e^a C_a\ ,\eea
which manifests the massive St\"uckelberg shift symmetry that can be
used to fix the gauge
\bea \chi&\stackrel{!}{=} &0\qquad \Rightarrow\qquad A\ =\
-\frac1{\overline{M}} e^a C_a\ .\eea
One notes that the massive shift symmetry remains well-defined also in the limit
$e^a\rightarrow 0\,$. 

As we shall see in Paper II, the above simple example has a direct generalization to the cases of mixed-symmetry massless fields in constantly curved backgrounds, wherein the dynamical potentials (that would be used in for example a standard first-order action) are the sum of a contractible field plus a remaining term given by background vielbeins contracted into a ``dynamical'' component of the Weyl zero$\,$-form (not necessarily the primary Weyl tensor).


\subsubsection{\sc Some generalities of unfolded integration}\label{Sec:444}


More generally, unfolded integration of \eq{0formcomp} up to
some finite level, say $0\leqslant \a\leqslant\tilde\ell$, introduces a finite
number of variables, say $X^{\bf p_\a}(\Th^{\ast
\a_r})\Th_{\a_r}\in\O^{{\bf p}_{\a_r}}(U)\otimes \Th_{\a_r}$ indicized by
$\a=-N(\tilde\ell),\dots,-1$ and $r=1,\dots,n_\a$, and with form
degrees $0\leqslant p_{\a_r}\leqslant H(\tilde\ell)$ for some finite
maximal degree $H(\tilde\ell)\,$. These variables together with
$\mC^{\bf 0}(\L;\overline M{}^2;\overline\Th)$ can be arranged into
spaces
\bea \mR(\tilde\ell;\L;\overline M{}^2;\overline\Th)&=&\mR^{\bf
H}\cup \mR^{\bf H-\bf 1}\cup \cdots\cup \mR^{\bf 0}\ ,
\\[5pt]
\mR^{\bf p}&:=&\O^{\bf p}(U)\otimes {\cal T}(p)\ ,\quad {\cal
T}(p)|_{\mm}\ =\ \bigoplus_{\a_r\;|\; p_{\a_r}=\,p} \Th_{\a_r}\
,\eea
that can be extended to the triangular module
$\mT=\bigoplus_{q\in\integ}\mR_q$ with $\mR_{_0}:=\mR$ and
$\mR_q=\bigcup_p\mR^{\bf p+\bf q}_q$ where
\bea \mR^{\bf p+\bf q}_q&:=&\O^{\bf p+\bf q}(U)\otimes {\cal T}(p)\
.\eea
Using the notation of Section \ref{Sec:Triang}, the extended
variable $Z_q=\sum_{\a_r} Z_q^{\a_r}\Th_{\a_r}\in \mT^+$ (where
$q\geqslant 0\,$, $Z_{_0}:=X$ and $Z_{_1}:=R\,$) obeys the
linearized equations
\bea R&:=&(\nabla+\s_{_0}(e))X\ \approx\ 0\ ,\qquad Z_{q+1}\ :=\
(\nabla+\s_q(e))Z_q\ \equiv 0\quad \mbox{for $q\geqslant 1$}\ ,\eea
where $\widetilde\nabla\equiv \nabla=d-\ft i2\,\o^{ab}\rho(M_{ab})\,$, $\r(M_{ab})\equiv \rho_q(M_{ab})$ being independent of
$q\,$, and $\s_q(e)=\s^-_q(e)+\s^+_q(e)$. The gauge transformations
with parameters in $\mT^-$ (\emph{i.e.} when $q\leqslant 0\,$, where
$\e_{_0}:=X\,$) read
\bea \delta_\e \e_q&:=& G_q\ =\ (\nabla +\s_{q-1}(e))\e_{q-1}\ ,
\qquad q\leqslant 0\ ,\eea
and $\d_\e(e+\o)\approx 0\,$. The resulting maps
$\s_q:\mR_q\rightarrow\mR_{q+1}$ ($q\in\integ$) have the expansions
\bea \s_q&=&\sum_{p-p'\geqslant -1}(\s_q)^{\bf p+\bf q+\bf 1}_{\bf
p'+\bf q}\ ,\qquad
 (\s_q)^{\bf p+\bf q+\bf 1}_{\bf p'+\bf q}~:~\mR^{\bf p'+\bf q}_{q}\longrightarrow \mR^{\bf p+\bf q+\bf 1}_{q+1}\ ,
\eea
where the ranges of $p$ and $p'$ are determined by ${\rm
deg}~(\s_q)^{\bf p+\bf q+\bf 1}_{\bf p'+\bf q}=p-p'+1\geqslant 0\,$.

The $\mg_{\lambda}$-transformations are represented in $\mT$ by
Cartan gauge transformations $\d_{\x,\L}$ with Killing parameters
obeying $\d_{\x,\L}(e+\o)\approx 0\,$. Thus
\bea \d_{\x,\L} Z_q&=& \ft{i}2 \L^{ab}\rho_q(M_{ab})Z_q+i \x^a
\rho_q(P_a|e) Z_q\ ,
\qquad q\geqslant 1\ ,\\[5pt]\d_{\x,\L} \e_q&=& \ft{i}2\,
\L^{ab}\rho_q(M_{ab})\e_q+i \x^a \rho_q(P_a|e) \e_q\ ,\qquad
q\leqslant 0\ ,\eea
where, as mentioned above, $\rho_q(M_{ab})=\rho(M_{ab})$ are
independent of $q\,$, and
\bea \r_q(P_a|e)&=&i\,{\partial\over\partial e^a}\s_q\ :\quad
\mR_q\rightarrow\mR_q\ \label{rhosigma}\eea
with the expansions
\bea \r_q(P_a|e)&=&\sum_{p-p'\geqslant 0} (\r_q(P_a|e))^{\bf p+\bf
q}_{\bf p'+\bf q}\ ,\qquad (\r_q(P_a|e))^{\bf p+\bf q}_{\bf p'+\bf
q}~:~\mR^{\bf p'+\bf q}_q\ \longrightarrow\ \mR^{\bf p+\bf q}_q\
,\eea
where $\rho(p|P_a):=(\r_q(P_a))^{\bf p+\bf q}_{\bf p+\bf q}$ are
$e^a$-independent representation matrices of form-degree $0\,$,
\emph{i.e.} $\rho(p|P_a):{\cal T}(p)\rightarrow {\cal T}(p)$, and
$\r_q^{\bf p-\bf p'}(p,p'|P_a|e):=(\r_q(P_a|e))^{\bf p+\bf q}_{\bf
p'+\bf q}$ with $p>p'$ are $e^a$-dependent Chevalley--Eilenberg
cocycles of positive form-degree $p-p'\,$. As discussed in Section \ref{Sec:cycles}, if $\l\neq 0$ such maps between two submodules can only exist if at least one of these is infinite-dimensional.

Integration of \eq{rhosigma} yields
\bea \s_q&=&\mu_q-i\int_0^1 dt ~e^a\rho_q(P_a|te)\ =\
\sum_{p\geqslant 0}(\mu_q)^{\bf p+\bf q}_{\bf p+\bf
q}-i\sum_{p-p'\geqslant 0} \ft{1}{ p-p'+1}\,e^a(\r_q(P_a|e))^{\bf
p+\bf q}_{\bf p'+\bf q}\ ,\eea
where thus $(\s_q)^{\bf p+\bf q}_{\bf p'+\bf q}$ with $p\geqslant
p'$ are integrals of the $e^a$-dependence in $(\rho_q(P_a|e))^{\bf
p+\bf q}_{\bf p'+\bf q}$, while
\bea \mu_q(p-1,p)&:=&(\mu_q)^{\bf p+\bf q}_{\bf p+\bf q}~:~ \mR^{\bf
p+\bf q}_q~\longrightarrow~ \mR^{\bf p+\bf q}_{q+1}\ ,\qquad
\mu_q(p-1,p)~:~{\cal T}(p)~\longrightarrow~{\cal T}(p-1)\qquad\eea
are massive integration constants of degree $0\,$ (see for example
the constant $\overline M$ in \eq{Massdefo}).

Cartan integrability amounts to that
\bea (\nabla+\s_{q+1})(\nabla+\s_q)\ \equiv\
0&\Leftrightarrow&\left\{ \ba{l} (\nabla\sigma_{q})+
\sigma_{q+1}\s_q\ \equiv\ \frac{i}{2}\, \lambda^2\,e^a e^b
\rho(M_{ab})\ ,
\\[5pt]
(\s_{q+1}+(-1)^{\s_q}\s_q)\nabla\ \equiv\ 0\ ,\ea\right.\eea
where the background field equations for $\nabla$ and $e^a$ have
been used. This implies that
\bea \s_q &\equiv & (-1)^{1+\s_{q-1}}\s_{q-1}\ =\
(-1)^{q(1+\s_{\!\circ})}\s_{_0}\ = \ \sum_{p-p'+1\geqslant
0}(-1)^{q(p-p')}(\s_{_0})^{\bf p+1}_{\bf p'}
\\[5pt]
&=& \sum_{p\geqslant 0}\Big((-1)^q\mu(p,p+1)-ie^a\rho(p|P_a)\Big)
-i\sum_{p>p'}\ft{(-1)^{q(p-p')}}{ p-p'+1}e^a\rho^{\bf p-\bf
p'}_{_0}(p,p'|P_a|e)\ , \qquad\eea
and that the independent element $\s_{_0}$ must obey the algebraic
equation
\bea \left[(-1)^{\s_{\circ}}\s_{_0}\right]\s_{_0} &\equiv &
\frac{i}{2}\, \lambda^2\,e^a e^b \rho(M_{ab})\ .\eea

The massive integration constants induce a maximal contractible
cycle $\mS_\m(\tilde\ell;\L;\overline{M}^2;\overline\Th)$ with
$e^a$-independent dimension which we refer to as the \emph{massively
contractible cycle}, \emph{viz.}
\bea \left.\mR(\tilde\ell;\L;\overline
M{}^2;\overline\Th)\right|_{\mg_\l}&=&
\mS_\mu(\tilde\ell;\L;\overline M{}^2;\overline\Th)\oplus
\mR'(\tilde\ell;\L;\overline{M}^2;\overline\Th)\ .\eea
The factorization under $\mg_\l$ of $\mR'$ depends on $e^a$
--- since disentangling its contractible cycles requires assumptions about the dual
vector frame $\th^a\,$.

If $e^a$ is non-degenerate then we refer to the maximal chain $\widetilde
\mR({\tilde\ell;\L;\overline{M}^2;\overline\Th})$ of dual submodules
in $\mR'$ as its \emph{potential module}, \emph{viz.}
\bea \left.\mR'(\tilde\ell;\L;\overline
M{}^2;\overline\Th)\right|_{\mg_\l}&=&
\widetilde\mR(\tilde\ell;\L;\overline M{}^2;\overline\Th)\supsetplus
\mC^{\bf 0}(\tilde\ell;\L;\overline{M}^2;\overline\Th)\ ,\eea
which extends up to some form-degree $\widetilde
H(\tilde\ell)\leqslant H(\tilde\ell)\,$. The resulting
$\s^-$-cohomology in $\mR'$ is a set of
\bea \mbox{dynamical gauge potentials}&:&
\left\{\varphi(\L;M^2;\Th)\right\}\ :=\ H^{q=0}(\s^-|\mT)~\cap
~\mR'\ ,\eea
which thus comprise the $\mm$-tensors in $\mR'$ that are
algebraically unconstrained and not subject to any algebraic shift
symmetries on-shell.

\subsubsection{\sc Skvortsov's $\miso(1,D-1)$-modules and obstructed $\L$-deformations}\label{Sec:445}

Recently Skvortsov~\cite{Skvortsov:2008vs} has given an $\miso(1,D-1)$ module
\bea \mR(\L\!=\!0;\Th)&:=&\widetilde
\mR_{_0}(\L\!=\!0;\Th)\supsetplus \mC^{\bf
0}(\L\!=\!0;\overline\Th)\ .\eea
providing an integration scheme that connects a massless Weyl tensor
$C(\overline\Th^\ast)$ in flat spacetime to the doubly
traceless Labastida tensor gauge field $\varphi(\Th^\ast)$ via a potential module
\bea \widetilde\mR_{q=0}(\L\!=\!0;\Th)\ :=\
\sum_{\a=-s_{_1}}^{-1}X^{\bf p_\a}(\Th^{\ast\a})\Th_{\a}\quad \in
\quad \bigoplus_{\a=-s_{_1}}^{-1} \O^{\bf p_\a}(U)\otimes \Th_\a\
,\eea
with $0< p_\a\leqslant p_{_B}\,$.
In the generalized holonomic gauge the dynamical field
$$\varphi(\L\!\!=\!0;\Th^\ast):=\mathbb P_{\Th}\left[i_{\th^{a_1}}\cdots
i_{\th^{a_{p_{_B}}}}X^{\a=-s_{_1}}\right]$$ can be identified as the Labastida field.

The system remains Cartan integrable and the local degrees of
freedom remain unchanged if $\Th_\a$ are replaced by
$\msl(D)$-types for $\a<0$ ($p_\a>0$). The trace parts form a
Cartan integrable subsystem without zero$\,$-form source, whose
contraction leads back to the original minimal system. Prior to
contracting the trace parts one has a dynamical metric-like
$\msl(D)$-tensor gauge field $\varphi(\Th^\ast)$ of the same
shape as the Labastida field. Hence the dynamical field equation
of the extended unfolded system must be the trace-unconstrained
Labastida equations of \cite{Francia:2002aa,Bekaert:2003az,Bekaert:2006ix}.

A key feature of the Skvortsov module is that in form-degrees $p>0$ it consists of
finite-dimensional smallest-type $\miso(1,D-1)$-irreps (see Section \ref{App:0} for
notation)
\bea {\cal T}(p)&:=& {\cal T}^-_{(p+1)}(\Th^-_{[p]})\ ,\qquad p>0\ ,
\label{skvgmod}\eea
where the smallest $\mm$-types $\Th^-_{[p]}$ depend on the overall spin $\Th$ in
accordance with~\cite{Skvortsov:2008vs}.

The smallest-type irrep ${\cal T}^-_{(i)}(\Th^-)$ can be deformed to $\mso(2,D-1)$
tensors iff $\Th^-$ is rectangular and $i={\rm height}(\Th^-)+1\,$. The irreps in
the Skvortsov module fulfil this criterion iff $\Th$ is rectangular, say
$\Th=[s_{_1};h_{_1}]\,$. Then also the twisted-adjoint
module ${\cal T}(\L\!\!=\!0;\overline M{}^2\!\!=\!0;\overline\Th)$ admits an uplift
to a twisted-adjoint $\mso(2,D-1)$-module
${\cal T}(\L;\overline M{}^2_{1};\overline\Th)\,$ with critical mass defined
by \eq{masslessmasses}. Hence there exists a ``vertical uplift'' $\mR'(\L;\Th)$ of
$\mR(\L\!=\!0;\Th)$ that requires only covariantizations
and critical mass terms without changing the field
content~\cite{Bastianelli:2008nm}, and with a smooth reverse
limit\footnote{In particular, if $\Th$ is rectangular of height
$h_{_1}=(D-2)/2$ then $\varphi(\L;\Th)$ is a conformal tensor field
which has a smooth limit from $\L\neq 0$ to $\L=0\,$. }
\bea \mbox{$\Th$ rectangular}&:&
\mR'(\L;\Th)\quad\stackrel{\l\rightarrow 0} {\longrightarrow}\quad
\mR(\L=0;\Th)\ .\eea

If $\Th$ has mixed symmetry, however, then the strictly massless Skvortsov system
cannot be trivially uplifted on its own. Instead, according to the conjecture by Brink, Metsaev and Vasiliev~\cite{Brink:2000ag} there exists a non-trivial extension by massless fields $\{\chi(\Th^{\prime\ast})\}_{\Th^\prime\in\S^1_{\rm
BMV}(\Th)}$ of lower rank such that the direct sum
$\mR_{\rm BMV}:=\bigoplus_{\Th^\prime\in\S^1_{\rm
BMV}(\Th)}\mR(\L=0;\Th^{\prime})\,$, admits a smooth deformation into constantly curved spacetime.

\subsection{\sc Unfolding the BMV conjecture}\label{Sec:BMV}

As found by Metsaev in \cite{Metsaev:1995re,Metsaev:1997nj}, a given
$\mso(D-1)$-spin of shape $\Th$ consisting of $B$ blocks yields $B$
inequivalent massless lowest-weight spaces
$\mD(e^I_{_0};\Th)$ of $\mso(2,D-1)\,$, each having a single singular vector
associated with the $I$th block of $\Th$ ($I=1,\dots,B\,$).
The corresponding Lorentz-covariant and partially gauge-fixed equations of
motion for a gauge field $\varphi(\L;M^2_I;\Th)$ were also given
in~\cite{Metsaev:1995re,Metsaev:1997nj} (the critical gauge-field mass follows the
critical mass $\overline M{}^2_I$ given in \eq{masslessmasses} for its
primary Weyl tensor). The partially massive
nature of the cases with $B>1$ later led Brink, Metsaev and Vasiliev
\cite{Brink:2000ag} to conclude that upon adding St\"uckelberg
fields $\{\chi(\L;\Th')\}_{\Th'\in\S^I_{\rm BMV}(\Th)}$ (associated with all
blocks except the $I$th one) the
resulting extended system must have a smooth flat limit in the
sense of counting local degrees of freedom.

Taking into account also the unitarity issue ---
only $\mD(e^1_{_0};\Th)$ is unitary --- BMV conjectured that
the fully gauge invariant action
$S^{\L}_{I}:=S[\varphi(\L;M^2_I;\Th),\{\chi(\L;\Th')\}]$ should have
the flat-space limit
\bea \mbox{BMV conjecture}&:&S^{\L}_{I}\quad \stackrel{\l\rightarrow
0} {\longrightarrow} \quad \sum_{\Th^{\prime}\in \S^I_{\rm
BMV}(\Th)} (-1)^{\e_{_I}(\Th')}S^{\L=0}[\varphi(\L=0,\Th^{\prime})]\ ,
\qquad\label{FTBMV}\\[5pt]
&&\S^I_{\rm BMV}(\Th)\ =\ \Th|_{\mso(D-2)}\setminus
\S_{{\tiny\mbox{$I^{\rm th}$ block}}}(\Th)\ ,\label{sigmaBMV}\eea
where: (i) $\S_{{\tiny\mbox{$I^{\rm th}$ block}}}(\Th)$ is the
subset of $\Th|_{\mso(D-2)}$ obtained by deleting at least one cell
in the $I^{\rm th}$ block; and (ii) the phase factors
$(-1)^{\e_{_I}(\Th')}$ are all positive iff $I=1\,$.
Group-theoretically, the BMV conjecture implies that
\bea \mbox{BMV contraction}&:& \mD(e^I_{_0};\Th)
\quad\stackrel{\l\rightarrow 0} {\longrightarrow}
\quad\bigoplus_{\Th^{\prime}\in \S^I_{\rm BMV}(\Th)}
(-1)^{\e_{_I}(\Th')}\mD(\L\!\!=\!0;M^2\!\!=\!0;\Th^{\prime})\ .\label{GTBMV}\eea

The dimensional reduction in (\ref{sigmaBMV}) and the fact that the
zero$\,$-forms carry the local unfolded degrees of freedom suggests
the following step-by-step unfolding of the BMV conjecture:

\begin{itemize}

\item[i)] unfold the tensor gauge field $\widehat\varphi(\widehat \Th)$ in
$\Real^{2,D-1}$ and foliate a region of $\Real^{2,D-1}$ with $AdS_D$
leaves of inverse radius $\l=1/L$ and with normal vector field $\x$
obeying $\x^2=-1\,$, which we shall refer to as the radial vector
field;

\item[ii)] set the radial Lie derivative
$({\cal L}_\x+\l\widehat\D)\widehat X=0\,$, where $\widehat\D$ are
scaling dimensions compatible with Cartan integrability, see Sec. \ref{Sec:Folia};

\item[iii)] constrain the shapes $\widehat\Th_{\widehat \a}$
($\widehat\a=0,1,\dots$) in the Weyl zero$\,$-form module $\widehat
\mC^{\bf 0}(\L\!\!=\!0;\overline M{}^2\!\!=\!0;\widehat{\overline{\Th}})$ in accordance with
(\ref{sigmaBMV}), \emph{i.e.} demand their $(p_{_I}+1)$st row to be
transverse to $\widehat\x$ where $p_{_I}=\bar{p}_{_I}-1=\sum_{J=1}^I h_{_J}\,$;

\item[iv)] demonstrate that the unfolded system in anti-de Sitter space time carries the massless degree of freedom $\mD(e^I_{_0};\Th)$ on the left-hand-side of \eq{GTBMV};

\item[v)] take the flat limit without fixing any massive shift
symmetries and show that the resulting unfolded system in flat
space carries the massless degrees of freedom on the
right-hand-side of \eq{GTBMV} and contains the corresponding
$D$-dimensional Skvortsov modules.

\end{itemize}
The above procedure is performed in Paper II.


\section{\sc \large On Local Degrees of Freedom in Unfolded Dynamics}\label{Sec:DOF}

The notion of ``local degrees of freedom'' differs between the standard and unfolded on-shell formulations of field theory. They essentially agree locally for standard propagating dynamical fields with unconstrained Weyl tensors. In this Section we also comment on the role of the cosmological constant and dual Weyl zero$\,$-forms for vertex-operator-like constructs in field theory in higher dimensions.


\subsection{\sc Fibrations and classical observables}\label{Sec:Obs}


The Lie derivatives ${\cal L}_\x=\{d,i_\xi\}$ along vector fields $\xi$ on the base manifold are realized on the constraint surface as Cartan gauge transformations with field-dependent parameters, \emph{viz.}
\bea {\cal L}_\xi X^\a & \equiv & \d_{i_\x(X)} X^\a+i_\xi R^\a\ \approx\ \d_{i_\x(X)} X^\a\ ,\qquad \x^\a(X)\ =\ i_\xi X^\a\ .\label{xiA}\eea
Conversely, by identifying a suitably defined generalized vierbein 1-form $E^A$ in the free differential algebra, a subset of the Cartan gauge symmetries, referred to as the local translations and associated with $\x^A$, can be traded for locally defined Lie derivatives. By furthermore declaring that only globally defined Lie derivatives are actual symmetries of the unfolded system\footnote{By partitioning the unity, any globally defined vector field can be written as a sum of locally defined vector fields with compact support.} it becomes possible to define free differential algebra invariants.

The first part of this definition, that we shall refer to as a \emph{fibration}, consists of a choice of base manifold ${\cal M}$ and a corresponding splitting
\bea X^\a&=&(\Omega^I;E^A,\Phi^{\a_0})\ ,\qquad \e^\a\ =\ (\L^I,\x^A)\ ,\eea
such that:
\begin{itemize}
\item[i)] $E^A=dX^M E_M{}^A:=E^A_{\a^{\bf 1}}(\Phi^{\a^{\bf 0}})X^{\a^{\bf 1}}\in \O^{\bf 1}(U)\otimes \Th^{\ast A}$ for generically invertible $E_M{}^A$, where $X^M$ are local coordinates on ${\cal M}$;
\item[ii)] $\d_{\L}$ form a subalgebra of the algebra of Cartan gauge transformations, referred to as the fiber rotations, with locally defined parameters $\L^I\in \O^{{\bf p}_I-\bf 1}(U)\otimes \Th^{\ast I}$ and fiber connection $\O^I$;
\item[iii)] $(E^A,\Phi^{\a^{\bf 0}})$ transform under fiber rotations in representations with well-defined invariants which we shall refer to, respectively, as the generalized types and fiber invariants\footnote{One sufficient criterion for a representation $\Th$ to have a well-defined quadratic invariant is that $\Th\cong \Th^\ast$ where $\Th^\ast$ is the dual of $\Th$. This can be obeyed for finite-dimensional as well as infinite-dimensional representations.
The latter is the case in Vasiliev's higher-spin gauge theory for symmetric tensor
fields and has been used in \cite{Sezgin:2005pv} to construct zero$\,$-form invariants, see Section \ref{Sec:HSINV}};
and
\item[iv)] the locally defined parameters $\x^A\in \O^{\bf 0}(U)\otimes \Th^{\ast A}$ are induced together with compensating $\L^I$ parameters from globally defined vector fields $\xi\in {\rm Vect}({\cal M})$ as in (\ref{xiA}).
\end{itemize}
In a fibration one can define $p$-form invariants as functions ${\cal C}^{\bf p}={\cal C}^{\bf p}(E,\Phi)$ that are (a) invariant under fiber rotations off-shell, \emph{i.e.} elements in $\O^{\bf p}({\cal M})$; and (b) closed on-shell, \emph{i.e.} $[{\cal C}]\in H^{\bf p}({\cal M})$ modulo $R^\a$, \emph{viz.}
\bea \d_{\L}{\cal C}&=&0\ ,\qquad d{\cal C}\ \approx\ 0\ .\label{globalform}\eea
We refer to an invariant ${\cal C}$ as topological if $d{\cal C}\equiv 0$, and dynamical if $d{\cal C}\equiv\!\!\!\!\!/\ 0$. The local symmetries of the fibration preserve the de Rham cohomology class $[{\cal C}]$ since $\d_\L{\cal C}=0$ and ${\cal L}_\xi{\cal C}\approx d(i_\x {\cal C})$ where $i_\x {\cal C}\in\O^{[p-1]}({\cal M})$ due to (iv). The $p$-form invariants are generalized Noether currents with associated conserved charges given by $\langle \S\mid {\cal C}\rangle:= \oint_\S {\cal C}$ where $\S\in H_p({\cal M})$ (modulo boundary conditions). The charges obey $\d_{\L}\langle \S\mid {\cal C}\rangle=0\approx \d_\xi\langle \S\mid {\cal C}\rangle$, and they are invariant under smooth deformations of $\S$, which is the essence of being a conserved charge. The charges are finite-dimensional integrals even if ${\cal M}$ is infinite-dimensional, though they may diverge on given classical solutions.


\subsection{\sc Local vs ultra-local degrees of freedom}\label{Sec:Loc}


If the unfolded system is Riemannian the generalized vielbein $E^A=(E^a,\cdots)$ and the on-shell system can be examined on a Riemannian submanifold ${\cal M}_D\subseteq {\cal M}$ with vielbein $e^a:=E^a|_{{\cal M}_D}$ (given in some ``frame''). Since $e^a$ appears in the $R^\a$ only through positive powers, the constraints $R^\a\approx 0$ can be analysed perturbatively in a local coordinate chart $U$ following
\begin{itemize}\item[(i)] the local approach based on first solving $\s^-$-cohomology and then analysing the resulting dynamical field equations subject to standard Cauchy initial conditions in $U$ and various boundary conditions on $\partial U$; or
\item[(ii)] the ultra-local approach based on directly integrating $R^\a\approx 0$ in $U$ subject to initial data for the zero$\,$-forms imposed at a point $x\in U$ and suitable gauge functions for the $p_\a$-forms with $p_\a\geqslant 1$.
\end{itemize}
While the former approach is well-adapted to standard Lagrangian formulations of field theory, the latter approach is more natural from the point-of-view of unfolded dynamics.

\subsubsection{\sc Local approach}
Assuming a perturbatively well-defined $\s^-$-cohomology (see Section \ref{Sec:sminus} and \ref{Sec:IntBI})  the variables in $\mR$ thus split into (i) St\"uckelberg fields which can be shifted away; (ii) auxiliary fields which are algebraically constrained; and (iii) dynamical fields which are thus algebraically unconstrained\footnote{The dynamical fields in general sit in $\mm$-types which can always be regrouped into $\msl(D)$-types subject to suitable (or trivial) trace constraints.} fields not subject to any shift symmetries. Let us denote the set of dynamical fields by
\bea {\cal S}_{\rm dyn}&=& \left\{\varphi(\L;M^2;\Th)\right\}\ .\eea
The constraints may also lead to dynamical field equations\footnote{In the unitarizable cases the field equations contain second-order hyperbolic kinetic terms.  Higher-derivative interactions may upset hyperbolicity and blur causality. These properties may, however, resurface eventually at the level of local observables. We thank F.~ Strocchi for illuminating comments on this issue.}.
The resulting on-shell content of $\varphi(\L;M^2;\Th)$ counts as local degrees of freedom only if the unfolding chain of $\varphi(\L;M^2;\Th)$ is sourced by a corresponding Weyl zero$\,$-form $X^{\bf 0}(\L;M^2;\overline\Th)$. If not then $\varphi(\L;M^2;\Th)$ is a frozen dynamical field, as for example the background vielbein in an unfolded rigid theory.

Conversely, a Weyl zero$\,$-form $X^{\bf 0}(\L;\overline M{}^2;\overline\Th)$ may source a set $\{\varphi_{_I}(\L;M^2_I;\Th_{_I})\}_{I=1}^P$ of dynamical fields in various dual pictures. Chiral dynamical fields arise if either $X^{\bf 0}$ is chirally projected or if some Chevalley--Eilenberg cocycle is projected while $X^{\bf 0}$ remains unprojected. The latter mechanism is realized in chiral Vasiliev-type four-dimensional higher-spin gauge theories in Euclidean or Kleinan signatures~\cite{Iazeolla:2007wt}.

\subsubsection{\sc Ultra-local approach}

If $Q^\a(X)={\cal O}(X^2)$, where $X$ comprises all unfolded variables including $e^a$, then it is possible to expand perturbatively around $X^\a=0$. The linearized equations of motion $dX^\a\approx 0$ imply that $X^\a$ carry no local degrees of freedom if
$p_\a>0\,$.
For $p_\a=0\,$, the integration of the field equations leaves us with constant
zero$\,$-forms in each coordinate chart $U\,$. Thus $\left\{X^\a\right\}$ can be reconstructed perturbatively in a coordinate chart $U$ from the initial datum $\left\{\Phi^{\a^{\bf 0}}\right\}|_{x\in U}$ and boundary conditions at $\partial U$.
This method incorporates all local degrees of freedom into $\mR^{\bf 0}$, facilitating the freezing of topological dynamical fields as well as chirality projections and duality extensions.

Given $\{\Phi^{\a^{\bf 0}}|_x\}$ and free boundary conditions a set of exact solutions are $\Phi^{\a^{\bf 0}}=\Phi^{\a^{\bf 0}}|_x$ and $X^{\a^{\bf p}}=0$ for $p>0$, which we refer to as ultra-local gauges.
Non-trivial $p$-forms with $p>0$ are switched on via gauge functions determined to some extent by boundary conditions. In the resulting local gauges the degrees of freedom are shared between $\left\{\Phi^{\a^{\bf 0}}|_x\right\}$ and $p$-forms with $p>0$. The latter to some extent spread the local degrees of freedom over the base manifold, where they can now be recuperated using zero$\,$-form charges as well as ``complementary'' charges of higher form-degree.

Since $p$-form charges $\langle \S\mid {\cal C}^{\bf p}\rangle$ with $p>0\,$
vanish on-shell if $\S\subset U$
(hence $\Sigma$ is trivial in $H_p({\cal M})$), the only locally available classical observables are the zero$\,$-form charges
\bea \langle x|{\cal C}^{\bf 0}\rangle&=&{\cal C}^{\bf 0}(\Phi^{\a^{\bf 0}}|_x)\ ,\qquad x\in U\ ,\eea
where $\langle x|{\cal C}^{\bf 0}\rangle$ is independent of the choice of $x$ on-shell.
Formally, these charges remain invariant under the gauge transformations between ultra-local and local gauges. This motivates the definition of the space of \emph{classical local degrees of freedom} of an unfolded system as
\bea {\cal S}_{\rm loc}&:=&\left\{{\cal C}^{\bf 0}(\Phi): \Phi\in \mR^{\bf 0}\,,\ {\cal C}^{\bf 0}\in{\cal S}_{\rm inv}\right\}\ ,\eea
where ${\cal S}_{\rm inv}$ is the set of all non-factorizable zero$\,$-form invariants.

The zero$\,$-form charges are given by infinite expansions in auxiliary zero$\,$-forms that are given by derivatives of physical fields on-shell. The existence of such charges rely crucially on a non-vanishing massive parameter that can be the cosmological constant but also the mass of a physical scalar field in flat spacetime. The definition of zero$\,$-form charges for strictly massless fields, on the other hand, requires a suitable extension of the Weyl zero$\,$-form by a dual ditto --- the ``vacuum expectation value'' and runaway modes --- to be discussed below.


\subsection{\sc Perturbative ultra-local analysis}


\subsubsection{\sc Expansion in Weyl zero-form, Riemannian and extended symmetries}

To examine how the local degrees of freedom are contained in $\mR^{\bf 0}$, we begin by considering the $\Phi$-expansion using the notation of Section \ref{Sec:Linfty}. Upon fixing the massive shift-symmetries $(\s_{_{-1}})^{\bf 0}_{\bf 0}:\mR^{\bf 0}_{_{-1}}\rightarrow \mR^{\bf 0}_{_0}$ the constraints \eq{Ralfa0} and \eq{Ralfa1} on the Weyl zero$\,$-form $\Phi$ and gauge connection $\widetilde A$ read
\bea d\Phi-i\rho(\widetilde A)\Phi&\approx&{\cal O}(\widetilde A\Phi^2)\ ,\qquad \Phi\in{\mC}^{\bf 0}_{_0}\ :=\ \frac{\mR^{\bf 0}_{_0}}{(\s_{_{-1}})^{\bf 0}_{\bf 0}\mR^{\bf 0}_{_{-1}}}\label{pert1}\\[5pt]
d\widetilde A+\widetilde A^2+\S(\widetilde A,\widetilde A;\widetilde C)&\approx&{\cal O}(\widetilde A^2\Phi^2,B\Phi)\ ,\qquad\widetilde A\in\widetilde\mg \ ,\label{pert2}\eea
where $\rho$ denotes the representation of the gauge Lie algebra $\widetilde \mg$ in ${\mC}^{\bf 0}_{_0}$ and $\widetilde C$ comprises the primary Weyl tensors of the connections in $\widetilde A$ which we assume are the leading zero$\,$-form sources of $\widetilde A$ (taking $\Phi$ and the variables $B$ in higher form-degrees to be weak fields).

For Riemannian systems $\widetilde \mg\cong\mg\subsetplus\mg'$ where $[\mg',\mg']$ may close into itself in which case $\widetilde \mg\cong\mg\oplus\mg'$ and $\mg'$ is an ``internal'' gauge algebra, or with $[\mg',\mg']\cap \mg\neq \emptyset$ in which case $\mg'$ is a non-trivial extension of $\mg$. In the latter case we assume that $[\mg,\mg']=\mg'$, inducing a level decomposition $\widetilde \mg|_{\mg}:=\bigoplus_{\ell}{\cal L}_\ell$ where ${\cal L}_{\ell=0}\cong \mg$. We write  $\widetilde A=\sum_\ell\widetilde A_\ell=\O+A'$ with $\O=\widetilde A_{\ell=0}=e+\o=-i(e^aP_a+\frac12 \o^{ab}M_{ab})$. One has the spin-$(2)$ covariantizations
\bea {\cal D}\Phi&:=&\nabla\Phi-ie^a\rho(P_a)\Phi\ ,\quad {\cal R}\ :=\ d\O+\O^2\ ,\quad {\cal D}A'\ :=\ dA'+ \O A' + A' \O \ ,\label{spin2cov1}\eea
where ${\cal R}:=-i(T^a P_a+\frac12 (R^{ab}+\l^2 e^a e^b)M_{ab})$ with
$T^a:=\nabla e^a=de^a+\o^{ab}e_b$ and $R^{ab}:=d\o^{ab}+\o^{ac}\o_{c}{}^b$, and ${\cal D}A'=dA'$ if $[\mg,\mg']=0$. Correspondingly, the primary Weyl tensor $\widetilde C=\sum_{\ell}C(\L;\overline M{}^2_{I_{\ell}};\overline{\th}_{\ell}):=C(2,2)+C'$ where
$\overline{\th}_{\ell}$ is the $\mathfrak{m}$-type of the primary
Weyl tensor associated with the field $\widetilde A_\ell\,$.

Eqs. \eq{pert1} and \eq{pert2} now read
\bea {\cal D}\Phi-i\rho(A')\Phi&\approx&{\cal O}(\widetilde A\Phi^2)\ ,\\[5pt]
{\cal R}+\mathbb P_{\mg}A^{\prime 2}+\S(e,e;C(2,2))&\approx&{\cal O}
(\widetilde A^2 \Phi^2,B \Phi)\\[5pt]
{\cal D}A'+\mathbb P_{\mg'}A^{\prime 2}+\S(e,e;C')&\approx&{\cal O}(e A' C',\Phi^2\widetilde A^2,B\Phi)\ .\eea

The Weyl zero$\,$-form module decomposes under $\widetilde\mg$ into perturbatively defined $\widetilde\mg$-multiplets, \emph{viz.}
\bea \left.\mC^{\bf 0}_{_0}\right|_{\widetilde\mg}&:=& \bigoplus_{\tiny\ba{c}{}\\[-15pt]\m\in{\cal S}_{\rm mult}\\[-2pt]\mbox{flavors $f$}\ea}\mC^{\bf 0}_{\m,f}+{\cal O}(\Phi^2)\ ,\qquad \mC^{\bf 0}_{\m}\ :=\ \O^{\bf 0}(U)\otimes {\cal T}_{\mu}\eea
where $f$ are external ``flavors'' and ${\cal T}_{\mu}$ are  $\widetilde\mg$-modules. Decomposing further under $\mg\cong \mg_\l$, assuming $\L\neq 0$, yields
\bea \left.\mC^{\bf 0}_{\m}\right|_{\mg}&:=&
\bigoplus_{(\overline{M}^2;\overline\Th ;c)\in\mu}
\mC^{\bf 0}(\L;\overline M^2;\overline\Th ;c)\ ,\qquad
\mC^{\bf 0}(\L;\overline M^2;\overline\Th ;c)\ :=\ \O^{\bf 0}(U)\otimes
{\cal T}(\L;\overline M^2;\overline\Th ;c)\ ,\eea
where $c$ are indices transforming under $\mg'$ and
${\cal T}(\L;\overline M^2;\overline\Th;c)$ are $\mg$-modules with mass $\overline M\,$
smallest $\mm$-type $\overline\Th$ (that may be finite-dimensional or twisted-adjoint)
and index $c\,$. Letting $\mC^{\bf 0}_{\rm \widetilde g}$ denote the direct sum of all $\mC^{\bf 0}_{\m,f}$ containing on-shell curvatures for $\widetilde A$, one has
\bea \left.\mC^{\bf 0}_{\rm \widetilde g}\right|_{\mg}&:=&\Big\{\bigoplus_{{\cal L}_\ell\subset\widetilde\mg}\mC^{\bf 0}(\L;\overline M{}^2_{I_{\ell}};\overline{\th}_{\ell})\Big\}\oplus\Big\{ \bigoplus_{\tiny\ba{c}{}\\[-15pt]\mbox{gauge}\\[-3pt]\mbox{ matter}\\[-3pt](\overline M{}^2;\Th)_{\kappa}\ea} \mC^{\bf 0}(\L;\overline M{}^2;\Th)_{\k}+{\cal O}(\Phi^2)\Big\}\ ,\eea
where the ``gauge matter'' is required for filling out the $\widetilde\mg$-multiplets, and may consist of dynamical fields with higher form-degree and/or higher spin (as for example in the case of the higher-spin gauge theories
in
$D=5$ and $D=7$ with extended supersymmetries considered
in~\cite{Sezgin:2001yf,Sezgin:2002rt}).
One may refer to the unfolded system as unified if $\mC^{\bf 0}_{\rm \widetilde g}$
is an irreducible $\tilde{\mg}$-module and gauge unified if in
addition $\widetilde\mg$ is irreducible.

\subsubsection{\sc Rigid, topological and gravity-like theories}

If all connections in $\widetilde A$ have non-vanishing Weyl tensors then we refer to the model as fully gauged, else partially gauged\footnote{Unified and non-chiral models are fully gauged which requires gravity for Riemannian systems, while \emph{e.g.} Yang-Mills theory is partially gauged since the unfolded background vielbein is frozen.}. In the latter case there exists a split
\bea &\widetilde\mg\ =\ \mg_{\rm top}\oplus\mg_{\rm col}\ ,\quad \widetilde A\ =\ A_{\rm top}+A_{\rm col}\ ,\eea
where we refer to $\mg_{\rm top}$ and $\mg_{\rm col}$ as the topological and
``color'' gauge algebras, respectively, and define
\bea \mbox{rigid and topological models}&:&  \mbox{$\mg\subseteq \mg_{\rm top}$ is non-compact and $\mg_{\rm col}$ is compact}\ ,\\[5pt]
\mbox{gravity-like models}&:& \mbox{$\mg\subseteq \mg_{\rm col}$ is non-compact}\ ,\eea
such that upon treating $\O=e+\o$ as a large field one has
\bea dA_{\rm top}+A_{\rm top}^2&\approx&{\cal O}(e A'_{\rm col}\Phi,\widetilde A^2\Phi^2,B\Phi)\ ,\\[5pt]
dA_{\rm col}+A_{\rm col}^2+\S_{\rm col}(e,e;C_{\rm col})&\approx&{\cal O}(e A'_{\rm col}\Phi,\widetilde A^2\Phi^2,B\Phi)\ ,\eea
where $A'_{\rm col}$ are the components of $A_{\rm col}$ that do not lie in $\mg$. It follows that $A'_{\rm col}={\cal O}(\Phi)$ perturbatively, so that
\bea dA_{\rm top}+A_{\rm top}^2&\approx&{\cal O}(\widetilde A^2\Phi^2)\ ,\qquad dA_{\rm col}+A_{\rm col}^2+\S_{\rm col}(e,e;C_{\rm col})\ \approx\ 0\ .\eea
In the leading order the $\mg_{\rm top}$-valued connection can be frozen locally by going to new variables
\bea \O&\approx&L^{-1}dL+{\cal O}(\widetilde\Phi^2)\ ,\qquad \Phi'\ :=\ \rho(L)\Phi\ ,\label{changevar}\eea
where $L$ is a local gauge function depending on boundary conditions at $\partial U$, and
\bea d\Phi'+\rho(A_{\rm col})\Phi'&\approx{\cal O}(\widetilde A\Phi^2)\ .\label{Phiprime}\eea
In rigid models the gauge function $L$ by definition remains well-defined at higher orders in the $\Phi$-expansion, and one may argue that the space of zero$\,$-form charges is given by $\mg_{\rm col}$-invariants, \emph{i.e.}
\bea \mbox{rigid models}&:& {\cal C}^{\bf 0}(\Phi)\ =\ I_{\rm col}[\Phi']+{\cal O}(\Phi^2)\ ,\qquad I_{\rm col}[\rho(\e)\Phi']\ \equiv\ 0\ \mbox{for all $\e\in\mg_{\rm col}$}\ .\eea
In topological models the gauge function $L$ is by definition obstructed at higher orders in the $\Phi$-expansion, and one may argue that the space of zero$\,$-form charges is given by $\widetilde \mg$-invariants, \emph{i.e.}
\bea \mbox{topological models}&:& {\cal C}^{\bf 0}(\Phi)\ =\ I[\Phi']+{\cal O}(\Phi^2)\ ,\qquad I[\rho(\e)\Phi']\ \equiv\ 0\ \mbox{for all $\e\in\widetilde\mg$}\ .\eea
Assuming that $\widetilde\mg$ is realized in unitarizable Weyl zero$\,$-form modules the extraction of zero$\,$-form charges thus leads to radically different invariant theories:
\bea \mbox{rigid models}&:& \mbox{invariants of finite-dimensional irreps}\ ,\\[5pt]
\mbox{topological/gravity-like models}&:&  \mbox{invariants of $\infty$-dimensional  $\mg$-modules}\ .\eea

We note that in rigid models the zero$\,$-form charges are manifestly $e^a$-independent, while some of the $p$-form charges with $p>0$ such as Noether currents require a non-degenerate vielbein. We also stress that the rigid models are manifestly diffeomorphism invariant prior to freezing the $\mg$-valued connection $\O$.

Physically speaking, the ``confinement'' of ``gravitational colors'' and the resulting  decrease in the number of local degrees of freedom should be a smooth transition from (i) a ``rigid phase'' at low energies in which $\O\approx L^{-1}dL+{\cal O}(\widetilde A\widetilde\Phi^2)$ makes sense for weak spin-2 Weyl tensor and graviton fields, and unconfined gravitational colors show up as particles with mass and spin; via (ii) an intermediate ``softly broken'' phase where still $\O\approx L^{-1}dL+{\cal O}(\widetilde A\widetilde\Phi^2)$ while gravitational colors starts getting confined into $\mg$-invariant $p$-form charges; to (iii) an unbroken phase at high energies in which $\O$ is expanded around $\O=0$ (with a weakly coupled ``dual'' description in terms of the unfolded Poisson sigma model) and all local degrees of freedom are confined into zero$\,$-form charges.

\subsection{\sc Free local degrees of freedom}

A special case of rigid theories are the free limits in which the representation matrices $\rho(\mg_{\rm col})\rightarrow 0$ so that $d\Phi'\approx0$ in the free limit of \eq{Phiprime}. The space ${\cal S}^{\rm free}_{\rm loc}$ of local free degrees of freedom of a linearized unfolded system can thus be defined as the space ${\cal T}^\ast(0)$ of integration constants for its zero$\,$-forms modulo the space of integration constants for the St\"uckelberg zero$\,$-forms, \emph{i.e.} the image $(\s_{_{-1}})_{\bf 0}^{\bf 0}{\cal T}^\ast(0)\subset {\cal T}^\ast(1)$ (see also Eq. \eq{Sberg}). In other words, taking into account what we have discussed so far,
\bea {\cal S}^{\rm loc}_{\rm free}\ \cong \ \frac{{\cal T}^\ast(0)}{(\s_{_{-1}})_{\bf 0}^{\bf 0}{\cal T}^\ast(1)}\ =\ \bigoplus_{(\overline M{}^2,\overline\Th)_{c,f}}{\cal T}^\ast(\L;\overline M{}^2,\overline\Th)_{c,f}\ ,\eea
where we note that the labeling using masses and smallest $\mm$-types is strictly speaking only making sense if $\L\neq 0$ while if $\L=0$ one needs to use additional discrete indices as discussed in Section \ref{Sec:Twadj}. We stress that ${\cal S}^{\rm loc}_{\rm free}$ contains the local degrees of freedom also in local gauges with non-trivial dynamical gauge fields. Thus, in order to establish whether free gauge fields carry unitary representations of $\mg$
\cite{Konshtein:1988yg,Konstein:1989ij}
it suffices, and actually simplifies greatly the analysis, to show that ${\cal S}^{\rm loc}_{\rm free}$ contains a unitarizable representation $\mD$ of $\mg$ as part of its spectral decomposition.


\subsection{\sc Zero-form charges in topological/gravity-like theories and role of $\L$}


In gravity-like and topological models the zero$\,$-form charges are built from invariant functions of the Weyl zero$\,$-form. The invariant theory differs radically between the self-dual ($|\L|+|\overline M{}^2|>0$) and strictly massless ($\L=\overline M{}^2=0$) cases. In the former case the zero$\,$-form charges are non-local functionals of the self-dual Weyl$\,$-zero form while in the latter case they are given by local functionals of the dual Weyl zero$\,$-form which is itself a non-local functional of the Weyl zero$\,$-form.


\subsubsection{\sc Exact zero$\,$-form charges in higher-spin gauge theory}
\label{Sec:HSINV}

Exact zero$\,$-form charges ${\cal C}^{\bf 0}_{{\rm HS};2N;\pm}$ have been
given~\cite{Sezgin:2005pv} for Vasiliev's full higher-spin gauge theories with
higher-spin algebras based on extensions of $\mso(2,3)$. The charges are given by
two types of potentially divergent traces ($\widehat {\rm Tr}_\pm$) of algebraic
powers of the full Weyl zero$\,$-form master field of Vasiliev's system. Similar
charges exist also for the Lorentzian and Euclidean theories based on extensions of
$\mso(1,4)$ and $\mso(5)$~\cite{Iazeolla:2007wt}. The full charges
${\cal C}^{\bf 0}_{{\rm HS};2N;-}$ are finite on at least one exact solution,
namely the $\mso(3,1)$-invariant solution~\cite{Sezgin:2005pv} and its Euclidean
``instanton'' continuation~\cite{Iazeolla:2007wt}, for which they obey the
``coherence'' relation
${\cal C}^{\bf 0}_{{\rm HS};2N;-}=({\cal C}^{\bf 0}_{{\rm HS};2;-})^N$. Their
perturbative weak-field expansion read
\bea {\cal C}^{\bf 0}_{{\rm HS};2N;-}&=&{\rm Tr}\left[(\Phi\star\pi(\Phi))^{N}\right]+{\cal O}(\Phi^{2N+1})\ ,\eea
where the Weyl zero$\,$-form master field $\Phi\in{\cal A}$, an associative unital $\star$-product algebra, and the trace operation ${\rm Tr}:{\cal A}\rightarrow \Comp$ is defined by ${\rm Tr}[X]=X|_{\1}$, the projection to the coefficient of $\1\in{\cal A}$.

\subsubsection{\sc Zero$\,$-form charges for self-dual free fields
($|\L|+|\overline M{}^2|>0\neq 0$)}

The bosonic higher-spin gauge theories generalize to signatures
$(2,D-1)$ and $(1,D-1)$ (and more general signatures as well).
Their unfolded systems admit the free limits
\bea
\Phi\rightarrow \sum_{s=0}^\infty \Phi(\L;s,s)\ ,\qquad
\Phi(\L;s,s)~\in~ \mC^{[0]}(\L;\overline M^2_{1};s,s)\ =\
\O^{\bf
0}(U)\otimes {\cal T}(\L;\overline M^2_{1};s,s)\ ,\eea
where
$\Phi(\L;s,s)$ are Weyl zero$\,$-forms for composite massless
spin-$(s)$ fields, with $s=0$ being the composite massless scalar
with $\overline M{}^2_{1}:=-4\e_{_0}\l^2$. Following the
enveloping-algebra approach to singletons and composite massless
fields~\cite{Angelopoulos:1997ij,Laoues:1998ik,Iazeolla:2008ix}
one has $\Phi\in{\cal A}$ given by
\bea {\cal A}\ \cong \
\frac{{\cal U}[\mg_\l]}{{\cal I}[V]}\ ,\quad V_{AB}\ :=\ \ft12\,
M_{(A}{}^C\star M_{B)C}-\ft{1}{D+1}\eta_{AB} C_{_2}[\mg]\ ,\quad
V_{ABCD}\ :=\ M_{[AB}\star M_{CD]}\ ,\eea
where ${\cal I}[V]$
is the two-sided ideal\footnote{One has ${\cal I}[V]\cong{\cal
I}[\mD(\e_{_0};(0))]$, the annihilator of the scalar singleton
$\mD_0=\mD(\e_{_0};(0))$ ($\e_{_0}=(D-3)/2$). The spectral
decomposition of the twisted-adjoint action on ${\cal A}$
contains the Flato- Fronsdal spectrum plus additional
compact-weight states forming a larger indecomposable module
\cite{Iazeolla:2008ix}.} generated by $\{~V_{AB}~,~V_{ABCD}~\}$
and $\star$ denotes the product in ${\cal U}[\mg]$ (reserving
juxtaposition for the symmetrized product). The twisted-adjoint
action is given by
\bea \r(Q) \Phi&=&Q\star
\Phi-\Phi\star\pi(Q)\ ,\qquad \pi(X\star Y)\ =\
\pi(X)\star\pi(Y)\quad\forall\; Q,X,Y\in {\cal U}[\mg]\ ,\eea
where
the automorphism $\pi$ is defined by $\pi(P_a):=-P_a$ and
$\pi(M_{ab})=M_{ab}$. It follows that
 \bea {\cal A}|_{\r(\mg)}&=&\bigoplus_{s=0}^\infty{\cal T}(\L;\overline
 M{}^2_{1};s,s)\ \cong \ \bigoplus_{s=0}^\infty{\cal T}^\ast(\L;\overline
 M{}^2_{1};s,s)\ ,\eea\\[-45pt]
 \bea \Phi(\L;s,s)&:=&\sum_{n=0}^\infty \frac{i^n}{n!}\phi^{a(n+s),b(s)}
 \Th_{a(n+s),b(s)}\ ,\\[5pt]
 \Phi^\ast(\L;s,s)&:=&\sum_{n=0}^\infty
 \frac{i^n}{n!}\phi^\ast_{a(n+s),b(s)} \Th^{\ast a(n+s),b(s)}\ ,\eea
with representation matrices
\bea \rho(P_a)\Th_{\a}&=&\l^2 (\hat \r^+_a)_{\a}^{(\a+1)}\Th_{(\a+1)}
+ (\hat \r^-_a)_{\a}^{(\a-1)}\Th_{(\a-1)}\ ,\label{types}\\[5pt]
\rho^\ast(P_a)\Th^{\ast\a}&=&- (\hat \r^-_a)_{(\a+1)}^{\a}\Th^{\ast(\a+1)}-\l^2
(\hat \r^+_a)_{(\a-1)}^{\a}\Th^{\ast(\a-1)}\ ,\label{dualtypes}\eea
and canonical inner products
\bea k_{\a\b}&:=&(\Th_\a,\Th_\b)_{\cal T}\ =\ \l^{-2n}\hat{\cal N}_{\a\b}\ ,
\qquad k^{\ast\a\b}\ :=\ (\Th^{\ast\a},\Th^{\ast\b})_{{\cal T}^\ast}\
=\ \l^{2n}\hat{\cal N}^{\ast\a\b}\ ,\eea
where $(\hat \r^+_a)_{\a}^{(\a+1)}$, $(\hat \r^-_a)_{\a}^{(\a-1)}$ and
$\hat{\cal N}^{\ast}_{\a\b}:=\left[\hat{\cal N}^{\ast \cdot\cdot}(\eta\ldots \eta)_{\ldots}\right]_{\a\b}=\hat{\cal N}_{\a\b}$ (with indices lowered by
$k_{\a\b}$) are independent of $\l\,$. {}From \eq{types} and \eq{dualtypes} it
follows that ${\cal T}(\L)\cong{\cal T}^\ast(\L)$ by the equivariant map
$\Th_\a\rightarrow \l^{-2n}\Th^\ast_\a\,$. On-shell
$\phi^{a(n+s),b(s)}\approx\mathbb{P}_{(n+s,s)}\nabla_{a_1}\cdots
\nabla_{a_n}C_{a(s),b(s)}\,$.

In the free limit the full zero$\,$-form charges ${\cal C}^{\bf 0}_{{\rm HS};2N}$
``fragmentize'' into elementary charges
\bea {\cal C}^{\bf 0}_{\L,{\rm free}}(s_{_1},\dots,s_{2N})&:=&
{\rm Tr}\left[\Phi(\L;s_{_1},s_{_1})\star\pi(\Phi(\L;s_{_2},s_{_2}))\star\cdots\star
\pi(\Phi(\L;s_{2N},s_{2N}))\right]\ .\label{eleminv}\eea
The quadratic charges can be identified as
${\cal C}^{\bf 0}_{\L;{\rm free}}(s,s)=(\Phi(\L;s,s),\Phi(\L;s,s))_{{\cal T}}\,$,
that immediately generalizes to
\bea \mbox{self-dual module ${\cal T}(\L;\overline M{}^2;\overline \Th)$}&:&
{\cal C}^{\bf 0}_{{\rm free};2}(\Phi)\ =\
k_{\a\b}\phi(\Th^{\ast\a})\phi(\Th^{\ast\b})\ .\eea
The higher-order invariants in \eq{eleminv} encode additional structure
coefficients of the algebra ${\cal A}$ and are related to correlators
$\langle V_{\Phi(\L;s_1,s_1)}\cdots V_{\Phi(\L;s_{2N},s_{2N})}\rangle$ in a
topological open string \`a la Cattaneo-Felder~\cite{Cattaneo:1999fm}
in the phase-space of the scalar singleton, providing a microscopic
framework for Vasiliev's oscillator formalism~\cite{Engquist:2005yt}.
What constitute the corresponding data for
general self-dual modules is an interesting problem.

\subsubsection{\sc Strictly massless limit}

In the case of free composite massless fields the quadratic zero$\,$-form charges
have the expansions
\bea {\cal C}^{\bf 0}_{\L,{\rm free}}(s,s)&=&\sum_{\a=0}\l^{-2n} \hat I_{n}
(s)\phi_{a(n+s),b(s)}\phi^{a(n+s),b(s)}\ ,\eea
for $\l$-independent $\hat I_{n}(s)$. These charges do not have a smooth flat
limit. More generally, one can see that
${\cal C}^{\bf 0}_{\L,{\rm free}}(s_{_1},\dots,s_{2N})$ have no flat limit.
Indeed, it is well-known that what we refer to as strictly massless
Weyl zero$\,$-forms do
not admit any perturbatively defined zero$\,$-form
charges, see e.g.~\cite{Brandt:1989et,Barnich:1995ap} and references therein.

\subsubsection{\sc Dual Weyl zero$\,$-forms in strictly massless cases }\label{Sec:dualWeyl}

Physically speaking, eqs. \eq{types} and \eq{dualtypes} shows that $\l\rightarrow 0$ is the ``strongly coupled'' limit of the oscillator realization of ${\cal T}$, in the sense that the classical part of the $\star$-product is scaled away, while it is at the same time the ``weakly coupled'' limit of the oscillator realization of its dual ${\cal T}^\ast$. Indeed, the dual zero$\,$-form charges
\bea {\cal C}^{\bf 0\ast}_{\L,{\rm free}}(s_1,\dots,s_{2N})&:=&{\rm Tr}\left[\Phi^\ast(\L;s_1,s_1)\star\cdots\star\pi(\Phi^\ast(\L;s_{2N},s_{2N}))\right]
\label{eleminv2}\eea
have finite flat limits ${\cal C}^{\bf 0\ast}_{\L\!=0,{\rm free}}(s_1,\dots,s_{2N})$. This suggests that strictly massless systems should be extended by dual Weyl zero$\,$-forms
 \bea &\Phi^\ast(\L\!\!=\!0)\ =\ \sum_{\a_r} {i^\a\over \a!} \phi^\ast(\Th_{\a_r}) \Th^{\ast\a_r}\ \in\ \O^{\bf 0}(U)\otimes {\cal T}^\ast(\L\!\!=\!0;\overline M{}^2\!\!=\!0;\overline\Th)\ ,&\\[5pt]& (\nabla-ie^a\rho^\ast(P_a))\Phi^\ast(\L\!\!=\!0)\ \approx\ 0\ .&\eea
Any non-factorizable $\mm$-invariant function $I_{1\dots N}:\overline \Th^\ast_{(1)}\otimes \cdots\otimes \overline \Th^\ast_{(N)}\rightarrow \Comp$ yields an elementary dual zero$\,$-form charge (``vacuum expectation value'')
\bea {\cal C}^{\ast\bf 0}_{\L \! = 0;{\rm free};I}(\Th_1,\dots,\Th_N)&=& I_{1\dots N}\left[\phi^\ast_0(\Th_1),\dots,\phi^\ast_0(\Th_N)\right]\ ,\qquad \nabla \phi^\ast_0(\Th)\ \approx\ 0\ .\label{VEVinv}\eea

For example, in the scalar sector $\Phi^\ast(\L\!=\!0;\overline M{}^2\!\!=\!0;(0))= \sum_{n=0}^\infty{i^n\over n!} \phi^\ast_{a(n)}\Th^{\ast a(n)}$, the elementary invariant ${\cal C}^{\bf 0\ast}_{\L\!\!=\!0;{\rm free}}(\1)=\phi^\ast_0$, where $\phi^\ast_0$ has the transformation rule $\d_\x\phi^\ast_0=0$ under local translations.
The physical scalar $\phi$ and the dual scalar $\phi^\ast$ obey
\bea \nabla^2 \phi&\approx 0\ ,\qquad \nabla\phi^\ast\ \approx 0\ .\eea
The on-shell content of $\Phi(\L\!\!=\!0)$ and $\Phi^\ast(\L\!\!=\!0)$ that is regular in ${\cal M}':=\Real^{1,D-1}\setminus\{x^2=0\}$ reads
\bea \phi^{a(n)}&\approx&\sum_{m=0}^\infty A^{a(n)b(m)} D_{b(m)}+ \sum_{m=0}^\infty  {\widetilde D}^{a(n)b(m)} \widetilde A_{b(m)}\ ,\\[5pt]
\phi^\ast_{a(n)}&\approx&\sum_{m=0}^n \widetilde A^{\ast}_{\{a(m)}D_{b(n-m)\}}+ \sum_{n=0}^\infty \D_{a(n)b(m)}A^{\ast b(m)}\ ,\eea
where (i) the harmonics $D_{a(n)}$ and ${\widetilde D}_{a(n)}$ are smooth functions in ${\cal M}'$ obeying $\nabla^{\phantom{(}}_{a\phantom{\{}}\! D_{b(n)}= \eta^{\phantom{(}}_{a\{b}D_{b(n-1)\}}$ and $\nabla_a \widetilde D_{b(n)}=\widetilde D_{ab(n)}$ and $D^{(n)}_{b(n)}|_{0}=0$ and
$\widetilde D_{b(n)}|_{\infty }=0$, reducing in Cartesian coordinates to
$D_{a(n)}\sim x_{\{a_1}\cdots x_{a_n\}}$ and $\widetilde D_{b(n)}\sim (x^2)^{-\ft12 n -\widehat\e_{_0}} x_{\{a_1}\cdots x_{a_n\}}$ where $\widehat\e_{_0}=\ft12(D-2)$; (ii) $\D_{a(n)}$ are distributions that are singular at $x^2=0\,$, their domain consisting
of functions that are smooth at $x^2=0$; and (iii) the coefficients $\{~A^{a(n)}\,\}$, $\{~\widetilde A^{a(n)}\,\}$, $\{~A^{\ast a(n)}\,\}$ and $\{~\widetilde A^{\ast a(n)}~\}$ are four  sets of integration constants spanning four separate $\miso(1,D-1)$ modules prior to taking into account any boundary conditions.

We propose to maintain the self-duality for $\L\neq 0$ in the
flat limit by defining
\bea &{\cal T}(\L)\ \cong\ {\cal
T}^\ast(\L)\quad \stackrel{\l\rightarrow 0}{\longrightarrow}\quad
\widehat {\cal T}(\L\!\!=\!0)\ :=\  {\cal T}(\L\!\!=\!0)~\uplus~
{\cal T}^\ast(\L\!\!=\!0)\ ,\eea
where ${\cal
T}(\L\!\!=\!0):=(\uplus_\a \Th^{a(n)})\uplus (\uplus_n
\widetilde \Th^{a(n)})$ and ${\cal
T}^\ast(\L\!\!=\!0):=(\uplus_n \Th^{\ast a(n)})\uplus
(\uplus_n \widetilde \Th^{\ast a(n)})$ and the dual pairing is
to be derived starting from
\bea \Th^\ast_0(S)&:=& \mbox{the
vacuum expectation value of $\phi_S$}\ ,\quad S~\in~{\cal
T}(\L\!\!=\!0)\ ,\label{VEVproposal}\eea
 where $\phi_S$ is the
field obtained by superposing the above mode functions with
coefficient $S\in {\cal T}(\L\!\!=\!0)$. The pairing
\eq{VEVproposal} is a ``strong-coupling'' relation in the sense
that the right-hand-side requires taking the (Euclidean)
$r\rightarrow \infty$ limit of $\phi_S$ starting from the
``initial datum'' $S$.

Physically speaking, one may think of a collection of mode functions
constituting a compact weight-space module of the form
\bea \left.{\rm Spec}~\widehat
{\cal T}(\L\!\!=\!0)\right|_{\mg_0}\ \cong \
{\mathfrak{W}}~\supsetplus~\mD~\supsetplus~{\mathfrak{U}}\ ,\qquad (\mD)^\ast\
\cong\ \mD\ ,\quad ({\mathfrak{W}})^\ast\ \cong\ {\mathfrak{U}}\
,\label{selfdual1}\eea
 where ${\rm Spec}~\widehat {\cal
T}(\L\!\!=\!0)$ and $({\rm Spec}~\widehat {\cal
T}(\L\!\!=\!0))^\ast$, respectively, carry the module structures
of the space of spacetime mode functions and the dual space of
polarization tensors times creation/annihilation operators (these
types of quantities thus carry Lorentz indices and compact
weights transforming in dual representations of $\miso(1,D-1)$).
One expects that (i) $\mD$ consists of normalizable wave-packages
given by superpositions of plane-waves $T_{a(n)}(p)\sim
p_{a_1}\cdots p_{a_n}$ with $p^2=0$; (ii) ${\mathfrak{W}}$ consists
of runaway solutions including the vacuum solution $\phi\approx
\phi_0$; and (iii) ${\mathfrak{U}}$ consists of singular solutions
including the static ``Coulomb-like'' solution $\phi \approx
r^{-2\e_{_0}} V_0$ ($\e_{_0}:=\ft12(D-3)$). A mathematical
argument for \eq{selfdual1} would consist of (1) identifying a
static ground state in ${\mathfrak{W}}$ from which ${\rm Spec}~\widehat
{\cal T}(\L\!\!=\!0)$ is generated by means of the
$\mg_0$-action; (2) use this action to define the canonical
bilinear form
$(\cdot,\cdot)_{{\rm Spec}~\widehat {\cal T}(\L \! = 0)}$;
(3) show that this form is non-degenerate on
${\mathfrak{W}}$ and vanishes on $\mD\,$.

In \cite{Iazeolla:2008ix} the analog of the above proposal for
composite massless fields in $AdS_D$ was examined in more detail,
and it was found that indeed
${\cal T}(\L<0)\cong {\mathfrak{W}}_{\L<0}\supsetplus \mD_{\L<0}$ where
$\mD_{\L<0}$ are the ``electric'' and
``magnetic'' lowest/highest-weight spaces (see next Section) and
${\mathfrak{W}}_{\L<0}$ is a ``lowest-spin'' module that is
unitarizable at least for composite massless scalar fields.


\subsubsection{\sc On zero$\,$-form charges in gravity with $\L\neq 0$ and $\L=0$}


Given the existence of zero$\,$-form charges in higher-spin gauge theory, it is natural to ask whether ${\cal C}^{\bf 0}_{\L;{\rm free}}(s_1,\dots,s_{2N})$ with all $s_i\leqslant 2$ admit perturbative corrections in the presence of gravity-like self-interactions, and if so, whether the resulting charges assume finite values on exact solutions. We propose that for systems of scalars and vectors interacting with gravity with finite $\L$ there exist sets of zero$\,$-form charges,
\bea \L\neq 0&:& {\cal S}^{\L}_{{\rm loc}}\ =\ {\cal S}^{\L;{\rm s=0}}_{{\rm loc}}~\cup~ {\cal S}^{\L;{\rm s=1}}_{\rm loc}~\cup~{\cal S}^{\L;{\rm s=2}}_{\rm loc}\ ,\eea
obtainable by perturbative ``dressing'' of the free-field zero$\,$-form charges given in \eq{eleminv}.

One may also entertain the idea that systems of the above kind with $\L=0$ admit non-trivial extensions by (interacting) dual Weyl zero$\,$-forms supporting sets of zero$\,$-form charges:
\bea \L=0&:& {\cal S}^{\L\!=\!0}_{{\rm loc}} \ =\ \left\{~{\cal C}^{\bf 0}_{\L=0;{\rm VEV}}(\Phi^\ast)\,\right\}~\cup~\left\{~{\cal C}^{\bf 0}_{\L=0;{\rm mixed}}(\Phi^\ast,\Phi)\,\right\}\ ,\eea
where ${\cal C}^{\bf 0}_{\L=0;{\rm VEV}}(\Phi^\ast)$ are obtainable by perturbative dressing of the invariants given in \eq{VEVinv}, and ${\cal C}^{\bf 0}_{\L=0;{\rm mixed}}(\Phi^\ast,\Phi)$ by perturbative dressing of the free-field duality relation $d (\Phi^\ast_{\rm free},\Phi_{\rm free})\approx0$ and other higher-order (non-factorizable) multi-linear forms $dI_{\rm mixed}(\underbrace{\Phi^\ast_{\rm free},\dots,\Phi^\ast_{\rm free}}_{\mbox{$N$ entries}};\underbrace{\Phi_{\rm free},\dots,\Phi_{\rm free}}_{\mbox{$N$ entries}})\approx 0$.

%
\subsection{\sc Spectral decomposition and harmonic expansion}
\label{Sec:Harm}

\subsubsection{\sc General set-up}

The spectral decomposition of a twisted-adjoint module ${\cal
T}(\L;\overline M{}^2;\overline \Th)=:{\cal{T}}|_{\mm}$ is an equivariant map
${\cal S}^{-1}$ from its defining $\mm$-covariant basis
$\left\{\Th_{\a_r}\right\}$ to a basis $\{\ket{\l}\}$ consisting
of $\mh$-types where $\mh$ is a compact subalgebra of $\mg_\l$.
To find the $\mh$-types one first reduces further under
\bea
\mh&\rightarrow &\ms\ :=\ \mh\cap\mm\ ;\qquad {\cal T}|_{\mm}\
\rightarrow\ {\cal T}|_{\ms}\ ;\qquad \l\ \rightarrow\ (\n,\th)\
,\eea
where $\ms$ is the spin-algebra; $\th$ are the common
spin labels of $\mm$ and $\mh$; and $\n$ is a complete set of
eigenvalues characterizing the representation $\mh/\ms$ on the
subspaces of ${\cal T}|_{\ms}$ with fixed spin $\th$. The maximal
compact subalgebras are
\bea \mbox{self-dual case
($|\L|+|\overline M{}^2|>0$)}&:& \mh\ =\ \left\{\ba{ll}
\mso(2)_E~\oplus~\mso(D-1)_{\ms}&\mbox{$\L\leq 0$}\ ,\\[5pt]
\mso(D)'_{J}&\mbox{$\L>0$}\ ,\ea\right.\\[5pt] \mbox{strictly
massless case ($|\L|=\overline M{}^2=0$)}&:& \mh\ =\
\mso(2)_E~\oplus~\mso(D-2)_{\ms}\ ,\eea
  where $E:=P_{_0}$,
$\ms$ is generated by $M_{rs}$ and $\mso(D)'$ is generated by
$J_{mn}=(M_{rs},P_r)$, and we note that $\mh$ is the maximal
compact subalgebra in the self-dual cases.

The $\mh$-types resulting from the spectral decomposition span a
$\mg_\l$-module ${\mathfrak{M}}$ referred to as the compact-weight
space, \emph{viz.}
\bea {\cal S}^{-1}&:&{\cal T}|_{\mm}\
\longrightarrow\ {\mathfrak{M}}\ :=\ {\cal T}|_{\mh}\ :=\
\bigoplus_\S{\mathfrak{M}}_{\S}\ ,\quad \left.
{\mathfrak{M}}_{\S}\right|_{\mh}\ =\ \bigoplus_{\l} \Comp\otimes
\ket{\l}_{\S}\ .\eea
where ${\mathfrak{M}}_{\S}$ are subspaces
forming separate $\mg_\l$-irreps (upon factoring out the
complement of ${\mathfrak{M}}_\S$ in ${\mathfrak{M}}$). We shall assume that
each ${\mathfrak{M}}_\S$ contains a reference state $\ket{\l^\S_0}_\S$,
referred to as the ground state, such that ${\mathfrak{M}}_\S$ is the
orbit of $\ket{\l^\S_0}_\S$ under ${\cal{U}}[\mg_\l]$. This state
generation is more straightforward for $\L\neq 0$ than for $\L=0$
since in the former case each ${\mathfrak{M}}_\S$ consists of a discrete
set of compact weights (while for fixed $\overline M{}^2$ and
$\overline\Th$ the labels $\S$ generically belong to a continuum
even some further assumptions have been made).

Assuming that ${\mathfrak{M}}_\S$ has a component, say $\,\theta_0^{\Sigma}\,$,
in $\overline{\Th}\vert_{\mathfrak{s}}\,$, the reference state can be chosen to be
\bea \ket{\l^\S_0}_\S&:=& \ket{\n^\S_0;\th^\S_0}_\S\ =\ f^\S_{\l^\S_0} \star
(\th^\S_0 |\overline\Th) \ ,\qquad f^\S_{\l^\S_0}~\in~{\cal{U}}^\S[\mg_\l]\ ,\eea
where (i) $(\th^\S_0 |\overline\Th)\in {\cal T}\vert_{\mm}$ is the projection of
$\overline\Th|_{\mathfrak{s}}$ onto the $\ms$-subtype
$\th^\S_0$; and (ii) $f^\S_{\l^\S_0} $, the spectral (reference) function of the
sector ${\mathfrak{M}}_\S$, belongs to an analyticity class
${\cal{U}}^\S[\mg_\l]$ of
${\cal{U}}[\mg_\l]$. These classes are nonpolynomial completions of
${\cal{U}}[\mg_\l]$  modulo
right-multiplication by the annihilator of $(\th^\S_0|\overline\Th)\,$,
into classes of
operators with symbols (defined by the symmetrized Poincar\'e--Birkhoff--Witt
product in ${\cal{U}}[\mg_\l]$)
given by analytic functions such that
\bea Q \star f&\in&{\cal{U}}^\S[\mg_\l]\qquad \mbox{for all
$Q\in{\cal{U}}[\mg_\l]$
and $f\in{\cal{U}}^\S[\mg_\l]$}\ .\label{embclass}\eea
Non-analyticity can only arise in enveloping-algebra variables that are $\ms$-singlets
since these are not protected
against becoming raised to fractional or negative integer powers
by the assumption that $\mathfrak{M}_{\Sigma}$ consists of $\ms$-types.
We refer to
$f^\S_{\l^\S_0}$ as (i) regular if its symbol is regular at $M_{AB}=0$ in which
case all states in ${\mathfrak{M}}_\S$ are reached from $(\th^\S_0|\overline\Th)$
by the action of
regular spectral functions; and (ii) irregular if its symbol is non-analytic at
$M_{AB}=0$. The orbit of an irregular spectral function may contain regular
$\mg_{\l}$-submodules giving rise to indecomposability.
The converse is not true, \emph{i.e.} orbiting a regular reference state may also yield
indecomposability~\cite{Iazeolla:2008ix}.
In the case of the scalar field, in general ${\mathfrak{M}}$ contains also sectors
${\mathfrak{M}}_\S$ whose reference states
are obtained by applying a spectral function to an $\ms$-tensor contained
in a descendant
$\Th_{\a_r}\in{\cal T}$ with $\a>0$~\cite{Iazeolla:2008ix}.

The idea is to diagonalize the action of the generators in $\mh/\ms$ using a set of
sectors ${\mathfrak{M}}_\S$ that is ``complete'' according to the (vaguely stated)
complementarity principle introduced in Section \ref{Sec:dualWeyl}.
Thus, prior to imposing any form of boundary conditions and/or reality conditions
on the Weyl zero$\,$-form, the complexified compact weight space is an
indecomposable $\mg_\l$-module.
Assuming the original twisted-adjoint module ${\cal T}|_{\mm}$
to be self-dual it is natural to seek a corresponding self-dual
compact-weight space (\emph{cf.} Eq. \eq{selfdual1}), \emph{viz.}
\bea {\widehat {\mathfrak{M}}}|_{\mg_\l}&=& {\mathfrak{W}}~\supsetplus~\mD~\supsetplus~{\mathfrak{U}}\ ,\qquad
({\mathfrak{W}})^\ast\ \cong\ {\mathfrak{U}}\ ,\quad \mD^\ast\ \cong\ \mD\ ,\label{selfdual2}\eea
where $\mD$ contains particles/anti-particles and ${\mathfrak{W}}$ and ${\mathfrak{U}}$ complementary sectors (runaway/singular solutions) --- so that one may view the indecomposability as an enveloping-algebra analog of the Unruh effect.

To be more precise, the aforementioned notion of completeness means that there should
exist an inverse of the spectral decomposition, called the harmonic expansion
\bea {\cal S}&:=& \bigoplus_{\S} {\cal S}^{{\cal T}}_{\S}\ ,\qquad {\cal S}^{{\cal T}}_{\S}\ :\ {\mathfrak{M}}_{\S}\ \longrightarrow\ {\cal T}\ ,\eea
whereby the Weyl zero$\,$-form becomes expanded as
$$X^{\bf 0}(\L;\overline M{}^2;\overline\Th|\S;L)=
{\cal{S}}^{\cal T}_{\S}\left[\sum_{\l\in{\mathfrak{M}}_\S} X^{\l}_{\S}
L^{-1}\ket{\l}_\S\right]\ ,\quad L:{\cal{M}}_D\rightarrow \mathfrak{g}_{\lambda}/\mathfrak{m}\ ,\quad e+\o~=~L^{-1}dL\,,$$ with component fields given by
\bea X^{\bf 0}(\Th^{\ast\a_r}|\S,L)&=&\sum_{\l\in{\mathfrak{M}}_\S} X^{\l}_{\S}~ D^{\a_r}_{\l,\S}(L)\ ,\qquad D^{\a_r}_{\l,\S}(L)\ :=\ \Theta^{*\a_r}
\left[L^{-1}\,{\cal {S}}^{\cal T}_{\Sigma}\ket{\l}_\S\right]\ ,\eea
where (i) $X^{\l}_{\S}$ are constants (to become creation and annihilation operators for states in the sector ${\mathfrak{M}}_\S$ upon second quantization); and (ii) $D^{\a_r}_{\l,\S}(L)$ are generalized spherical harmonics carrying $\mm$-indices $\a_r$ as well as compact indices $\l$. These reduce to polarization tensors times plane waves when $\L=0\,$ for a subset of the ${\mathfrak{M}}_\S\,$ (namely, in $\mathfrak{D}\,$).

The generalized spherical harmonics require embeddings of the $\mm$-tensors $\Th_{\a_r}$ into ${\mathfrak{M}}_\S$, which amounts to an embedding function $\Psi_{(\th^{\prime\S}_0|\overline\Th)}$ such that
\bea \ket{\th^{\prime\S}_0|\overline\Th}_\S&:=&\Psi_{(\th^{\prime\S}_0|\overline\Th)} \ket{\n^{\prime\S}_0;\th^{\prime\S}_0}_\S\ :=\ \sum_{\nu}\Psi_{(\th^{\prime\S}_0|\overline\Th);~\nu}~\ket{\nu;\th^{\prime\S}_0}_\S\ ,\qquad \Psi_{(\th^{\prime\S}_0|\overline\Th)}\in {\cal{U}}_\S[\mg_\l]\ ,\label{embedding}\eea
where (i) $\th^{\prime\S}_0$ is an $\ms$-subtype of $\overline\Th$
;
(ii) $\ket{\nu^{\prime\S}_0;\th^{\prime\S}_0}_\S$ is corresponding reference state in
${\mathfrak{M}}_\S$ (that need not be the ground state); (iii) $\{\ket{\nu;\th^{\prime\S}_0}_\S\}$ is a basis for all states in ${\mathfrak{M}}_\S$ of $\ms$-type $\th^{'\Sigma}_0$; (iv) $\Psi_{(\th^{\prime\S}_0|\overline\Th);~\nu}$ are complex coefficients (that can always be taken to be real by a choice of basis); and (v)
${\cal{U}}_\S[\mg_\l]$ is the analyticity class of the embedding function (whose definition is an analog of \eq{embclass}).

We stress that the requirement of an embedding is a necessary criterion for determining whether a given module ${\mathfrak{M}}_\S$ arises in a spectral decomposition of a given ${\cal T}$. Sufficient criteria requires a deeper understanding of the boundary conditions and related complementarity issues that we have touched upon above.


\subsubsection{\sc The case of $\L<0$}


In the case of $\L<0$ (in what follows $\mg:=\mso(2,D-1)$) the
spectral decomposition of ${\cal T}(\L;\overline M{}^2;\overline
\Th)$, and the corresponding harmonic expansion of the primary
Weyl tensor $C(\overline M{}^2;\overline\Th)$, first requires
that one assigns the lowest $\mm$-type a definite $\pi$-parity,
\emph{viz.}
\bea \pi(\overline\Th)&=&(-1)^{\e_\pi(\overline
\Th)}\overline\Th\ ,\qquad \e_\pi(\overline\Th)\ \in\ \{0,1\}\
.\eea
The $\mh$-types, that we denote by $T^\S_{\nu;\th}$, are
then the solutions to
  \bea \rho(E) T^\S_{\nu;\th}& =& \nu\; T^\S_{\nu;\th} \ ,\qquad T^\S_{\nu;\th}\ :=\
\rho(f^\S_{\nu;\th}(E)) (\th|\overline\Th)  \ ,\quad
f^\S_{\nu;\th}(E)\ =\ \sum_{n\;\in\; p_0+{\mathbb{N}}} f^\S_{\nu;\th;n} E^n\
,\label{energycond}\eea
 where (i) $(\th| \overline\Th)$ is the
embedding of the $\ms$-plet $\th$ into the smallest $\mm$-type $\overline\Th\in
{\cal T}$ (ignoring the special case arising for the scalar field as noted
above) containing $\th$; (ii) $f^\S_{\nu;\th}(E)$ is the
spectral function with analyticity class determined by $p_0$
($p_0=0$ for regular spectral functions).

Drawing on the results of \cite{Iazeolla:2008ix} for composite
massless scalars, we expect at least two independent solutions
$f^\S_{\nu;\th}(z)$ for each fixed $\th$ and $\nu\in\Comp$, say
$N_{\rm reg}$ regular and $N_{\rm irreg}$ irregular solutions.
For every $\mh$-type with energy $\nu$ there is a corresponding
$\mh$-type with energy $-\nu$ obtained by applying the $\pi$-map,
that is
\bea
T^\S_{-\nu;\th}&=&(-1)^{\e^\S(\nu;\th)}~\pi(T^\S_{-\nu;\th})\ =\
(-1)^{\e^\S(\nu;\th)+\e_\pi(\overline\Th)}
~\rho(f^\S_{\nu;\th}(-E))~(\th|\overline\Th)\ ,\eea
 where
$\e^\S(\nu;\th)$ depends on the normalizations of
$T^\S_{\pm\nu;\th}$ and monodromies in the $E$-plane that arise
for non-integer $p_0$. There is also the parity
\bea
\e\left(T^\S_{\nu;\th}\right)&:=& |\th|+[({\rm Re}\,\nu)]\quad
\mbox{mod $2$}\ ,\eea
that is preserved by the action of
regular elements in ${\cal{U}}[\mg]$. Thus, restricting to real
$\mu$, one has
\bea \mbox{$\L<0$}&:& {\mathfrak{M}}|_{\mg}\ =\
\int_0^1 d\mu \bigoplus_{{\e=\pm}}\left({\mathfrak{M}}^{\rm
reg}_{\m;\e}\subsetplus {\mathfrak{M}}^{\rm irreg}_{\m;\e}\right)\
,\eea
where $\mu\in[0,1[$ labels a continuum of sectors in
which $\mu~:=~\nu-[\nu]$.

For each value of $\overline M{}^2$, $\overline \Th$ and $\th$ there is a special value of $\mu$ for which the compact weights $T^{\S_\m}_{e^\pm_0;\th_0}$ with energies ($\e_{_0}:=\ft12(D-3)$)
\bea e^\pm_{_0}&=&1+\e_{_0}\pm \D_{_0}\ ,\quad  \D_{_0}\ :=\ \sqrt{(1+\e_{_0})^2+C_{_2}[\mg_\l|\overline M{}^2;\overline \Th]-C_{_2}[\ms|\th]}\eea
are candidate lowest-weight states (and their image under $\pi$ are candidate highest-weight states). {}From the embedding condition \eq{embedding} it follows that if $C(\L;\overline M{}^2;\overline\Th)$ is massive then it contains
$(1+\pi)\mD(e^\pm_{_0};\overline \Th)^+$ (see Paper II for a detailed analysis). One refers to $\mD( e_{_0};\overline \Th)$, which is a unitary module, as the physical lowest-weight space, and $\mD(\widetilde e_{_0};\overline \Th)$ as its shadow. The former space contains the mode-functions obeying Dirichlet conditions at the boundary of $AdS_D$ while the latter space contains the mode-functions obeying Neumann conditions.

The physical module can be embedded, or glued, into the
shadow module by an element in a suitable analyticity class
${\cal U}^C[\mg]$ of ${\cal U}[\mg]$. For example, the physical
lowest-energy state $\ket{ e_{_0};(0)}$ of a scalar field can be
reached from the lowest-energy state of its shadow as follows
($x~:=~ \d^{rs} L^+_r L^+_s$):
\bea &\ket{e_{_0};(0)}\ =\
x^{\D_{_0}}\ket{\widetilde e_{_0};(0)}\ ,\quad \D_{_0}\ =\
\sqrt{(1+\e_{_0})^2+L^2 \overline M^2}\ .&\eea
The above gluing
generalizes to arbitrary spins as (using Howe-dual notation) %
\bea \ket{e_{_0};\overline\Th}&=&\sum_{p} \sum_{\tiny\ba{c}\\
[-15pt]\{p_i\}\\ [-2pt]\sum_i
p_i=p\ea}f^{e_{_0};\overline\Th}_{\{p_i\}} x^{\D_{_0}-p}\prod_i
(L^{+(i)}L^+_{(i)})^{p_i}\ket{\widetilde e_{_0};\overline\Th}\
,\label{glue}\eea
where the coefficients are fixed by $L^-_r
\ket{e_{_0};\overline\Th}=0$. One notes that the above
transformation is regular for special masses (which are in
general not related to the critical masses). Thus, in the above
sense, one has
\bea \mbox{massive case}&:& {\mathfrak{M}}\ \ \supset
\ \mD_{C}\ :=\ (1+\pi)\left[\mD(e_{_0};\overline
\Th)^+~\subsetplus^C~ \mD(\widetilde e_{_0};\overline
\Th)^+\right]\ ,\eea
 where $\subsetplus^C$ refers to the
analyticity class ${\cal U}^C[\mg]$ defined by the \eq{glue}.

In the critically massless limits, there arises a primary Bianchi
identity, say in block $I$, and $C(\L;\overline
M{}^2_I;\overline\Th)$ develops a vanishing (multiple) curl below the $I$th
block.
Its  integration yields a gauge field $\varphi(\L;
M{}^2_I;\Th)\,$. The $\mm$-types are
\bea
\Th &=& \Big( [s_{_1};h_{_1}],\dots,[s_{_B};h_{_B}]\Big)\ ,
\label{fieldtype}
\\[5pt]
\overline\Th &=& \Big( [s_{_1};h_{_1}],\dots , [s_{_{I-1}};
h_{_{I-1}}],[s_{_I};h_{_I}+1],[s_{_{I+1}};h_{_{I+1}}-1],[s_{_{I+2}};h_{_{I+2}}],\dots ,
[s_{_B};h_{_B}]\Big) \ .
\label{Weyltype}
\eea
Correspondingly, singular vectors appear in
$\mD(e^\pm_{_0};\Th)$, resulting in the extended module
structure\footnote{We are thankful to E.~Skvortsov for
illuminating discussions of this issue.}:
\bea
&\mD_{C,\varphi}\ :=\ (1+\pi)\left[\mD(e^{I,{\rm
gauge}}_{_0}+1;\overline \Th')^+ \subsetplus \mD(e^{I,{\rm
el}}_{_0};\Th)^+\subsetplus^C \mD(e^{I,{\rm magn}}_{_0};\overline
\Th)^+\subsetplus \mD(\widetilde e^{I,{\rm
el}}_{_0};\Th)^+\right]\ ,\qquad\label{extmodule}\eea
where
$\Th'$ is obtained by deleting one cell from the $I$th block of
$\Th$ and the energy levels are given by ($p_{_I}\ :=\
\sum_{J=1}^I h_{_J}$):
\bea \mbox{``gauge'' LWS}&:& e^{I,{\rm
gauge}}_{_0} \ =\ s_{_I}+D-1-p_{_I}\ , \\[5pt] \mbox{``electric''
LWS}&:& e^{I,{\rm el}}_{_0} \ =\ s_{_I}+D-2-p_{_I}\ ,
\label{MetsCritMass}\\[5pt] \mbox{''magnetic'' LWS}&:& e^{I,{\rm
magn}}_{_0}\ =\ 1+p_{_I}-s_{_{I+1}}\ ,\\[5pt] \mbox{``shadow''
LWS}&:& \widetilde e^{I,{\rm el}}_{_0} \ =\ 1-s_{_I}+p_{_I}\
.\eea
The electric spaces are unitary iff $I=1$ and the
magnetic spaces are non-unitary for all $I$ except for sporadic
cases with low spin in dimensions $D=4$ and $D=5$.

As we shall see in Paper II, the resulting harmonic expansions read:
\bea \mbox{critically massless Weyl tensor}&:& \mD_{C}\ =\ (1+\pi)\left[\mD(e^{I,{\rm el}}_{_0};\Th)^+\subsetplus^C \mD(e^{I,{\rm magn}}_{_0};\overline \Th)^+\right]\ ,\\[5pt]
\mbox{critically massless gauge field}&:&\mD_{\varphi}\ =\ (1+\pi)\left[\mD(e^{I,{\rm gauge}}_{_0};\overline \Th')^+  \subsetplus^C \mD(\widetilde e_{_0};\Th)^+\right]\ .\eea
One may also speculate that the structure is part of a
generalization of the compact weight-space analog of the spacetime
$\s^-$-$\,$cohomology found in \cite{Iazeolla:2008ix,Metsaev:2008fs}.
We defer further details to future work \cite{WIPZhen}.

The harmonic expansion of the Weyl zero$\,$-form thus contains mode-functions with three types of complementary asymptotic behaviors: (i) fall-off/runaway behaviors at the boundary
of spacetime, (ii) singular behaviors close to a point and (iii) periodicity in
time. Boundary conditions are linear combinations of (i) and (ii) enforced by ``gluings'' of power series expansions in various Euclidean distances. These boundary conditions correspond
to finiteness of various combinations of conserved charges.

The case of composite massless fields was examined in \cite{Iazeolla:2008ix}. It was found that runaway mode-functions with divergent Noether charges fill
modules ${\mathfrak{W}}_{C}$, referred to in \cite{Iazeolla:2008ix} as lowest-spin modules, in which the energy operator is unbounded from above \emph{and} below. These modules contain static ground states generating the indecomposable structure $\mD_{C}\subsetplus \mW_{C}$. It was found that the non-degenerate bilinear form on ${\mathfrak{M}}_{C}$ (in which ${\mD_C}$ are null states) unitarizes $\mW_{C}$ at least for the composite scalar.

It was also argued that the states in ${\mathfrak{M}}_{C}$ have finite zero$\,$-form charges (except for the static ground state whose zero$\,$-form charges are logarithmically divergent). The particle-like states in $\mD_C$, on the other hand, have divergent zero$\,$-form charges whose regularization requires a map to projectors in the quantum-mechanical model defining the fiber part of Vasiliev's equations.
It was proposed in \cite{Iazeolla:2008ix} that this regularization method may make sense in the full higher-spin gauge theory, such that the zero$\,$-form charges of \cite{Sezgin:2005pv} have the following properties
\bea \mbox{higher-spin models}&:& \mbox{${\cal C}_{[0]}$ converge for perturbative initial data in $\mW$ and $\mD$}\ ,\label{HSprop}\eea
and one may further speculate that if a specific lower-spin model can be embedded into a higher-spin model by a nonlinear consistent truncation then also
\bea
\mbox{lower-spin models}&:& \mbox{${\cal C}_{[0]}$ converge for perturbative initial data in $\mW$ and $\mD$}\ .\eea
Physically speaking, the standard Noether charges (obtained from Noether potentials 
in the case of a gauge theory) are sensitive to the fall-off behaviour at the boundary while zero$\,$-form charges are more sensitive to the local derivatives at a point. Thus the latter may be formally divergent and require a regularization
when evaluated on the solutions in $\mD$ which have to ``curve'' faster in the bulk than runaway solutions in order to fall off at the boundary to yield finite Noether charges. The runaway solutions, on the other hand, curve more slowly in order to render the zero$\,$-form charges finite, and hence do not fall off fast enough at the boundary leading instead
to infinite Noether charges.

The above proposal also rhymes well with the fact that higher-spin gauge theories have local interactions that are ``exotic'' (see \cite{Boulanger:2008tg} for a recent discussion) in the sense that their canonical perturbative expansion is given by a derivative expansion headed by ``top-vertices'' covered by inverse powers of $\L$ whose regularization also seems to require the algebraic form of the interactions provided by Vasiliev's formulation.

\section{\sc \large Conclusion}
\label{Sec:Conclusions}

In the present Paper I we discussed some properties of unfolded
dynamics that will be used in the companion Paper II in which
we derive the equations of motion for free tensor fields in
$AdS_D\,$, thereby providing an unfolded formulation of the BMV
mechanism.

In the present paper we already provided the group-theoretic
structure for the twisted-adjoint Weyl module associated with
arbitrary-shaped tensor fields propagating in $AdS_D\,$. In other words, in terms
of Lorentz-covariant Harish-Chandra module of the non-compact
$AdS_D$-algebra, we worked out the structure of the
infinite-dimensional module associated with the generalized Weyl
tensors. An explicit oscillator realization is given in Paper II,
where we explicitly integrate the zero$\,$-form Weyl module using
appropriate modules in higher form degrees.

We have also discussed the notion of local degrees of freedom realized in
unfolded dynamics as vertex-operator-like algebraic functions of the
Weyl zero$\,$-form and its dual.
Their constructions for mixed-symmetry fields goes beyond the scope
of Paper II and we plan to return to it later.

\section*{Acknowledgments}
We are thankful to K.~Alkalaev and E.~Skvortsov for enlightening comments and
E.~Skvortsov for several useful remarks on an early version of this paper. We also
wish to thank G.~Barnich, F.~Bastianelli, X.~Bekaert, A.~Campoleoni, P.P.~Cook,
J.~Demeyer, F.A.~Dolan, J.~Engquist, D.~Francia, M.~G\"unaydin, S.~Leclercq,
T.~Poznansky, A.~Sagnotti, E.~Sezgin, D.~Sorokin, Ph.~Spindel, F.~Strocchi,
E.~Tonni, M.A.~Vasiliev and P.~West for discussions.
A.~Sagnotti is also thanked for his encouragements and support.
This work is supported in part by the EU contracts
MRTN-CT-2004-503369 and MRTN-CT-2004-512194
and by the NATO grant PST.CLG.978785. The research of P.S. was supported in part by
Scuola Normale Superiore and by the MIUR-PRIN contract 2007-5ATT78

\providecommand{\href}[2]{#2}\begingroup\raggedright\endgroup


\end{document}